%% file: main.tex
\input{preamble.tex}

\usepackage{xcolor}

\title{Massive de Sitter correlators at one loop}
\author{Amara McCune}
\affiliation{Anthropic STEM Fellows Program, San Francisco, CA, USA}
\emailAdd{amaramccune@gmail.com}
\abstract{Fermions and massive fields charged under an unbroken symmetry couple to the inflaton only in pairs, so their signals in primordial correlators begin at one loop. Templates for these signals are mostly late-time estimates. Starting from a representation of massive de~Sitter correlators as finite Mellin--Barnes integrals, we compute the exact squeezed coefficients of these signals and their corrections away from the squeezed limit. The leading coefficient factorizes into a hard part and a soft vertex. Late-time estimates treat the hard part exactly but approximate the soft vertex by its late-time limit, whereas we compute both exactly. Without a chemical potential, the exact coefficient differs from the late-time one by a universal factor that depends only on the masses in the loop and the coupling at the soft vertex. For the heavy Standard Model loops, the exact signals are ten to thirty times larger than their late-time estimates. With a chemical potential, the ratio depends on the model, and at our benchmarks it reaches a factor of thirty for helical vectors and about a hundred for the fermion triangle, where we agree with recent exact results. Away from the squeezed limit the leading term soon fails, mainly through a drift in the phase of the oscillation, which we derive for scalar and fermion loops.}

\begin{document}
\maketitle
\raggedbottom

\input{sec1.tex}

\input{sec2.tex}
\input{sec3.tex}

\input{sec4.tex}

\input{sec5.tex}

\input{secR.tex}

\input{sec7.tex}

\input{sec8.tex}

\input{sec9cmp.tex}

\input{sec9.tex}

\appendix
\input{appccsun.tex}

\input{appB.tex}

\input{appC.tex}

\input{appDirac.tex}

\input{appE.tex}

\input{bib.tex}
\end{document}

%% file: preamble.tex
\documentclass[a4paper,11pt]{article}
\usepackage{jheppub}
\usepackage{amsmath,amssymb,amsthm,mathtools}
\usepackage{bm}
\usepackage{mathrsfs}
\usepackage{graphicx}
\usepackage{booktabs,array,multirow}
\usepackage{enumitem}
\usepackage{xcolor}

\numberwithin{equation}{section}
\theoremstyle{plain}

\theoremstyle{definition}

\theoremstyle{remark}

\newcommand{\dd}{\mathrm{d}}
\newcommand{\ii}{\mathrm{i}}

\renewcommand{\Re}{\operatorname{Re}}

\DeclareMathOperator{\Li}{Li}

\newcommand{\bl}{\bm{\ell}}
\newcommand{\bk}{\mathbf{k}}

%% file: sec1.tex
\section{Introduction}\label{sec:intro}

Primordial correlators are sensitive to physics at the Hubble scale of inflation. Fields with masses of order $H$ are produced by the expanding background, and their exchange leaves an imprint on the correlators of the inflaton that is often called the cosmological collider signal~\cite{ref1,BaumannGreen2012,NoumiYamaguchiYokoyama2013,ref2,ref3,ArkaniHamed2020,Craig2024}. The imprint is cleanest in the squeezed limit, where one momentum, $k_3$, is much smaller than the other two ($k_1\approx k_2$). There a field of mass $m$ contributes a term proportional to $(k_3/k_1)^{3/2\pm\ii\mu}$, with $\mu=(m^2/H^2-\frac94)^{1/2}$. Such a term is non-analytic in the momentum ratio (it oscillates in $\ln(k_3/k_1)$), and measuring it would give direct access to the mass and spin of the heavy field. Its size, however, is suppressed by the Boltzmann factor $e^{-\pi\mu}$.

For many well-motivated fields the leading signal does not come from a single exchange. If a field is charged under a symmetry that remains unbroken, every coupling to the inflaton contains it twice. The same holds for a fermion, which enters only through bilinears, so the signals of both first appear at one loop. The Standard Model fields are an example, as long as electroweak symmetry remains unbroken during inflation~\cite{ref4,ref22}. If it is broken, the Higgs can be exchanged at tree level, and the other Standard Model fields become heavy~\cite{ref21,ref76,ref77,ref79,ref81}. The heavy states of grand unification are another~\cite{Kumar2019,ref73}. Realistic templates for such fields require one-loop diagrams with several massive lines, often with spin or with a chemical potential that lifts the Boltzmann suppression~\cite{ref17,ref23}. Most of these templates are late-time estimates, in which the massive lines are replaced by their leading behavior outside the horizon~\cite{ref4,ref78,ref17,ref20,AokiStrumia2026}. This approximation has known limitations. Already at tree level it gives the wrong power of the mass in the signal~\cite{ref17}, and numerical and exact loop calculations differ from it~\cite{ref80,ref5,ref25}. The size of the error, however, has not been determined in general. This matters because searches of the cosmic microwave background and of galaxy surveys fit these templates to the data~\cite{Meerburg2017,Sohn2024,Cabass2025,Kumar2026planck,Kumar2026scalars,Philcox2026dissect,Philcox2026lim}. For fermion loops with a chemical potential, Ref.~\cite{AokiStrumia2026} recently found the signal to be orders of magnitude below earlier estimates~\cite{ref20,ref21,ref23}, and the bubble and the triangle have since been computed exactly~\cite{HanLeiQinYin2026,Song2026}. When the vertex of the soft leg has no derivatives, the signal vanishes entirely~\cite{SilentHeavy2026}.

The part of a one-loop correlator that is non-analytic in the soft momentum, the nonlocal signal, follows from tree-level data through a factorization theorem~\cite{ref9,ref43,Tong2022}. The rest of the correlator is less constrained. It contains a local signal and a background, and both depend on the masses and couplings of every line in the loop. In this work we compute one-loop correlators beyond their nonlocal signals. Loops in inflation have a long history~\cite{Weinberg2005,SenatoreZaldarriaga2010,MarolfMorrison2010a}, and a variety of methods have recently been brought to bear on massive loops, including the spectral decomposition~\cite{ref5}, dimensional regularization~\cite{ref6}, split representations~\cite{ref7}, on-shell and integration-by-parts techniques~\cite{ref13,ref14}, Feynman-like parameterizations~\cite{Herderschee2025}, Euler integrals continued from the sphere~\cite{KatsevichSun2026}, and direct numerical integration~\cite{ref80,ref16}. For conformally coupled internal lines the triangle is even known in closed form~\cite{ref11,ref12}, and such lines can also be treated with dressing rules and differential equations~\cite{ref47,Chowdhury2026,Sachs2026}. At tree level, massive correlators are by now under good control~\cite{ArkaniHamed2020,Sleight2020,SleightTaronna2020,ref53,Liu2025dispersive,ref52,ref51,Werth2024,Gasparotto2025,Baumann2026diffeq,Grafe2026gluing}.

Our starting point is the representation we developed in Ref.~\cite{ref26}, which is built on the propagator of Sleight and Taronna written as a Mellin--Barnes integral~\cite{SleightTaronna2021a,ref34}. With it, tree-level and one-loop wavefunction coefficients become Mellin--Barnes integrals whose contours do not depend on the masses. There we used it to compute the bubble and the triangle directly at physical momenta. Here we take the representation further and ask three questions of each one-loop diagram and each type of line. First, which pieces does the diagram contain? Second, what weight does each piece receive in the correlator that is observed? Third, when can the coefficients of the pieces be written in closed form? Finally, we ask how large the squeezed and collapsed signals are, and when the late-time templates get them right. We find that the templates treat the hard part of a diagram correctly, but not its soft vertex. Here the hard part consists of the vertices that carry the large momenta, and the soft vertex is the one at which the soft external leg ends. For lines that carry no chemical potential the error is a single factor, which we give in closed form and which is between 10 and 31 for the Standard Model loops of heavy fields (Table~\ref{tab:summary}). A chemical potential removes this universality, and the ratio in Table~\ref{tab:summary} then ranges from $0.12$ to $113$. Beyond the squeezed limit, even the exact leading term soon fails.

The paper is organized as follows. Section~\ref{sec:block} collects the two ingredients that all of our results rely on, the loop block and the weight of each piece in the correlator. Section~\ref{sec:bubble} treats the bubble. At equal masses it is known to split into a nonlocal signal, a local signal, and a background, each in closed form~\cite{ref5}. At unequal masses both signals are known, but the background has been left with an undetermined constant~\cite{ref10}. We give all three pieces at any two masses (heavy or light) and at general vertex powers, as single convergent sums over tree-level exchanges, and we show that renormalization fixes the background with no free constant. The renormalized bubble at two masses has very recently also been written as a spectral integral~\cite{Bassett2026}. We close the section with spinning lines.

In Section~\ref{sec:close} we turn to a second, complementary route. For any diagram with two vertices, the loop integrals reduce to one integral over the invariant distance between the vertices, against an elementary kernel. Such a reduction is known for loops without external legs~\cite{MarolfMorrison2010a,ref8}, and we extend it to correlators at general kinematics (for a spectral route to the same kernel, see Ref.~\cite{Bassett2026}). It gives an independent evaluation of the bubble and shows when the coefficients have closed forms. The same steps give the massive two-loop sunset, whose leading signal follows from factorization~\cite{ref43}, and banana diagrams at higher loops.

Section~\ref{sec:triangle} adds a third and a fourth vertex. The massive triangle has been studied at equal masses~\cite{ref15} and integrated numerically~\cite{ref16}, and Ref.~\cite{ref26} gives its pair-frequency signal at general masses to all orders in the squeezed limit. We extend this signal to every shape. The box has closed forms only in limits, and its collapsed limit takes the factorized form of Ref.~\cite{ref9}.

The remaining sections ask how large the resulting signals are, and what happens when the line carries a chemical potential or is a fermion. In Section~\ref{sec:regions} we show that the leading coefficient of a squeezed or collapsed signal separates into a hard factor and the soft vertex, and that late-time estimates are exact in the first but not in the second. This was recently observed for the fermion triangle~\cite{Song2026}, and it holds for any one-loop diagram. Three cases then arise. When both ends of every soft line are hard (as in the collapsed limit), the late-time coefficient is exact. When the soft lines end on the soft vertex and the lines respect the de Sitter boosts, the late-time estimate misses by a universal factor, which we give in closed form for lines of any spin. In Section~\ref{sec:SM} we apply it to the Standard Model background of the cosmological collider~\cite{ref4,ref78}, that is, to the loops of the Standard Model fields themselves, and we also obtain their bispectra at every shape. When a chemical potential breaks the boosts, no such factor exists, and the soft vertex must be computed. We then follow each signal to larger $k_3/k_1$. When the lines carry a chemical potential, even the exact leading term is accurate to ten percent only for $k_3/k_1\lesssim0.08$, and the error is largely a phase. If the chemical potential of a fermion is large, the signal oscillates instead in the ratio of $k_3$ to the total energy. We derive this from the form of the fermion modes, for the bubble and for the dominant part of the triangle. Section~\ref{sec:compare} compares the shapes of all of the loops and their reach in surveys.

A chemical potential lifts the Boltzmann suppression, and it changes the weights of the pieces but not their exponents. The loop of helical vectors is among the largest loop signals proposed~\cite{ref17}, and a natural source of a parity-odd trispectrum. At tree level, with scale invariance and the Bunch--Davies vacuum, the parity-odd scalar trispectrum vanishes for scalar exchanges and is factorized in general~\cite{CabassParity2023,StefanyszynTongZhu2024}. At one loop it need not vanish, and the loop is then the leading parity-odd contribution~\cite{LeeMcCullochPajer2023}. Parity-odd trispectra are being searched for in galaxy surveys and in the cosmic microwave background~\cite{HouSlepianCahn2022,Philcox2022,Philcox2023cmb,HouDESI2025}. The helical loop has been computed at leading late-time order~\cite{ref17}, as a series for the enhanced helicity~\cite{ref18}, and numerically~\cite{ref80,Niu2022gw,Niu2022tri}. In Section~\ref{sec:chem} we obtain its pair-frequency signal exactly, summed over all nine helicity pairs and with the hard helical-vector line restored at every shape, and we give its collapsed trispectrum in closed form. We also give the charged-scalar loop with a chemical potential on its vertex~\cite{ref19} in closed form, and its bispectrum at every shape. Fermion loops have been computed at leading late-time order~\cite{ref20,ref24,AokiStrumia2026} (for their use as a probe of leptogenesis, see Ref.~\cite{CuiXianyu2022}), and without a chemical potential their signals are known exactly~\cite{ref25}. With an axial chemical potential, the bubble~\cite{HanLeiQinYin2026} and the triangle~\cite{Song2026} have just been computed exactly. In Section~\ref{sec:dirac} we obtain both independently, with the soft vertex in position space. The late-time soft vertex misses the triangle by up to two orders of magnitude and the bubble by up to one, and we extend the triangle beyond the squeezed limit.

Table~\ref{tab:summary} compares the published templates for these models with our exact results. In the collapsed trispectrum both ends of each soft line are hard, so its late-time coefficient is exact (Section~\ref{sec:regions.split}).

\begin{table}[!htbp]
\centering
\footnotesize
\renewcommand{\arraystretch}{1.15}
\begin{tabular}{@{}>{\raggedright\arraybackslash}p{0.23\textwidth}>{\raggedright\arraybackslash}p{0.25\textwidth}>{\raggedright\arraybackslash}p{0.26\textwidth}>{\raggedright\arraybackslash}p{0.17\textwidth}@{}}
\toprule
Loop & Template & Parameters & $|C_{\rm exact}/C_{\rm template}|$\\
\midrule
Higgs portal (Sec.~\ref{sec:SM}) & late-time~\cite{ref78} & $\mu_H=\frac12$, $1$, $2$, $4$; $\nu_H=1.45$ & $10.1$, $11.6$, $14.8$, $20.3$; $1.03$\\
$W/Z$ (Sec.~\ref{sec:SM}) & late-time, as in Ref.~\cite{ref78}\textsuperscript{a} & $\mu=1$, $2$, $4$ & $11.6$, $14.8$, $20.3$\\
Top quark (Sec.~\ref{sec:SM}) & late-time, first term\textsuperscript{b} & $m=1$, $2$ & $26$, $31$\\
Charged scalars (Sec.~\ref{sec:chem.vertex}) & stationary phase at large $\lambda$~\cite{ref19} & $(\lambda,\mu)=(40,4)$, $(20,2)$, $(10,1)$ & $0.98$, $0.93$, $0.83$\\
\midrule
Helical-vector bubble (Sec.~\ref{sec:chem.bubble}) & late-time~\cite{ref17} & $\kappa=\mu=1$, $2$, $3$ & $6.1$, $16.7$, $30.0$\\
Helical-vector triangle (Sec.~\ref{sec:chem.hard}) & late-time, hard line contracted~\cite{ref17} & eight points from $(\mu,\kappa)=(2,3)$ to $(20,20.3)$ & $0.12$ to $11.5$\\
Fermion bubble (Sec.~\ref{sec:dirac.bubble}) & late-time pair term\textsuperscript{c} & $(\mu,\kappa)=(0.5,0.25)$, $(1,0.5)$, $(1,1)$, $(2,2)$ & $13$, $8.9$, $2.5$, $0.41$\\
Fermion triangle (Sec.~\ref{sec:dirac.triangle}) & late-time~\cite{AokiStrumia2026} & $(\mu,\kappa)=(1,1.3)$, $(1,5)$, $(1,10)$, $(2,20)$ & $2.4$, $37$, $113$, $67$\\
\bottomrule
\end{tabular}
\caption{Published templates against our exact squeezed coefficients, with masses and chemical potentials in units of $H$. The last column is the modulus of the exact coefficient over that of the template. \textsuperscript{a}The late-time coefficient computed as in Ref.~\cite{ref78}, whose printed coefficient differs from ours by a rational function of the mass (Section~\ref{sec:SM.wz}). \textsuperscript{b}The template of Ref.~\cite{ref78} has no term at the leading power, so the ratio is that of the first term to its late-time estimate. \textsuperscript{c}No template has been published, so we use the late-time pair term of Section~\ref{sec:dirac.bubble}.}
\label{tab:summary}
\end{table}

We conclude in Section~\ref{sec:concl}, and the longer formulas are collected in the appendices. Throughout, we have checked the closed forms against independent numerical evaluations wherever possible, and we quote the level of agreement alongside each result. The code that reproduces the numbers of Sections~\ref{sec:regions}--\ref{sec:compare} is part of Ref.~\cite{ref28}.

%% file: sec2.tex
\section{Building blocks}\label{sec:block}

In this section we collect the two ingredients on which the rest of the paper relies. The first is the loop block, the Gamma-function factor into which the loop momentum integrates, together with the single pole that this integral crosses. The second is the cut structure, which determines the weight of each pole family in the observable correlator. The representation itself, its convergence, and the cutting rule are derived in Ref.~\cite{ref26}, so we will only recall what is needed below.

\subsection{Setup}\label{sec:block.setup}

We use Hubble units throughout and work on the rotated time contour. There, the mode function of a field of mass $m$ on the principal series is proportional to $t^{3/2}K_\nu(kt)$, with $\nu=\ii\mu$ and $\mu=(m^2/H^2-\frac94)^{1/2}$. Consider a diagram with $n$ vertices and $L$ massive lines. Its stripped wavefunction coefficient (with couplings and symmetry factors removed) is
\begin{align}
\psi_n &= \int_0^\infty \prod_{v=1}^{n} \dd t_v\; t_v^{r_v-1}\, e^{-x_v t_v} \prod_{l=1}^{L} \mathcal{G}_{\nu_l}(y_l;t_{v_l},t_{v_l'})\,, \qquad \mathcal{G}_\nu(y;t,t') = K_\nu(y\,t_>)\,I_\nu(y\,t_<)\,, \label{eq:psin}\\
r_v &= \tfrac32\deg(v) + n_v - 3\,. \label{eq:rv}
\end{align}
Here $t_\gtrless$ are the maximum and minimum of the two times, and $y_l$ is the energy of the line $l=(v_l,v_l')$. The vertex energy $x_v$ is the sum of the magnitudes of the external momenta at $v$. The exponent $r_v$ is the mass-independent power of time at a vertex with $n_v$ conformally coupled legs and $\deg(v)$ line ends. It collects a factor $t^{3/2}$ from each line end, a factor $t\,e^{-kt}$ from each conformally coupled leg, and $t^{-4}$ from the measure. If a massless leg is instead coupled through $\dot\phi$, it contributes $k^2t_v^2e^{-kt_v}$, and each such leg increases $r_v$ by one.

A vertex thus enters only through $r_v$ and the lines it joins, so our results apply to any operator with that structure, and a model enters only through its choice of vertices, lines, and couplings. Operators that differ by an integration by parts, or by a redefinition of the heavy fields (such as the use of their equations of motion), give the same correlators of the inflaton. This is because the heavy fields decay at late times, so the boundary terms of these manipulations vanish~\cite{Craig2024}. We use this freedom in three places. For the charged scalars of Section~\ref{sec:chem.vertex}, a field redefinition moves the rolling inflaton into a phase of the vertex that breaks their $U(1)$. For the fermions of Section~\ref{sec:dirac}, the coupling to the vector current drops out. Their axial chemical potential is likewise equivalent, up to the chiral anomaly, to a phase of the mass that rotates with the inflaton, which is how it arises in Ref.~\cite{AokiStrumia2026}. Conversely, a given model need not contain every operator at its order in power counting. For example, the minimal basis of the abelian gauge-Higgs-inflaton theory contains $F_{\mu\nu}\tilde F^{\mu\nu}(\partial\phi)^2$ and $F_{\mu\rho}F_\nu{}^\rho\,\partial^\mu\phi\,\partial^\nu\phi$ at the same order as the $F_{\mu\nu}F^{\mu\nu}(\partial\phi)^2$ of the gauge-boson loop in Section~\ref{sec:SM}~\cite{Craig2024}. We compute the operators of each model as given, and Section~\ref{sec:regions.universal} states which operators the universal factor covers.

Following Sleight and Taronna~\cite{Sleight2020,SleightTaronna2020,SleightTaronna2021a,ref34}, we write $\mathcal{G}_\nu$ as a double Barnes integral in which the two times enter only through powers,
\begin{equation}\label{eq:GBarnes}
\mathcal{G}_\nu(y;t,t')=\frac12\int\!\!\int\frac{\dd s\,\dd s'}{(2\pi\ii)^2}\,\frac{\Gamma(-s)\,\Gamma(-s')\,\Gamma(b)\,\Gamma(1-b)}{\Gamma(1+\nu+s)\,\Gamma(1+\nu+s')}\Big(\frac y2\Big)^{2b-2}t^{\nu+2s}\,t'^{\,\nu+2s'}\,,
\end{equation}
with $b=1+\nu+s+s'$. For $t>t'$, closing the contour of $s'$ on the poles of $\Gamma(-s')$ gives the series of $I_\nu(yt')$, and the remaining integral is the Barnes form of $K_\nu(yt)$. Each vertex integral, $\int\dd t\,t^{r_v+\Sigma_v-1}e^{-x_vt}$ with $\Sigma_v$ the sum of the powers of $t$ from its lines, then yields a Gamma function, as in the partial Mellin--Barnes method~\cite{ref18,ref67,XianyuZang2024,ref9}. To be specific, we assign one Mellin variable $s_{v,l}$ to each vertex--line incidence. We then write $b_l=1+\nu_l+s_{v,l}+s_{v',l}$ on the line $l=(v,v')$ and $\Sigma_v=\sum_{l\ni v}(\nu_l+2s_{v,l})$ at the vertex $v$. In these variables the coefficient becomes~\cite{ref26}
\begin{equation}\label{eq:master}
\begin{split}
\psi_n = \Big(\frac12\Big)^{L} \prod_{(v,l)} \int \frac{\dd s_{v,l}}{2\pi\ii} \prod_{\text{lines } l} \Gamma(b_l)\,\Gamma(1-b_l)\,\Big(\frac{y_l}{2}\Big)^{2b_l-2} \hspace{8em}&\\
\times \prod_{\text{vertices } v}\left[\prod_{l\ni v}\frac{\Gamma(-s_{v,l})}{\Gamma(\nu_l+s_{v,l}+1)}\right]\Gamma(r_v+\Sigma_v)\; x_v^{-(r_v+\Sigma_v)}\,.&
\end{split}
\end{equation}
For a tree, the integral converges absolutely on straight contours with $\Re s_{v,l}<0$, $0<\Re b_l<1$, and $\Re(r_v+\Sigma_v)>0$, as proven in Ref.~\cite{ref26}. Such contours exist as long as all the $r_v$ are positive. A vertex without external legs ($x_v=0$) instead retains a delta function in its Mellin variables (this rule holds for $0<r_v<\deg(v)$, see the Supplemental Material of Ref.~\cite{ref26}). Every Gamma argument is an integer combination of Mellin variables shifted by the $\nu_l$ and $r_v$, so closing the contours places the residues on finitely many lattices. On an open set of real kinematics, $\psi_n$ is therefore a finite sum of hypergeometric series, each a non-integer power times a series with integer steps. Each such series collects the residues of one family of poles of the integrand. Following Refs.~\cite{ref10,ref25,HanLeiQinYin2026}, we call it a pole family, or simply a family.

\subsection{The loop block}\label{sec:block.block}

A loop diagram differs from a tree only in that its line energies depend on the loop momentum $\bl$. At fixed $\bl$, everything above applies. Each line then enters Eq.~\eqref{eq:master} through the power $(y_l/2)^{2b_l-2}=(y_l/2)^{-2\delta_l}$ alone, where $\delta_l\equiv1-b_l$ is a constant at each point of the contours. The loop integral thus reduces to a flat-space integral over powers of the line energies. For two lines with energies $y_1=|\bl|$ and $y_2=|\bk-\bl|$, this is a standard result~\cite{ref54,ref31},
\begin{equation}\label{eq:C2}
\int\!\dd^3\ell\;y_1^{-2\delta_1}\,y_2^{-2\delta_2}=k^{3-2\mathcal{S}}\,C_2(\delta_1,\delta_2)\,,\qquad C_2=\pi^{3/2}\,\frac{\Gamma(\tfrac32-\delta_1)\,\Gamma(\tfrac32-\delta_2)}{\Gamma(\delta_1)\,\Gamma(\delta_2)}\,\frac{\Gamma(\mathcal{S}-\tfrac32)}{\Gamma(3-\mathcal{S})}\,,
\end{equation}
with $k=|\bk|$ and $\mathcal{S}=\delta_1+\delta_2$. This is the block used in Ref.~\cite{ref26}. For the spinning, helical, and Dirac lines of Sections~\ref{sec:spin}, \ref{sec:chem}, and~\ref{sec:dirac}, it is more convenient to label the block by the powers of the line energies, with $\hat C_2(p_1,p_2)\equiv C_2\big(\frac{1-p_1}{2},\frac{1-p_2}{2}\big)/2\pi$, so that $\int\dd^3\ell\,y_1^{p_1-1}y_2^{p_2-1}=2\pi\,k^{\,p_1+p_2+1}\hat C_2(p_1,p_2)$. Three lines similarly give the triangle block $C_3$, a two-fold Mellin--Barnes integral at physical momenta~\cite{ref26}. The masses enter only through the values of the $\delta_l$, so the block is the same function for every mass.\footnote{Every loop integral in a wavefunction coefficient or a correlator carries the measure $\dd^3\ell/(2\pi)^3$ throughout this paper. Integrals that only define a block or a moment, such as Eq.~\eqref{eq:C2}, are written with the bare measure $\dd^3\ell$ and carry their factors explicitly. Reference~\cite{ref26} uses the bare measure, so its loop coefficients are $(2\pi)^3$ times ours.}

Compared to a tree, the block contains one new Gamma function, $\Gamma(\mathcal{S}-\frac32)$, and the divergence of the loop sits at its first pole. The momentum integral itself converges only for $\Re\mathcal{S}>\frac32$. Once the block is inserted, however, the Mellin--Barnes integral converges only for $\Re\mathcal{S}<\frac32$, for the bubble as well as for the triangle with conformally coupled legs. The bound comes from the vertex powers and from one direction of power-law decay along the contours~\cite{ref26}. Inserting the block thus continues the loop integral across the pole at $\mathcal{S}=\frac32$. Consequently, a comoving cutoff $\Lambda$ commuted through the contours picks up the residue of that pole times $\ln\Lambda$. A cutoff fixed in physical units corresponds to $\Lambda=a\Lambda_{\rm UV}$, and in the presence of long modes it must also follow the local geometry~\cite{FumagalliInomata2026}. Neither choice affects the nonlocal signals, and Section~\ref{sec:bubble.scheme} relates the cutoff to dimensional regularization. For the bubble with vertex exponents $r_1,r_2$, this term has the contact shape $\Gamma(R)E^{-R}\ln\Lambda/8\pi^2$, with $R=r_1+r_2-3$ and $E=x_1+x_2$. For the triangle, it is instead a momentum-independent coefficient times a secular logarithm, which cancels in the correlator~\cite{ref26}. We return to the bubble in Section~\ref{sec:bubble}. With $\dot\pi$ legs, by contrast, the vertex powers are larger and the two convergence regions overlap (for the triangle the contours need only $\Re\mathcal{S}<\frac52$), so no pole is crossed~\cite{ref26}.

\subsection{Pole families and their cut weights}\label{sec:block.weights}

The amplitude of a collider signal is set by the behavior of a pole family at its vertex-energy cut. As we now show, this behavior is captured by a single number per family. Consider a vertex $v$ and let $\mathcal{C}$ be the set of lines meeting at it. We close the contour of each $s_{v,l}$ with $l\in\mathcal{C}$ to the right, onto the poles of either $\Gamma(-s_{v,l})$ (sign $\sigma_l=-1$) or $\Gamma(1-b_l)$ (sign $\sigma_l=+1$). On these poles $r_v+\Sigma_v$ equals $r_v-\sum_{l\in\mathcal{C}}\sigma_l\nu_l$ up to integers and the remaining Mellin variables, and successive poles shift it by two. The family with sign vector $\sigma$ thus comes with a single power of $x_v$ modulo even integers,
\begin{equation}\label{eq:families}
\psi_n=\sum_\sigma x_v^{\rho_\sigma}f_\sigma(x_v)+f_0(x_v)\,,\qquad \rho_\sigma=-r_v+\sum_{l\in\mathcal{C}}\sigma_l\nu_l\pmod 2\,,
\end{equation}
where $f_\sigma$ is a power series in $x_v^{-2}$ (whose coefficients depend on the other energies) and $f_0$ collects the integer powers left over from the poles of the vertex factor. Since every power $x_v^{\rho_\sigma-2m}$ jumps by the same factor across the negative $x_v$ axis, the family as a whole jumps by $W_\sigma|x_v|^{\rho_\sigma}f_\sigma$, with the cut weight
\begin{equation}\label{eq:weight}
W_\sigma=2\ii\sin(\pi\rho_\sigma)\,.
\end{equation}
Equivalently, continuing the family around $x_v=0$ multiplies it by $e^{2\pi\ii\rho_\sigma}$, so each family has a definite monodromy at any mass.

The simplest example is a single line at an end vertex with $r_v=\frac12$, for which $\rho_\pm=-\frac12\pm\ii\mu$. The two monodromies are then $-e^{\mp2\pi\mu}$, and the weights are $W_\pm=-2\ii\cosh\pi\mu$. The ratio of the two monodromies, $e^{-2\pi\mu}$, is the thermal weight at the Gibbons--Hawking temperature. Since an amplitude goes as the square root of this ratio, we refer to $e^{-\pi\mu}$ as the Boltzmann factor. With $|\Gamma(\ii\mu)|^2=\pi/(\mu\sinh\pi\mu)$, the observable amplitude of the nonlocal signal becomes $e^{-\pi\mu}(1+\bar n)$, where $\bar n=(e^{2\pi\mu}-1)^{-1}$ is the Bose--Einstein occupation. The weights relevant for the scalar loops are listed in Table~\ref{tab:weights}. As we will see, spin shifts $r_v$ (Section~\ref{sec:spin}), a chemical potential modifies the connection data of the line (Section~\ref{sec:chem}), and statistics shifts the exponents by half a unit (Section~\ref{sec:dirac}).

\begin{table}[t]
\centering
\footnotesize
\begin{tabular}{llll}
\toprule
object & cut & $-r_v$ & weights $W_\sigma$\\
\midrule
tree line, end vertex & one line & $-\frac12$ & $W_\pm=-2\ii\cosh(\pi\mu)$\\
bubble vertex & two lines & $-2$ & $W_{\sigma_1\sigma_2}=-2\sinh\pi(\sigma_1\mu_1{+}\sigma_2\mu_2)$\\
 & & & (equal orders: $W_{++}=-W_{--}=-2\sinh2\pi\mu$, $W_{\pm\mp}=0$)\\
triangle vertex & two lines & $-1$ & $-2\ii\sin\pi(\sigma_A\nu_A{+}\sigma_C\nu_C)$\\
\bottomrule
\end{tabular}
\caption{Cut weights of Eq.~\eqref{eq:weight} for the scalar diagrams of this paper, with every cut line on the principal series, $\nu_l=\ii\mu_l$. The bubble vertex has two conformally coupled legs, and the triangle vertex has one.}
\label{tab:weights}
\end{table}

\subsection{From wavefunction coefficients to correlators}\label{sec:block.cut}

What is observed, however, is an equal-time in-in correlator~\cite{Weinberg2005,ref50} rather than $\psi_n$ itself. The relevant propagator is then the Feynman kernel, which differs from the ordered kernel by a thermal term,
\begin{equation}\label{eq:GF}
G^F_\nu(y;t,t') = K_\nu(y\,t_>)\Big[I_\nu-\tfrac{\ii}{\pi}\,e^{\ii\pi\nu}K_\nu\Big](y\,t_<) = K_\nu(yt_>)\big[(1+\bar n)\,I_\nu-\bar n\,I_{-\nu}\big](yt_<)\,.
\end{equation}
The cutting rule of Ref.~\cite{ref26} expresses the stripped correlator as a signed sum over sets of cut lines. Removing the cut lines splits the diagram into connected components. Each component lies on a single branch $\varepsilon_B=\pm$ of the in-in contour, and the two ends of every cut line lie on opposite branches. These assignments are exactly the Schwinger--Keldysh labelings of the vertices, grouped by their cut lines~\cite{ref26}. Each term of the sum is a product of the coefficients $\Phi_B$ of the components, built on Feynman kernels and conjugated on the minus branch,
\begin{equation}\label{eq:cutrule}
F \;=\; \sum_{\text{cut sets }\mathcal{C}}\;\;\sum_{\substack{\text{branches }\varepsilon_B\text{, opposite}\\ \text{across every cut line}}}\;\;\prod_{\text{components }B}\Phi_B^{(\varepsilon_B)}\,,\qquad \Phi^{(+)}=\Phi\,,\quad \Phi^{(-)}=\Phi^*\,.
\end{equation}
Each cut line contributes its two amputated ends, $K_\nu(yt)K_\nu(yt')$, together with a factor $1/\pi$, whereas each uncut line gives a sign $-1$ (see the Supplemental Material of Ref.~\cite{ref26} for the normalization and the proof, and Refs.~\cite{Goodhew2021,Tong2022} for related cutting rules). In addition, a vertex with $n_v$ conformally coupled legs multiplies its component by the phase $\ii^{\,n_v}$. If all vertex phases are real, the contribution without cut lines reduces to $(-1)^L2\Re\psi_n^F$. Complex phases arise, for instance, when each vertex of the triangle carries one conformally coupled leg. Each phase is then conjugated together with its own component.

For a loop, it is convenient to organize the same sum line by line. This is the form we will use for the bubble and the triangle. At fixed loop momentum and fixed rotated times, the branch propagators of the line $l$ are $G^{++}=-F_l$, $G^{--}=-\bar F_l$, and $G^{+-}=G^{-+}=C_l$. Here $F_l$ is the Feynman kernel of Eq.~\eqref{eq:GF} and $C_l=\frac1\pi K_{\nu_l}(y_lt)K_{\nu_l}(y_lt')$ is the cut kernel, with the factor $(tt')^{3/2}$ stripped. For a ring of $L$ lines with real vertex factors, the sum over branch assignments then becomes a trace of $2\times2$ matrices~\cite{ref35,ref34},
\begin{equation}\label{eq:trace}
S_L\equiv\sum_{a\in\{\pm\}^L}\prod_lG^{a_la_{l+1}}=\mathrm{Tr}\prod_{l=1}^LT_l\,,\qquad T_l=-\tfrac12\big(\mathcal{G}^+_l+\mathcal{G}^-_l\big)\mathbb{1}+C_l\big(\ii\cos(\pi\nu_l)\,\sigma_z+\sigma_x\big)\,,
\end{equation}
where $\mathcal{G}^\pm_l=K_{\nu_l}(y_lt_>)I_{\pm\nu_l}(y_lt_<)$ are the ordered kernel and its shadow. Expanding the trace gives a sum over sign patterns $\epsilon\in\{\pm\}^L$ (one sign per line) with weights in closed form. For heavy lines,
\begin{equation}\label{eq:patterns}
S_L=(-1)^L\sum_\epsilon W_\epsilon\prod_l\mathcal{G}^{\epsilon_l}_l\,,\qquad W_\epsilon=\prod_{v=1}^L\sinh\!\Big[\frac\pi2\big(\epsilon_{v-1}\mu_{v-1}+\epsilon_v\mu_v\big)\Big]\prod_l\frac{\epsilon_l}{\sinh\pi\mu_l}\,,
\end{equation}
with $W_{-\epsilon}=W_\epsilon$. At equal masses only the two uniform patterns survive, and we recover the familiar form $S_L=(-1)^L2\Re\prod_l\mathcal{G}_l^+$.

The situation is somewhat different for light lines. For a line of real order $0<\nu_l<\frac32$, the trace form holds unchanged. So do the weights of Eq.~\eqref{eq:patterns}, with $\mu_l\to-\ii\nu_l$ (each $\sinh$ becomes a $\sin$, and the factors of $\ii$ cancel). At equal orders the sum is then $(-1)^L\big[\prod_l\mathcal{G}^+_l+\prod_l\mathcal{G}^-_l\big]$, whatever the common order. However, this equals $(-1)^L2\Re\prod_l\mathcal{G}^+_l$ only when the order is imaginary or an integer. Writing the correlator as twice the real part of the ordered product is thus correct on the principal series, whereas for light lines the shadow product must be kept. As noted in Ref.~\cite{ref35}, this is because complex conjugation exchanges the two branches only for heavy fields. Similarly, vertex factors whose phase differs between the two branches (as for a non-derivative coupling to a conformally coupled leg) shift the arguments of the sines in Eq.~\eqref{eq:patterns} by the vertex phase. In that case the twice-real-part form fails even at equal masses.

%% file: sec3.tex
\section{The bubble at any two masses}\label{sec:bubble}

We begin with the bubble, the simplest one-loop diagram and the prototype for the diagrams that follow. At equal masses, its decomposition into a nonlocal signal, a local signal, and a background is known~\cite{ref5}. At unequal heavy masses both signals are known, but the background was left with an undetermined constant~\cite{ref10}. In this section we complete this picture. For any two masses (heavy or light) and general vertex powers, we show that the coefficients of all three pieces are single sums over tree-level exchanges, and that renormalization fixes the background with no free constant. We close with spinning lines and the tensor bubble.

\subsection{Nonlocal signal, local signal, and background}\label{sec:bubble.families}

\begin{figure}[t]
\centering
\includegraphics{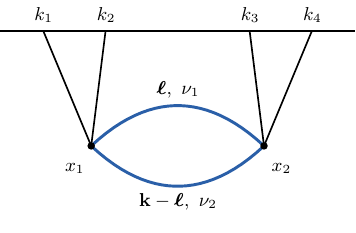}
\caption{The bubble. The external momentum $k$ runs through two lines of orders $\nu_1$ and $\nu_2$, which carry $\ell$ and $k-\ell$ and join vertices of energies $x_1$ and $x_2$.}
\label{fig:bubble}
\end{figure}

Consider two lines of orders $\nu_1,\nu_2$ connecting a pair of vertices, whose energies and exponents are $x_1,x_2$ and $r_1,r_2$ (see Fig.~\ref{fig:bubble}). Inserting the block of Eq.~\eqref{eq:C2}, we obtain the loop-integrated coefficient
\begin{equation}\label{eq:bubbleMB}
\begin{aligned}
\psi_2^{\mathrm{loop}} = \frac14\prod_{(v,j)}\int\frac{\dd s_{v,j}}{2\pi\ii}&\prod_v\left[\prod_{j\ni v}\frac{\Gamma(-s_{v,j})}{\Gamma(\nu_j+s_{v,j}+1)}\right]\frac{\Gamma(r_v+\Sigma_v)}{x_v^{\,r_v+\Sigma_v}}\\
&\times\prod_j\Gamma(b_j)\Gamma(1{-}b_j)\;\frac{2^{2\mathcal{S}}\,k^{3-2\mathcal{S}}}{(2\pi)^3}\;C_2(\delta_1,\delta_2)\,,
\end{aligned}
\end{equation}
with $b_j=1+\nu_j+s_{1,j}+s_{2,j}=1-\delta_j$ and $\Sigma_v=\nu_1+\nu_2+2s_{v,1}+2s_{v,2}$. For simplicity we omit the symmetry factor $\frac12$ of identical lines, here and in everything derived from Eq.~\eqref{eq:bubbleMB}. We denote by $\psi_2^{\rm MB}$ the integral taken on contours with $\frac12<\Re\mathcal{S}<\frac32$ (between the first two poles of $\Gamma(\mathcal{S}-\frac32)$). As discussed in Section~\ref{sec:block.block}, this integral continues the loop integral. A comoving cutoff at $|\bl|=\Lambda$ makes the relation precise,
\begin{equation}\label{eq:cutoff}
\psi_2^{\mathrm{loop}}\big|_\Lambda = \psi_2^{\mathrm{MB}} + \frac{\Gamma(R)}{8\pi^2}\,E^{-R}\ln\Lambda + P_0 + o(1)\,,\qquad R=r_1+r_2-3\,,\quad E=x_1+x_2\,.
\end{equation}
This extends the cutoff identity of Ref.~\cite{ref26} to general vertex exponents, with $P_0$ the cutoff-independent term left behind by the crossed pole. The coefficient of the logarithm follows from the flat-space limit of the loop, as we show in Section~\ref{sec:bubble.scheme}.

Remarkably, the loop integral leaves exactly three pole families. The reason is that every exponent of $k$, $x_1$, and $x_2$ in Eq.~\eqref{eq:bubbleMB} depends on the Mellin variables only through the vertex sums $\sigma_v=s_{v,1}+s_{v,2}$. A family is then fixed by two pole conditions on $(\sigma_1,\sigma_2)$, and the two remaining directions are integrated into its coefficients. Three sequences of poles can occur. We denote by $R_v$ the poles of the numerators $\Gamma(-s_{v,j})$, by $V_v$ those of the vertex factor $\Gamma(r_v+\Sigma_v)$, and by $U$ the ultraviolet poles of $\Gamma(\mathcal{S}-\frac32)$ at $\mathcal{S}=\frac32-n$. These $U$ are the UV poles of Refs.~\cite{ref10,ref25,HanLeiQinYin2026}. The $R_v$ play the role of their spectrum poles, since the families built on them carry the masses of the lines. For $k<x_2<x_1$, once the logarithm of Eq.~\eqref{eq:cutoff} is removed, the cutoff loop tends to $A+F_0+F_{\mathrm{BG}}$ with ($\bar\nu=\nu_1+\nu_2$)
\begin{equation}\label{eq:threefamilies}
\begin{aligned}
A:R_1\wedge R_2\,,&\qquad k^{\,3+2\bar\nu+2m_1+2m_2}\;x_1^{-r_1-\bar\nu-2m_1}\;x_2^{-r_2-\bar\nu-2m_2} &&\qquad\text{(nonlocal signal)},\\
F_0:R_1\wedge U\,,&\qquad k^{\,2n}\;x_1^{-r_1-\bar\nu-2m_1}\;x_2^{3-r_2-2n+\bar\nu+2m_1} &&\qquad\text{(local signal)},\\
F_{\mathrm{BG}}:V_2\wedge U\,,&\qquad k^{\,2n}\;x_1^{-(R+l+2n)}\;x_2^{l} &&\qquad\text{(background)},
\end{aligned}
\end{equation}
where $m_v$, $n$, and $l$ are non-negative integers. On the residues of $U$ the kinematic dependence becomes $(x_1/x_2)^{2\sigma_2}$, so the $\sigma_2$ contour closes to the left onto $V_2$. This is why closing every contour to the right misses $F_{\rm BG}$. The terms with $n=0$ in $F_0$ and $F_{\rm BG}$ arise instead from the crossed-pole term $P_0$. No other family can occur, since every other pair of pole sequences closes either on an empty region or on a zero of $1/\Gamma(\nu_j+s_{v,j}+1)$.

On the principal series, $A$ oscillates in the product of the vertex energies and $F_0$ in their ratio, so these are the nonlocal and local signals of Refs.~\cite{ref5,ref9,ref10}. Every exponent of $F_{\rm BG}$, on the other hand, is an integer independent of the masses. The background is thus analytic in every momentum, with the masses entering only through its coefficients. In what follows we derive the coefficients of all three pieces as single sums.

\begin{figure}[t]
\centering
\includegraphics{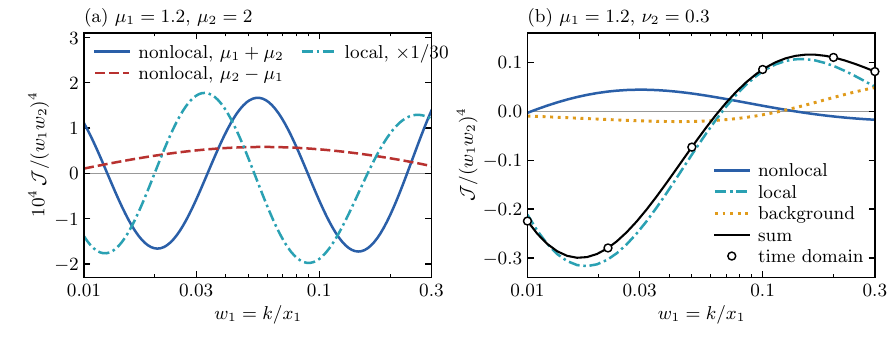}
\caption{The pieces of the bubble in the correlator, for the $\phi'^2\sigma^2$ vertices of Ref.~\cite{ref5} ($r_v=4$), along $w_2=k/x_2=\frac12$ against $w_1=k/x_1$. The unit is $\mathcal{J}/(w_1w_2)^4$, with $\mathcal{J}=\frac12k^5F$, where $F$ is the stripped correlator of Eq.~\eqref{eq:cutrule}. These are the units of Ref.~\cite{ref5} (Section~\ref{sec:bubble.scheme}). (a) Two heavy lines, $\mu_1=1.2$ and $\mu_2=2$. Shown are the nonlocal signal at the sum and at the difference frequency, and the local signal (scaled by $1/30$). The background (not shown) lies between $-0.11$ and $-0.03$ in these units. (b) One heavy and one light line, $\mu_1=1.2$ and $\nu_2=0.3$, with the sum of the three pieces (black line). In both panels the background is in $\overline{\rm MS}$ at $\mu_R=1$. The sum agrees with an independent time-domain evaluation of the cutoff loop, converted by Eq.~\eqref{eq:cutMS}, to $10^{-10}$ or better (circles in (b), and at six points for (a)).}
\label{fig:families}
\end{figure}

Figure~\ref{fig:families} shows the three pieces for two mass assignments not covered by the equal-mass results. With two heavy lines (panel a), the sum and difference frequencies of the nonlocal signal have comparable amplitudes, as expected from their common Boltzmann factor~\cite{ref10}, and both are far smaller than the local signal. The background, which has no Boltzmann factor, dominates the correlator at these masses. With one light line (panel b), the light order enters as a real power, and the local signal dominates. Only the background depends on the subtraction (through the contact term of Eq.~\eqref{eq:MSbar}), whereas the two signals are scheme independent.

\subsection{The families as a sum over trees}\label{sec:bubble.trees}

We now turn to the coefficients, all three of which follow from a single sum over tree-level exchanges. The ordered kernel $K_\nu(yt_>)I_\nu(yt_<)$ is $(t_1t_2)^{-3/2}$ times the bulk-to-bulk propagator of dimension $\Delta=\frac32+\nu$ in Euclidean anti-de Sitter space, so the loop integral of two kernels is the Fourier transform of a product of two propagators at the same pair of points. By an identity due to Fitzpatrick and Kaplan~\cite{ref30,ref33} (extended to in-in de Sitter propagators at unequal masses by Sleight and Taronna~\cite{ref34}), such a product is a sum of single propagators of dimensions $\Delta_1+\Delta_2+2n$ with known coefficients $a_{\Delta_1,\Delta_2}(n)$. Reference~\cite{ref26} used this identity in Mellin form to write the bubble as a sum over single exchanges. The identity also underlies the bubble functions of Refs.~\cite{ref5,ref35,ref36}, and in Ref.~\cite{ref5} the equal-mass signal already appears as a sum over exchanges. Applied to the ordered momentum-space kernel, it gives the complete coefficient at unequal masses. With $[z]\equiv\Gamma(z)$ we find
\begin{equation}\label{eq:product}
\int\frac{\dd^3\ell}{(2\pi)^3}\,K_{\nu_1}(y_1t_>)I_{\nu_1}(y_1t_<)\,K_{\nu_2}(y_2t_>)I_{\nu_2}(y_2t_<) = \frac{(t_1t_2)^{-3/2}}{\pi^2}\sum_{n\ge0}G_n\,K_{\nu_n'}(kt_>)I_{\nu_n'}(kt_<)\,,
\end{equation}
with $\nu_n'=\bar\nu+\frac32+2n$ and the weights
\begin{equation}\label{eq:Gn}
G_n=\frac{[\tfrac32{+}n]\,[\tfrac32{+}\nu_1{+}n]\,[\tfrac32{+}\nu_2{+}n]\,[\tfrac32{+}\bar\nu{+}n]\,(\tfrac32{+}\bar\nu{+}2n)\,[1{+}\bar\nu{+}2n]}{n!\,[1{+}\nu_1{+}n]\,[1{+}\nu_2{+}n]\,[1{+}\bar\nu{+}n]\,[3{+}\bar\nu{+}2n]}\,,
\end{equation}
where $G_n=\pi^2a_{\Delta_1,\Delta_2}(n)$ at $\Delta_j=\frac32+\nu_j$. For $t_1\neq t_2$ the sum converges geometrically and locally uniformly in the orders, so the identity continues from real positive orders to the principal series. At $t_1=t_2$ the sum diverges logarithmically (since $4G_n\to2n$), reproducing the ultraviolet divergence of the loop.

Under the vertex integrals, each term of Eq.~\eqref{eq:product} becomes a tree-level exchange of order $\nu'_n$ with vertex exponents $r_v-\frac32$. Each such exchange in turn splits into three terms. We write the kernel $K_{\nu}(kt_>)I_{\nu}(kt_<)$ of the tree as $K_{\nu}(kt_2)I_{\nu}(kt_1)$ plus a time-ordered remainder supported on $t_1>t_2$, which is analytic in $k^2$ and in $x_2$. The connection formula $K_\nu=\pi(I_{-\nu}-I_\nu)/(2\sin\pi\nu)$ then turns the first term into a product of two $I_{\nu'_n}$ moments and a product of an $I_{\nu'_n}$ and an $I_{-\nu'_n}$ moment. In terms of the moment
\begin{equation}\label{eq:Imoment}
\hat I_\nu(\rho,x)=\int_0^\infty\dd t\,t^{\rho-1}e^{-xt}I_\nu(kt)=\frac{(k/2)^\nu[\rho+\nu]}{[1+\nu]\,x^{\rho+\nu}}\;{}_2F_1\Big(\tfrac{\rho+\nu}{2},\tfrac{\rho+\nu+1}{2};1{+}\nu;\tfrac{k^2}{x^2}\Big),
\end{equation}
the first two families are
\begin{gather}\label{eq:AF0}
A=-\sum_{n\geq0}\frac{G_n}{2\pi\sin\pi\nu'_n}\,\hat I_{\nu'_n}(r_1{-}\tfrac32,x_1)\,\hat I_{\nu'_n}(r_2{-}\tfrac32,x_2)\,,\\
F_0=\sum_{n\geq0}\frac{G_n}{2\pi\sin\pi\nu'_n}\,\hat I_{\nu'_n}(r_1{-}\tfrac32,x_1)\,\hat I_{-\nu'_n}(r_2{-}\tfrac32,x_2)\,.\nonumber
\end{gather}
When $2\bar\nu$ is not an integer, the powers of the three terms differ by non-integers. These terms are then precisely the three families of Eq.~\eqref{eq:threefamilies}, confirming that there is no fourth. Both sums converge geometrically for $k<x_2<x_1$, so each family is a single series of products of two Gauss functions.

The background follows from the same sum over trees, although its coefficients require a little more care. It comes entirely from the time-ordered remainders. For a tree of order $\nu$ the remainder is $\frac{\pi}{2\sin\pi\nu}[I_{-\nu}(kt_1)I_\nu(kt_2)-I_{-\nu}(kt_2)I_\nu(kt_1)]$ on $t_1>t_2$. We expand each $I_{\pm\nu}$ in its series and $e^{-x_2t_2}$ in powers of $x_2$, integrate $t_2$ from zero to $t_1$, and then integrate $t_1$ over the half-line. With $\Gamma(1+\nu)\Gamma(1-\nu)=\pi\nu/\sin\pi\nu$, the monomial $(k/2)^{2n}(-x_2)^l\,\Gamma(R+l+2n)\,x_1^{-(R+l+2n)}/l!$ then has the coefficient
\begin{equation}\label{eq:phinl}
\begin{aligned}
\varphi_{nl}(\nu)&=\frac1{2\nu}\sum_{p+q=n}\frac{1}{p!\,q!\,(1-\nu)_p\,(1+\nu)_q}\Big[\frac1{z_l+\nu+2q}-\frac1{z_l-\nu+2p}\Big]\\
&=\sum_{j=0}^{n}\frac{1}{\big[\nu^2-(z_l+2j)^2\big]\,(n{-}j)!\,j!\,(1{+}z_l{+}2j)_{n-j}\,(1{-}z_l{-}2j)_j}\,,\qquad z_l=r_2-\tfrac32+l\,.
\end{aligned}
\end{equation}
The second line is the partial-fraction expansion of the first, since $\varphi_{nl}$ is even in $\nu$ and its only poles lie at $\nu=\pm(z_l+2j)$. We have checked the first line against a direct integration of the remainder, to $10^{-12}$ or better. Under the sum over trees with the weights $G_t$, each partial fraction gives the same function,
\begin{equation}\label{eq:S0}
S_0(z)\equiv\lim_{N\to\infty}\Big[\sum_{t=0}^{N}\frac{4G_t}{{\nu'_t}^2-z^2}-\tfrac12\ln N\Big],
\end{equation}
whose summand is that of a very-well-poised ${}_9F_8$ at unit argument with vanishing parameter excess (hence the logarithmic divergence, since $4G_t\to2t$). A difference $S_0(z)-S_0(z')$ is a convergent very-well-poised ${}_{11}F_{10}$ at unit argument. Collecting the factors of Eq.~\eqref{eq:product}, the monomial $k^{2n}x_1^{-(R+l+2n)}x_2^l$ of $F_{\rm BG}$ has the coefficient
\begin{equation}\label{eq:ladder}
\frac{(-1)^{n+l}\,[R{+}l{+}2n]}{2^{3+2n}\,\pi^{3/2}\,n!\,l!\,[\tfrac32{+}n]}\;\Phi_{nl}\,,\qquad \Phi_{nl}=(-1)^n n!\,(\tfrac32)_n\sum_{t\ge0}4G_t\,\varphi_{nl}(\nu'_t)\,.
\end{equation}
For $n\geq1$ this reads
\begin{equation}\label{eq:Phinl}
\Phi_{nl}=(-1)^n n!\,(\tfrac32)_n\sum_{j=0}^{n}\frac{S_0(z_l+2j)}{(n{-}j)!\,j!\,(1{+}z_l{+}2j)_{n-j}\,(1{-}z_l{-}2j)_j}\,.
\end{equation}
The weights in Eq.~\eqref{eq:Phinl} sum to zero, so the logarithm drops out and every $\Phi_{nl}$ with $n\geq1$ is a finite sum of ${}_{11}F_{10}$ functions. The coefficients with $n=0$, by contrast, diverge logarithmically, with a coefficient independent of $l$. Their divergent parts sum over the row to $\Gamma(R)E^{-R}$, the contact shape anticipated in Section~\ref{sec:block.block}. We give $S_0$ in closed form in Section~\ref{sec:close}, on the lattice of points where the bubble requires it.

\subsection{Renormalization and scheme constants}\label{sec:bubble.scheme}

It remains to fix the coefficients with $n=0$, which encode the renormalization. As we now show, once these are fixed, the subtracted loop contains no free constant. For other regularizations of loops in de Sitter space, see Refs.~\cite{ref6,Premkumar2024}. We define the subtracted loop by
\begin{equation}\label{eq:psifin}
\psi_2^{\rm fin}\equiv\lim_{\Lambda\to\infty}\Big[\psi_2^{\rm loop}\big|_\Lambda-\frac{\Gamma(R)}{8\pi^2}E^{-R}\ln(\Lambda/x_1)\Big]\,.
\end{equation}
Its coefficients with $n=0$ are $\Phi^{\rm cut}_{0l}=S_0(z_l)+\frac12\psi^{(0)}(R{+}l)$, with $\psi^{(0)}$ the digamma function. For the coefficients of $F_{\rm BG}$ with $n=0$, we instead adopt the same sum with the digamma function replaced by one,
\begin{equation}\label{eq:rung}
\Phi_{0l}\equiv S_0(z_l)+\tfrac12\,.
\end{equation}
The difference $\frac12[\psi^{(0)}(R{+}l)-1]$ then sums over the row to a contact term, so that
\begin{equation}\label{eq:fin}
\psi_2^{\rm fin}=A+F_0+F_{\rm BG}+\frac{\Gamma(R)}{8\pi^2}E^{-R}\big[\psi^{(0)}(R)-1-\ln(E/x_1)\big]\,,
\end{equation}
which contains no free constant. Together with Eqs.~\eqref{eq:AF0}, \eqref{eq:Phinl}, and~\eqref{eq:rung}, this expresses the subtracted loop as a sum over trees with no integral left. We have checked Eq.~\eqref{eq:fin} against a direct time-domain evaluation of the cutoff loop. The two agree to $10^{-11}$ or better at equal and unequal orders (including orders of opposite sign, which give the difference frequency), as well as with one light line. With both lines on the complementary series, the result agrees with direct quadrature to $10^{-10}$.

These families differ from the $\overline{\rm MS}$ result of Ref.~\cite{ref5}, in dimensional regularization at a scale $\mu_R$ in Hubble units, only by a contact term, whose constant we can find in closed form. If we expand $F_{\rm BG}=\sum_jc_j\,w_1^{R+j}w_2^{-j}+\dots$ on the lattice ($w_i=k/x_i$), our coefficients with the choice of Eq.~\eqref{eq:rung} differ from theirs only by the contact term $(k/E)^R$. In the units of Ref.~\cite{ref5}, where $\mathcal{J}=\frac12k^R[\psi_2^{\rm loop}(\ii\mu)+\psi_2^{\rm loop}(-\ii\mu)]$,\footnote{Reference~\cite{ref5} writes $\mathcal{J}=\kappa_{\rm XZ}k^R[\psi_2(\ii\mu)+\psi_2(-\ii\mu)]$ with $\kappa_{\rm XZ}=1/(8\pi^2)$, for loop coefficients normalized to $(2\pi)^2$ times ours.} we find
\begin{equation}\label{eq:MSbar}
c_j-c_j^{\overline{\rm MS}}(\mu_R)=\frac{\Gamma(R)}{8\pi^2}\,\ln\frac{2}{\mu_R}\;(-1)^j\binom{R+j-1}{j}\,,
\end{equation}
so that the choice of Eq.~\eqref{eq:rung} is $\overline{\rm MS}$ at $\mu_R=2$ for every $R$ and every mass. At the vertices of Ref.~\cite{ref5} ($r_v=4$, so $R=5$), the offset at $\mu_R=1$ is $\delta=3\ln2/\pi^2=0.2106914783\ldots$.

This constant follows from the flat-space limit of the loop. In the ultraviolet each line takes its flat-space form, and in flat space a cutoff $\Lambda_{\rm ph}$ on the spatial loop momentum and $\overline{\rm MS}$ at scale $\mu_R$ differ by $[\ln(2\Lambda_{\rm ph}/\mu_R)-1]/8\pi^2$. Since the comoving cutoff is the physical cutoff $\Lambda t$ at time $t$~\cite{SenatoreZaldarriaga2010}, this constant integrates under the vertex integral (whose ultraviolet weight is $t^{R-1}e^{-Et}$) to
\begin{equation}\label{eq:cutMS}
\psi_2^{\rm loop}\big|_\Lambda-\psi_2^{\overline{\rm MS}}(\mu_R)=\frac{\Gamma(R)}{8\pi^2}E^{-R}\Big[\ln\frac{2\Lambda}{E\mu_R}+\psi^{(0)}(R)-1\Big]\,,
\end{equation}
with $\mu_R$ in Hubble units and $\Lambda$ and $E$ comoving (the digamma function is the time average of $\ln t$). Comparing with Eq.~\eqref{eq:fin} confirms that Eq.~\eqref{eq:rung} is $\overline{\rm MS}$ at $\mu_R=2$. Equation~\eqref{eq:cutMS} also relates $\overline{\rm MS}$ to the minimal subtraction of $\ln(\Lambda/E)$ at any vertex powers. The two differ by $\Gamma(R)E^{-R}[\ln(\mu_R/2)+1-\psi^{(0)}(R)]/8\pi^2$, which at $R=3$ is the offset used in Ref.~\cite{ref26}, up to its normalization.

Finally, since the loop block does not depend on the masses, the masses enter the bubble only through exponents, and its singularities lie on a mass-independent locus. This is in line with the description of massive diagrams, at tree level and at any loop order, as generalized Euler integrals, that is, GKZ hypergeometric functions~\cite{GKZ1990,CaloroMcFadden2023,GrimmHoefnagels2025,Herderschee2025,Grafe2026gluing,KatsevichSun2026}, whose singularities can be located by Landau and D-module analyses~\cite{ref38,ref12,KatsevichSun2026}.

\subsection{Two masses in the squeezed limit}\label{sec:bubble.twomass}

Unlike a tree line, a loop can contain two species. Its squeezed signal is then a doublet at the sum and the difference of the two mass parameters. After the loop integral, all four lines $|{\pm}\mu_1{\pm}\mu_2|$ of the in-in correlator share the same Boltzmann factor~\cite{ref9,ref10}. Here we give their relative amplitude in closed form. In the correlator, the difference lines come from the patterns of Eq.~\eqref{eq:patterns} with $\epsilon_1\neq\epsilon_2$. Relative to the sum lines, they are weighted by $[1-\cosh\pi(\mu_1{-}\mu_2)]/[\cosh\pi(\mu_1{+}\mu_2)-1]$, a ratio that is thermally small at fixed internal energies. Consider now one soft leg, $x_2=k_3$. The loop integral over the soft region turns the soft-vertex factor into a line-dependent moment $\mathcal{M}_\sigma$, and the difference-to-sum amplitude ratio becomes
\begin{equation}\label{eq:Rratio}
\mathcal{R}=\frac{\cosh\pi|\mu_1{-}\mu_2|-1}{\cosh\pi(\mu_1{+}\mu_2)-1}\,
\left|\frac{\Gamma\bigl(r_1{+}\ii|\mu_1{-}\mu_2|\bigr)}{\Gamma\bigl(r_1{+}\ii(\mu_1{+}\mu_2)\bigr)}\right|
\left|\frac{\mathcal{M}_{+-}}{\mathcal{M}_{++}}\right|,
\end{equation}
where $r_1$ ($r_2$) denotes the exponent at the hard (soft) vertex. The moment is the loop integral of the soft vertex with both lines amputated, weighted by the powers that the hard vertex leaves on the two lines,
\begin{equation}\label{eq:twomoment}
\begin{aligned}
\mathcal{M}_\sigma&=\int\dd^3\ell\,\Bigl(\frac{y_1}{2}\Bigr)^{\sigma_1\nu_1}\Bigl(\frac{y_2}{2}\Bigr)^{\sigma_2\nu_2}\!\int_0^\infty\!\dd t\;t^{\,r_2-1}e^{-k_3t}K_{\nu_1}(y_1t)K_{\nu_2}(y_2t)\\
&=\pi^2\,2^{2-r_2}\,k_3^{\,3-r_2}\Bigl(\frac{k_3}{2}\Bigr)^{\Sigma_\sigma}\frac{\Gamma(\frac32{+}\sigma_1\nu_1)\,\Gamma(\frac32{+}\sigma_2\nu_2)\,\Gamma(r_2{+}\Sigma_\sigma)\,\Gamma(r_2{-}3{-}\Sigma_\sigma)}{\Gamma(3{+}\Sigma_\sigma)\,\Gamma(r_2{-}1)}\,,
\end{aligned}
\end{equation}
with $\Sigma_\sigma=\sigma_1\nu_1+\sigma_2\nu_2$, $y_1=|\bl|$, and $y_2=|\bk_3-\bl|$. The integrals converge for $r_2>3$, and otherwise we define $\mathcal{M}_\sigma$ by continuation in $r_2$. This is legitimate because, with a momentum cutoff, the divergent terms are analytic in $k_3$, while the two lines are not.

\begin{figure}[t]
\centering
\includegraphics{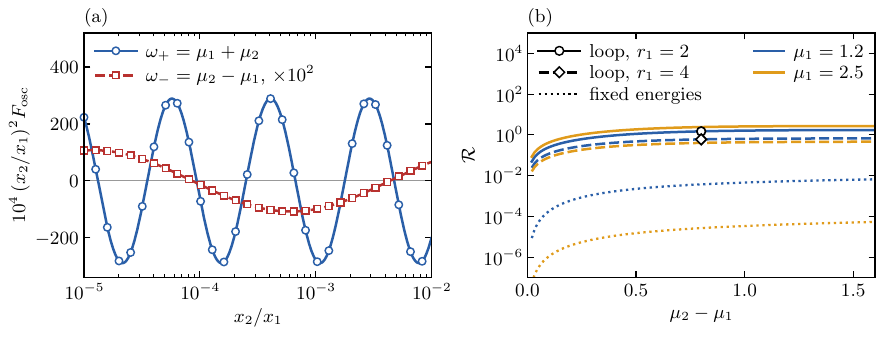}
\caption{The two-mass doublet of the in-in bubble. (a) The oscillatory part of $(x_2/x_1)^2F$ against the ratio $x_2/x_1$ of the soft and hard vertex energies, at fixed internal energies for $(\mu_1,\mu_2)=(1.2,2)$ and $r_1=r_2=2$, with the difference line multiplied by $10^2$. Curves are the two lines of a fit to a direct evaluation, and symbols are the evaluation with the other line and the higher terms removed. The ratio of the two amplitudes equals the product of the thermal and Gamma-function ratios of Eq.~\eqref{eq:Rratio}, $3.7312\times10^{-3}$, to $2\times10^{-7}$. (b) The difference-to-sum ratio $\mathcal{R}$ against the splitting at $\mu_1=1.2$ and $2.5$. Dotted curves are at fixed internal energies, and solid and dashed curves are Eq.~\eqref{eq:Rratio} after the loop integral, for $(r_1,r_2)=(2,2)$ and $(4,2)$. Symbols are loop-integrated scans with an independent time-domain code at a splitting of $0.8$ (circle for $r_1=2$, diamond for $r_1=4$).}
\label{fig:twomass}
\end{figure}

Figure~\ref{fig:twomass} shows how the loop integral raises the relative weight of the two lines from thermally small to order one. At fixed internal energies the moment ratio is one, and $\mathcal{R}=3.7\times10^{-3}$ at $(\mu_1,\mu_2)=(1.2,2)$ with $r_1=2$. After the loop integral the moment ratio is large ($|\mathcal{M}_{+-}/\mathcal{M}_{++}|=399$ at $r_2=2$), and $\mathcal{R}$ rises to order one. At $r_1=r_2=2$ the ratio takes the closed form
\begin{equation}\label{eq:Rclosed}
\mathcal{R}=\tanh\frac{\pi|\mu_1{-}\mu_2|}{2}\,\coth\frac{\pi(\mu_1{+}\mu_2)}{2}\,\left[\frac{4+(\mu_1{+}\mu_2)^2}{4+(\mu_1{-}\mu_2)^2}\right]^{1/2},
\end{equation}
which equals $1.49$ at these masses. It grows with the splitting and exceeds one once the splitting passes a few tenths ($0.43$ at $\mu_1=1.2$). A harder vertex on the hard side lowers the ratio, to $0.60$ at the same masses for $r_1=4$. For a conformally coupled leg on a cubic soft vertex ($r_2=1$), the leading moments vanish and the doublet starts at higher order in $k_3/k_1$. As a check, loop-integrated scans with an independent time-domain code reproduce $\mathcal{R}$ to $0.25\%$ at a splitting of $0.8$ for both values of $r_1$ (symbols in Fig.~\ref{fig:twomass}(b)). Together with the decay of the signal in $k_3/k_1$, the ratio $\mathcal{R}$ can help distinguish a two-mass loop from several tree-level lines~\cite{XianyuZang2024,ref49,Belrhali2026klf,ChenXianyu2026,Ahmed2026}. The moment of Eq.~\eqref{eq:twomoment} will reappear in Section~\ref{sec:regions} as the soft vertex of every squeezed bubble.

\subsection{Spinning lines and the tensor bubble}\label{sec:spin}

So far every line has been a scalar. We now ask what spin, the simplest change of line, does to the pieces and their weights. The chemical potential and statistics follow in Sections~\ref{sec:chem} and~\ref{sec:dirac}. As is well known~\cite{ref2,ref3}, the top helicities $\lambda=\pm s$ of a field of spin $s$ obey the scalar mode equation with a shifted mass parameter,
\begin{equation}\label{eq:mus}
\mu_s^2=m^2-\big(s-\tfrac12\big)^2\qquad(s=1,2,\dots)\,,
\end{equation}
so that the order $\nu_s=\ii\mu_s$ replaces $\nu$ in every Gamma function of the line. In the present representation spin has two further effects. First, counting powers of the vertex time for the cubic vertex of lowest derivative order between $\sigma_{i_1\cdots i_s}$ and two conformally coupled scalars shows that each end of a spin-$s$ line raises $r_v$ by $s$ (from $\frac12$ for a scalar to $\frac32$ for a transverse vector and $\frac52$ for a tensor). Second, the polarization sum multiplies the integrand by a polynomial in the external momenta that contains no Mellin variable. Since spinning lines have been treated in partial Mellin--Barnes form in Refs.~\cite{ref67,ref53}, we do not repeat the details. For integer $s$ the pieces lie on the scalar lattice shifted by whole units, and the cut weights are those of Table~\ref{tab:weights} at the shifted exponent, $W_\pm=-(-1)^s\,2\ii\cosh\pi\mu_s$.

For the massive spin-2 field of Fierz and Pauli~\cite{ref61}, the constrained helicities $\pm1$ and $0$ can be solved in terms of the same Bessel pair through derivative dressings. Using the Bessel recurrences, each derivative becomes a sum over the neighboring orders $\nu\pm1$. A constrained helicity is thus a finite sum of scalar-type terms with shifted vertex exponents. Inside the horizon the helicity-0 polarization bilinear grows as $\frac23(q/m)^4$ (the flat-space longitudinal growth), while helicity $\pm1$ grows as $(q/m)^2$ and the transverse-traceless modes remain bounded. The small-mass enhancement of a massive spin-2 loop thus comes entirely from the constrained sector. In de Sitter space the helicity-0 mode is a ghost for $0<m^2<2H^2$, the Higuchi bound~\cite{ref63,ref62}, and $m^2=2H^2$ is the partially massless point~\cite{ref64,ref65,ref66}. The transverse-traceless lines below are unaffected.

\paragraph{The tensor bubble.}\label{sec:spin.tensor}

The simplest loop in which spin changes the analytic structure is the bubble with two transverse-traceless tensor lines. We take lines of order $\nu$ (with $\nu=\ii\mu_2$ for massive spin 2 and $\nu=\frac32$ at the massless endpoint) between non-derivative vertices $\phi^n\gamma_{ij}\gamma_{ij}$ with $n$ conformally coupled legs. Such a vertex describes, for example, pair production of a massive spin-2 field coupled to inflaton fluctuations. It differs from the graviton loop of general relativity, whose vertex contains two inverse metrics and raises the vertex exponent by four~\cite{ref74,ref75}. Since each tensor end counts as $t^{3/2}$, we have $r_v=n\equiv r$, and the object is
\begin{equation}\label{eq:TTbubble}
\psi_2^{\rm TT}=\int\frac{\dd^3\ell}{(2\pi)^3}\;T_2(c)\int_0^\infty\!\!\int_0^\infty\dd t_1\,\dd t_2\,(t_1t_2)^{\,r-1}\,e^{-x_1t_1-x_2t_2}\prod_{j=1,2}K_\nu(y_j\,t_>)\,I_\nu(y_j\,t_<)\,,
\end{equation}
with $c=\hat{\mathbf{y}}_1\cdot\hat{\mathbf{y}}_2$ the cosine of the opening angle between the two lines. Summing over the helicities of each line gives the transverse-traceless projector, and the vertices contract the two projectors into
\begin{equation}\label{eq:T2}
T_2(c)=\Pi^{\rm TT}_{ij,kl}(\hat{\mathbf{y}}_1)\,\Pi^{\rm TT}_{ij,kl}(\hat{\mathbf{y}}_2)=\frac{c^4+6c^2+1}{4}\,,\qquad T_2(\pm1)=2\,.
\end{equation}
The tensor bubble is thus the scalar bubble of Section~\ref{sec:bubble}, with its loop integrand weighted by $T_2(c)$.

The ultraviolet coefficient can be read off before any contour is closed. For $y_j\gg x_v$ the uniform expansions of $K_\nu$ and $I_\nu$~\cite{ref29}, in which the order first appears at relative order $1/(y_jt)$, reduce the product of the two kernels to $e^{-(y_1+y_2)|t_1-t_2|}/(4y_1y_2t_1t_2)$. On the shell $\ell\gg k$ the lines are antiparallel, $c\to-1$. For $r>\frac32$ a comoving cutoff then gives
\begin{equation}\label{eq:TTUV}
\psi_2^{\rm TT}\big|_\Lambda=c_{\rm UV}\ln\Lambda+\text{finite}\,,\qquad c_{\rm UV}=\frac{\Gamma(2r-3)\,T_2(-1)}{8\pi^2\,E^{\,2r-3}}\,,\qquad E=x_1+x_2\,,
\end{equation}
which is twice the scalar coefficient and independent of the mass (as for the vector bubble~\cite{ref6}).

The polarization weight acts on the pieces through one rational factor. Expanded in $y_1^2$, $y_2^2$, and $k^2$, the polynomial $(2y_1y_2)^4\,T_2$ is a sum of monomials. Each monomial shifts the exponents $\alpha=(1-p_1)/2$ and $\beta=(1-p_2)/2$ of the loop block by integers, and the shifted blocks sum to
\begin{equation}\label{eq:varrho}
\int\dd^3\ell\,T_2(c)\,y_1^{p_1-1}y_2^{p_2-1}=2\pi\,k^{\,p_1+p_2+1}\,\varrho(\alpha,\beta)\,\hat C_2(p_1,p_2)\,,
\end{equation}
with
\begin{equation}\label{eq:varrhodef}
\varrho=\frac{16\alpha^2\beta^2+16\alpha\beta-12\alpha^2-12\beta^2+27}{8\,\alpha(\alpha+1)\,\beta(\beta+1)}\,.
\end{equation}
On the ultraviolet surface $\alpha+\beta=\frac32$ the factor $\varrho$ is identically $T_2(-1)=2$, consistent with Eq.~\eqref{eq:TTUV}. We have checked Eq.~\eqref{eq:varrho} against direct integration over the loop momentum at three pairs of powers, to $10^{-11}$ or better. The three pieces of the scalar bubble keep their exponents, with $\varrho$ entering only their coefficients.

The tensor bubble also contains pieces with no scalar counterpart. They arise because the poles of $\varrho$ at $\alpha,\beta\in\{0,-1\}$ coincide with zeros of the loop-block factor $1/(\Gamma(\alpha)\Gamma(\beta))$. In the scalar bubble these zeros cancel every pole of the time-ordering factors $\Gamma(1-b_j)$. In the tensor bubble, however, the poles $b_j=1$ and $b_j=2$ survive on each line, because the transverse-traceless projector is not analytic at zero line momentum. With one line on a surviving pole and the other on its rising poles, the powers are $k^{3+2\nu_j+2m}$ and $k^{5+2\nu_j+2m}$, which involve the order of one line only. With both lines on surviving poles the powers are $k^3$, $k^5$, and $k^7$. At the massless order $\nu=\frac32$ the single-line powers meet the ultraviolet poles at $k^{2n}$ with $n\ge4$, and the two cancel, leaving terms $k^{2n}\ln k$. A term $k^6\ln k$ is absent because the numerator of $\varrho$ vanishes at $(\alpha,\beta)=(0,-\frac32)$.

%% file: sec4.tex
\section{Two vertices and any number of lines}\label{sec:close}

In the previous section we wrote the bubble as a sum over trees. In this section we show that, for diagrams with two vertices, the same loop integral can also be written as a single integral over the invariant distance between the vertices. This form serves two purposes. First, it evaluates the bubble independently of the Mellin--Barnes route and fixes the local term by which the families differ from a subtraction in position space. Second, it makes clear where the last integral can be done in known functions. We will find closed forms at conformal coupling, for two massless lines, and for the coefficients of the background on a lattice of points at any mass. We will also find that the two-loop sunset splits into the same three pieces as the bubble, with coefficients that are again sums over trees. Since the form holds for any number of lines, it also gives banana diagrams at higher loops, and we evaluate the three- and four-loop cases in Section~\ref{sec:close.banana}. Reference~\cite{Bassett2026} also starts from the bubble kernel as a function of the de~Sitter invariant. It attaches the external points through the Mehler--Fock transform of the kernel, whereas we integrate over the invariant distance directly.

\subsection{One integral over the invariant distance}\label{sec:close.Z}

We begin with a single line. Its ordered kernel has a classical resolution in which the line momentum $y$ appears only through a flat propagator~\cite{ref29},
\begin{equation}\label{eq:flatres}
K_\nu(yt_>)\,I_\nu(yt_<)=\int_0^\infty\dd\eta\;\frac{\eta\,J_\nu(\eta t)\,J_\nu(\eta t')}{\eta^2+y^2}\,,\qquad\Re\nu>-1\,.
\end{equation}
With $\rho$ the comoving separation of the two end points, the spatial Fourier transform of $1/(\eta^2+y^2)$ is the Yukawa potential $e^{-\eta\rho}/4\pi\rho$. The remaining integral over $\eta$ then gives a Legendre function~\cite{ref29},
\begin{equation}\label{eq:gnu}
\int\frac{\dd^3y}{(2\pi)^3}\,e^{\ii\mathbf{y}\cdot\bm{\rho}}\,K_\nu(yt_>)I_\nu(yt_<) = \frac{g_\nu(Z)}{4\pi^2\,(tt')^{3/2}}\,,\qquad g_\nu\equiv -Q'_{\nu-\frac12}\,,\quad Z=\frac{t^2+t'^2+\rho^2}{2tt'}\,,
\end{equation}
where $Q$ is the Legendre function of the second kind. Up to normalization, this is the bulk-to-bulk propagator of Euclidean anti-de Sitter space, written as a function of the invariant distance~\cite{ref30}. The point $Z=1$ corresponds to coincident end points. Apart from the conformal factor, a line thus depends on a single variable, and its mass enters only through the degree of the Legendre function. For example, $g_{1/2}=1/(Z^2-1)$ for a conformally coupled line, while $g_{3/2}=Z/(Z^2-1)-\frac12\ln\frac{Z+1}{Z-1}$ for a massless one.

Consider now two vertices with energies $x_1,x_2$, joined by $n$ lines of arbitrary orders $\nu_l$ with total momentum $k$. The momentum integrals of such a diagram form a convolution, which becomes a pointwise product in position space. Since all $n$ lines share the same two end points, this product depends on $Z$ alone. In the measure $\prod_{l<n}\dd^3\ell_l/(2\pi)^3$ and at the vertex exponents $r_1=r_2=\frac{3n-2}{2}$, we find
\begin{gather}
\psi_n=2\,(2\pi)^{-2(n-1)}\int_1^\infty\dd Z\;G_n(Z)\,K(Z)\,,\qquad G_n(Z)=\prod_{l=1}^{n}g_{\nu_l}(Z)\,,\label{eq:Zform}\\
D(Z)=E^2+2x_1x_2(Z-1)+k^2(Z^2-1)\,,\qquad K(Z)=\frac{1}{2k}\ln\frac{k\sqrt{D}+k^2Z+x_1x_2}{(x_1+k)(x_2+k)}\,,\nonumber
\end{gather}
so that $\partial_ZK=1/(2\sqrt{D})$ and $K(1)=0$. In other words, the kernel $K$ is elementary and knows nothing about the fields, while the density $G_n$ contains all of the line content and none of the kinematics.

The derivation is short. In position space the $n$ lines give $G_n(Z)/[4\pi^2(t_1t_2)^{3/2}]^n$, so at the exponents above the vertex integrals carry $(t_1t_2)^{-2}$. We write $t_{1,2}=\lambda e^{\pm u/2}$, so that $\rho^2=2\lambda^2(Z-\cosh u)$, and do the angular integral of $e^{-\ii\mathbf{k}\cdot\bm{\rho}}$. The powers of $\lambda$ then combine into $\dd\lambda/\lambda$, and the scale integral of $e^{-\lambda(x_1e^{u/2}+x_2e^{-u/2})}\sin k\rho$ gives an arctangent. The remaining integral over $|u|<\operatorname{arccosh}Z$ vanishes at $Z=1$. Its derivative in $Z$ is elementary, and it equals $\pi/\sqrt{D}$ for every ordering of $x_1$, $x_2$, and $k$. These steps require only $\Re\nu_l>-\frac12$ on each line.

Evaluating loop integrals as the Fourier transform of a product of position-space propagators is a known device. For diagrams without external legs, it reduces de Sitter loops at arbitrary masses to a single integral over the invariant distance~\cite{MarolfMorrison2010a,ref8}. Equation~\eqref{eq:Zform} extends this reduction to diagrams with external legs at general kinematics. In de Sitter space the device has been applied to conformally coupled lines~\cite{ref39,ref40}, to massless lines~\cite{ref41,ref42}, and to massive diagrams in the squeezed limit~\cite{ref5,ref9}. In anti-de Sitter space, diagrams with two vertices and external legs have been organized by the geodesic distance through their spectral transform~\cite{Carmi2019,Carmi2021,Carmi2024}.

Raising a vertex exponent by one acts on the kernel as $-\partial_{x_v}$, and the derivatives of $K$ are algebraic in $Z$. As a result, the one-integral form with an elementary kernel holds at $r_v=\frac{3n-2}{2}+j_v$ for every integer $j_v\geq0$. At other exponents the kernel is still a single integral over $u$, but we have not found it in elementary form. Changing the fields, on the other hand, changes only the density. We use this in Section~\ref{sec:close.closed} for two massless lines and in Section~\ref{sec:dirac} for the Dirac pair. A line with a chemical potential, however, has no flat-propagator form (see Section~\ref{sec:chem}), so the helical-vector bubble is not covered.

\subsection{Subtractions and the local term}\label{sec:close.sub}

For $n\geq2$ the density grows as $(Z-1)^{-n}$ at coincidence, which in these variables is the ultraviolet divergence of the loop. Since $K$ is even in $k$ and its Taylor coefficients in $k^2$ vanish to increasing order at $Z=1$, the divergence sits in the first $n-1$ coefficients, which are local. With $e=Z-1$, $s=\sqrt{E^2+2x_1x_2e}$, and
\begin{equation}\label{eq:K0K2}
K_0=\frac{e}{E+s}\,,\qquad K_2=-\frac{e^2\,(3Ee+6E+es+6s)}{6\,s\,(E+s)^3}\,,
\end{equation}
the subtracted bubble and two-loop sunset become
\begin{align}
\psi_2(k)-\psi_2(0)&=\frac{1}{2\pi^2}\int_1^\infty\dd Z\;g_{\nu_1}g_{\nu_2}\,\big[K-K_0\big]\,,&r_v&=2\,,\label{eq:Zbub}\\
\psi(k)-\psi(0)-k^2\partial_{k^2}\psi(0)&=\frac{1}{8\pi^4}\int_1^\infty\dd Z\;G_3\,\big[K-K_0-k^2K_2\big]\,,&r_v&=\tfrac72\,,\label{eq:Zsun}
\end{align}
for heavy, light, and mixed orders at every $x_1,x_2,k>0$.

The scheme enters through the subtracted terms. In these variables a natural choice is the finite part at coincidence, in which we cut the integral at $Z-1=\epsilon$ and drop the term in $\ln\epsilon$. For the bubble we thus define
\begin{equation}\label{eq:psiren}
\psi_2^{\rm ren}\equiv\frac{1}{2\pi^2}\,\mathrm{FP}\!\int_1^\infty\dd Z\;g_{\nu_1}(Z)\,g_{\nu_2}(Z)\,K(Z)\qquad(r_v=2)\,,
\end{equation}
and at $r_v=2+j_v$ we define $\psi_2^{\rm ren}$ through $j_v$ derivatives $-\partial_{x_v}$ at each vertex. It is natural to ask how this subtraction compares with the comoving cutoff of Section~\ref{sec:bubble}. We find that the two differ by a local term in closed form, $\psi_2^{\rm fin}-\psi_2^{\rm ren}=(8\pi^2E)^{-1}[\ln(x_1/E)+\frac12(\ln2-1)]$ at $r_v=2$. When this is combined with Eq.~\eqref{eq:fin}, the logarithms cancel and the families are related to the finite part by a pure constant,
\begin{equation}\label{eq:local}
\psi_2^{\rm ren}=A+F_0+F_{\rm BG}-\frac{\Gamma(R)}{8\pi^2}\,E^{-R}\big(\gamma_E+\tfrac12+\tfrac12\ln2\big)\,.
\end{equation}
Since the families and the contact term both obey the derivative rule, Eq.~\eqref{eq:local} holds at every $r_v=2+j_v$, within the range of Section~\ref{sec:bubble} ($x_2<x_1$, $|\Re\nu_j|<\frac32$, and $\bar\nu\notin\mathbb{Z}$). Together with Eq.~\eqref{eq:MSbar}, it also gives the $\overline{\rm MS}$ bubble directly from the integral over $Z$,
\begin{equation}\label{eq:renMS}
\psi_2^{\overline{\rm MS}}(\mu_R)=\psi_2^{\rm ren}+\frac{\Gamma(R)}{8\pi^2}\,E^{-R}\big(\gamma_E+\tfrac12-\tfrac12\ln2+\ln\mu_R\big)\,.
\end{equation}
Equivalently, the $\overline{\rm MS}$ coefficients with $n=0$ of $F_{\rm BG}$ are $\Phi^{\overline{\rm MS}}_{0l}=S_0(z_l)+\frac12(1-\ln2)+\frac12\ln\mu_R$, which differ from Eq.~\eqref{eq:rung} by the same constant for every $l$ and every mass.

\begin{figure}[t]
\centering
\includegraphics{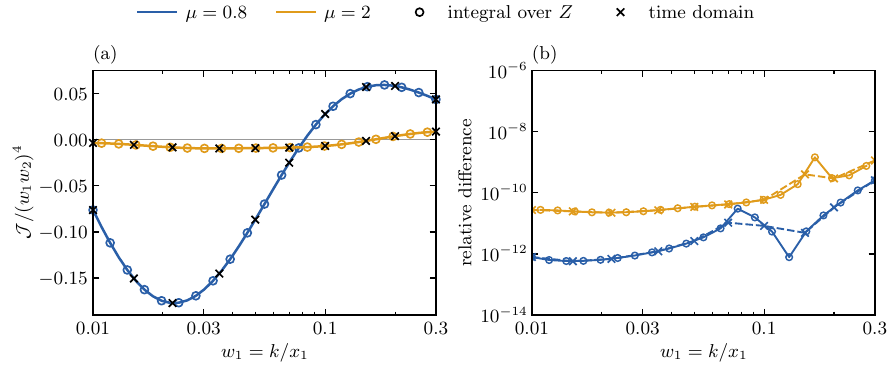}
\caption{The bubble at $r_v=4$ and equal orders $\nu=\pm\ii\mu$, along $w_2=k/x_2=\frac12$, in units of $\mathcal{J}/(w_1w_2)^4$ with $\mathcal{J}=\frac12k^5[\psi_2(\ii\mu)+\psi_2(-\ii\mu)]$ (the units of Ref.~\cite{ref5}, as in Eq.~\eqref{eq:MSbar}) and in $\overline{\rm MS}$ at $\mu_R=\mu$. (a) The sum of the three families with Eq.~\eqref{eq:rung} (lines), the integral over $Z$ of Eq.~\eqref{eq:psiren} with the derivatives taken under the integral and converted by Eq.~\eqref{eq:renMS} (circles), and an independent time-domain evaluation of the cutoff loop converted by Eq.~\eqref{eq:cutMS} (crosses). (b) Relative differences of the two evaluations from the families. The integral over $Z$ and the time-domain evaluation agree with each other to $10^{-12}$. The slow rise at large $w_1$ comes from the truncation of the family series.}
\label{fig:Zcheck}
\end{figure}

We test these relations in Fig.~\ref{fig:Zcheck} at the vertices of Ref.~\cite{ref5}. At every point, the integral over $Z$, the families, and an independent time-domain evaluation agree to $10^{-9}$ or better. This is a genuine test, because the integral over $Z$ is obtained in position space rather than by transforming the Mellin--Barnes integrand of Section~\ref{sec:bubble}. The comparison also shows that the derivatives in the definition of $\psi_2^{\rm ren}$ can be taken under the integral before the finite part.

\subsection{The two-loop sunset}\label{sec:close.sunset}

\begin{figure}[t]
\centering
\includegraphics{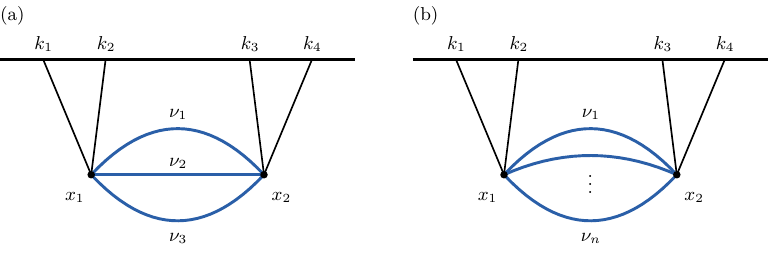}
\caption{(a) The two-loop sunset, with three lines of orders $\nu_1,\nu_2,\nu_3$ between vertices of energies $x_1$ and $x_2$. (b) A banana diagram with $n$ lines (Section~\ref{sec:close.banana}).}
\label{fig:sunset}
\end{figure}

The two-loop sunset (Fig.~\ref{fig:sunset}a) splits into the same three pieces as the bubble, as we now show. The essential point is that its two momentum integrals over powers of the three line energies give a single ratio of Gamma functions. With $z_l=2-2b_l$ and $\bar z=z_1+z_2+z_3$, the iterated two-line block gives~\cite{ref31,ref44}
\begin{equation}\label{eq:sunblock}
\int\dd^d\ell_1\,\dd^d\ell_2\;y_1^{-z_1}\,y_2^{-z_2}\,y_3^{-z_3}=\pi^d\,k^{\,2d-\bar z}\,\frac{\Gamma\big(\frac{\bar z-2d}{2}\big)}{\Gamma\big(\frac{3d-\bar z}{2}\big)}\prod_{i=1}^{3}\frac{\Gamma\big(\frac{d-z_i}{2}\big)}{\Gamma\big(\frac{z_i}{2}\big)}\,,
\end{equation}
which converges for $2d<\Re\bar z<3d$ and $0<\Re z_i<d$ and is continued elsewhere by its Gamma functions. The sunset block thus keeps no Mellin variable. We consider vertex exponents $r_1,r_2$ and three lines of orders $\nu_l$, with $\bar\nu=\nu_1+\nu_2+\nu_3$, in the measure $\dd^3\ell_1\,\dd^3\ell_2/(2\pi)^6$. For $k<x_2<x_1$, closing the contours as in Section~\ref{sec:bubble.families} then gives the same three kinds of family,
\begin{equation}\label{eq:sunfam}
\begin{aligned}
A&:R_1\wedge R_2\,,&\qquad&k^{\,6+2\bar\nu+2p+2q}\;x_1^{-r_1-\bar\nu-2p}\;x_2^{-r_2-\bar\nu-2q}&\qquad&\text{(nonlocal signal)},\\
F_0&:R_1\wedge U\,,&&k^{\,2n}\;x_1^{-r_1-\bar\nu-2p}\;x_2^{6-r_2-2n+\bar\nu+2p}&&\text{(local signal)},\\
F_{\rm BG}&:V_2\wedge U\,,&&k^{\,2n}\;x_1^{-(r_1+r_2-6+l+2n)}\;x_2^{l}&&\text{(background)},
\end{aligned}
\end{equation}
where $p$, $q$, $n$, and $l$ are non-negative integers. On the principal series the nonlocal signal oscillates in $\ln(x_1x_2/k^2)$ and the local signal in $\ln(x_2/x_1)$, both at the single frequency $\mu_1+\mu_2+\mu_3$. The leading nonlocal signal of this diagram was obtained at arbitrary masses from a factorization theorem valid at all loop orders~\cite{ref43}, and for conformally coupled lines the sunset is known to act as a single exchange of shifted dimension~\cite{ref40}. Here we give the coefficients of all three pieces at arbitrary masses.

For this we first note that the two momentum integrals fuse three amputated lines into a single amputated line of order $3+\bar\nu$,
\begin{equation}\label{eq:fusion}
\int\dd^3\ell_1\,\dd^3\ell_2\prod_{l=1}^{3}\Big(\frac{y_l}{2}\Big)^{\nu_l}K_{\nu_l}(y_lT)=\frac{16\pi^3\prod_l[\tfrac32{+}\nu_l]}{[\frac92{+}\bar\nu]}\,\Big(\frac{k}{2}\Big)^{3+\bar\nu}\,T^{-3}\,K_{3+\bar\nu}(kT)\,,
\end{equation}
which follows from the product of the three position-space transforms and converges for $\Re\nu_l>-\frac32$. More generally, by applying the product identity of Section~\ref{sec:bubble.trees} twice (an iteration noted in Ref.~\cite{ref34}), we can write the sunset at fixed times as a sum over tree-level exchanges,
\begin{equation}\label{eq:suntrees}
\begin{aligned}
P(t,T)&=\frac{4\pi^2}{(tT)^3}\sum_{N\geq0}A_N\,K_{\nu''_N}(kT)\,I_{\nu''_N}(kt)\,,\\
A_N&=16\sum_{n=0}^{N}G_n(\nu_1,\nu_2)\,G_{N-n}(\nu_1{+}\nu_2{+}\tfrac32{+}2n,\nu_3)\,,
\end{aligned}
\end{equation}
with $\nu''_N=3+\bar\nu+2N$ and $G_n$ the weights of Eq.~\eqref{eq:Gn} at the indicated orders. The sum converges geometrically for $t<T$, and its leading term as $t\to0$ reproduces Eq.~\eqref{eq:fusion}. The time integrals of each term are those of a tree-level exchange with vertex exponents $r_v-3$. Consequently, each family of the sunset is a sum over $N$ of the same family of these trees, just as for the bubble. For the background this sum converges from $n=2$ on.

The ultraviolet divergences are local, and they reside in the terms of the background with $n=0$ and $n=1$. The ultraviolet poles now come from the factor $\Gamma(\mathcal{S}-3)$ of the assembled integrand. The contour integral continues the loop integrals across its first two poles, where the sum over trees diverges quadratically and logarithmically. With the two vertex times split as $|t-T|>\delta$, the singular terms go as $\delta^{-2}$, $\ln^2\delta$, and $\ln\delta$. The squared logarithm is the nested divergence of a one-loop subdiagram (as for the flat-space sunset in four dimensions). Since all of these terms are polynomial in $k$ of degree zero and two, the two signals and the terms of the background with $n\geq2$ are independent of the scheme.

We now turn to the correlator. In the cutting rule of Eq.~\eqref{eq:cutrule}, cutting one or two lines leaves a single component, so the ends of each cut line would lie on the same branch. Only the uncut term and the triple cut therefore contribute. We expand each Feynman kernel by Eq.~\eqref{eq:GF} and each cut kernel at the earlier time by the connection formula of Section~\ref{sec:bubble.trees}. For real vertex phases, at fixed times $t<T$ and fixed line energies, the two terms then sum to $\sum_\sigma B_\sigma\prod_lK_{\nu_l}(y_lT)\,I_{\sigma_l\nu_l}(y_lt)$, where the sum runs over all eight sign vectors $\sigma$ and
\begin{equation}\label{eq:sunbracket}
B_\sigma=-\prod_l\hat N_l^{\sigma_l}-\prod_l\hat N_l^{-\sigma_l}+2\prod_l\frac{-1}{2\sin\pi\sigma_l\nu_l}=-\,\sigma_1\sigma_2\sigma_3\,\frac{\sinh\pi(\sigma_1\mu_1{+}\sigma_2\mu_2{+}\sigma_3\mu_3)+\ii}{4\sinh\pi\mu_1\,\sinh\pi\mu_2\,\sinh\pi\mu_3}\,.
\end{equation}
Here $\hat N_l^+=1+\bar n_l$ and $\hat N_l^-=-\bar n_l$, with $\bar n_l=\ii e^{\ii\pi\nu_l}/(2\sin\pi\nu_l)$ the occupation of Eq.~\eqref{eq:GF}. The right-hand side also holds for light lines, with $\mu_l\equiv-\ii\nu_l$. Consequently, the family $\sigma$ of the correlator is the corresponding family of the sunset with $\nu_l\to\sigma_l\nu_l$, times $B_\sigma$. We stress that no complex conjugate is to be added, since the sum over $\sigma$ already contains it. Note also that the correlator contains the four frequencies $|\pm\mu_1\pm\mu_2\pm\mu_3|$, whereas the wavefunction coefficient contains only the sum frequency. Each sign vector has the cut weight $W_\sigma=2\ii\sin\pi\rho_\sigma$ with $\rho_\sigma=-r_v+\sum_l\sigma_l\nu_l$. The ultraviolet poles displace the local signal from the nonlocal signal by the even integer six, so the two have the same weight. For the bubble, by contrast, the odd displacement reverses the sign.

The collider coefficients then follow from Eq.~\eqref{eq:fusion}. For a soft vertex of energy $x_2$, with all three lines amputated and weighted by the powers that the hard vertex leaves, the moment is a single time integral of the fused line,
\begin{equation}\label{eq:sunmoment}
\begin{aligned}
\mathcal{M}^{\rm sun}_\sigma&=\int\dd^3\ell_1\,\dd^3\ell_2\prod_l\Big(\frac{y_l}{2}\Big)^{\sigma_l\nu_l}\int_0^\infty\dd t\,t^{\,r_2-1}e^{-x_2t}\prod_lK_{\nu_l}(y_lt)\\
&=\frac{16\pi^{7/2}\prod_l[\tfrac32{+}\sigma_l\nu_l]}{[\frac92{+}\Sigma_\sigma]}\,\frac{k^{2c_\sigma}\,[r_2{-}3{+}c_\sigma]\,[r_2{-}3{-}c_\sigma]}{(x_2+k)^{r_2-3+c_\sigma}\,[r_2{-}\frac52]}\\
&\quad\times{}_2F_1\Big(r_2{-}3{+}c_\sigma,\,c_\sigma{+}\tfrac12;\,r_2{-}\tfrac52;\,\frac{x_2-k}{x_2+k}\Big),
\end{aligned}
\end{equation}
with $\Sigma_\sigma=\sum_l\sigma_l\nu_l$ and $c_\sigma=3+\Sigma_\sigma$. The integrals converge for $r_2>6+\Re\Sigma_\sigma$, and otherwise we define the moment by continuation in $r_2$. With one soft leg ($x_2=k=k_3$) the Gauss function equals one, and the squeezed correlator contains
\begin{equation}\label{eq:sunsq}
F\supset\sum_\sigma B_\sigma\,\frac{[r_1{+}\Sigma_\sigma]}{\prod_l[1{+}\sigma_l\nu_l]}\;\frac{\prod_l[\tfrac32{+}\sigma_l\nu_l]\,[r_2{+}\Sigma_\sigma]\,[r_2{-}6{-}\Sigma_\sigma]}{4\pi^{5/2}\,[\frac92{+}\Sigma_\sigma]\,[r_2{-}\frac52]\,(2k_3)^{r_2-3}}\;\Big(\frac{k_3}{2}\Big)^{3+\Sigma_\sigma}x_1^{-(r_1+\Sigma_\sigma)}\,.
\end{equation}
In the collapsed limit of a four-point function both vertices are hard, and the signal is instead the leading term of the nonlocal signal,
\begin{equation}\label{eq:suncol}
F\supset\sum_\sigma B_\sigma\,\frac{k^6}{512\,\pi^3\,x_1^{r_1}x_2^{r_2}}\Big(\frac{k^2}{4x_1x_2}\Big)^{\Sigma_\sigma}\frac{[r_1{+}\Sigma_\sigma]\,[r_2{+}\Sigma_\sigma]\,[-3{-}\Sigma_\sigma]}{[\frac92{+}\Sigma_\sigma]}\prod_l\frac{[\tfrac32{+}\sigma_l\nu_l]}{[1{+}\sigma_l\nu_l]}\,.
\end{equation}
Both coefficients are ratios of Gamma functions. As a check, the moment of Eq.~\eqref{eq:sunmoment} agrees with direct quadrature to $10^{-16}$. The coefficient of Eq.~\eqref{eq:suncol} agrees with a fit of the integral over $Z$ of Eq.~\eqref{eq:Zsun} to $3\times10^{-9}$ on the principal series and to $7\times10^{-9}$ with one light line. This fit also tests Eq.~\eqref{eq:Zsun} itself.

At fixed internal energies, the difference frequencies are suppressed relative to the sum frequency by the ratio of their brackets. After the two loop integrals, however, all four frequencies share the triple Boltzmann factor $e^{-\pi(\mu_1+\mu_2+\mu_3)}$, so their relative size is set by the Gamma functions and depends on the vertex exponents. For example, at $r_2=\frac72$ and $r_1=\frac{11}{2}$ (a soft vertex with two conformally coupled legs and a hard vertex with $\dot\pi$ legs), the difference-to-sum ratios of Eq.~\eqref{eq:sunsq} lie between 0.54 and 0.85 for $1\leq\mu_l\leq3$. A light line on the complementary series contributes a real power rather than an oscillation, and the formulas above hold unchanged.

\subsection{Banana diagrams at higher loops}\label{sec:close.banana}

Since Eq.~\eqref{eq:Zform} holds for any number of lines, the same single integral gives the wavefunction coefficients of banana diagrams (Fig.~\ref{fig:sunset}b) at every loop order. For conformally coupled and massless lines these act as single exchanges of shifted dimension~\cite{ref40,ref42}, and for massive lines the leading nonlocal signal is known at all loop orders~\cite{ref43}. With $n$ lines the density grows as $(Z-1)^{-n}$ at coincidence, so the first $n-1$ Taylor coefficients in $k^2$ are subtracted, as in Section~\ref{sec:close.sub}. Independently, the part of each ordered kernel that is not analytic in $y^2$, integrated with the $n$-line block
\begin{equation}\label{eq:nblock}
\int\prod_{l=1}^{n-1}\dd^3\ell_l\;\prod_{l=1}^n(y_l^2)^{-\delta_l}=\pi^{\frac{3(n-1)}{2}}\,k^{3(n-1)-2\mathcal{S}}\,\frac{\Gamma\big(\mathcal{S}-\frac{3(n-1)}{2}\big)}{\Gamma\big(\frac{3n}{2}-\mathcal{S}\big)}\prod_{l=1}^n\frac{\Gamma(\frac32-\delta_l)}{\Gamma(\delta_l)}\,,
\end{equation}
(with $\mathcal{S}=\sum_l\delta_l$, which reduces to Eq.~\eqref{eq:sunblock} at $n=3$), gives the nonlocal signal as a series $k^{3(n-1)+2\bar\nu}\sum_jT_jk^{2j}$ with Gamma-function coefficients and leading term
\begin{equation}\label{eq:bananaT0}
T_0=\frac{1}{2^{4n-3}\,\pi^{\frac{3(n-1)}{2}}\,x_1^{r_1}x_2^{r_2}}\Big(\frac{1}{4x_1x_2}\Big)^{\bar\nu}\,\Gamma(r_1{+}\bar\nu)\,\Gamma(r_2{+}\bar\nu)\,\frac{\Gamma\big(-\bar\nu-\frac{3(n-1)}{2}\big)}{\Gamma\big(\frac{3n}{2}+\bar\nu\big)}\prod_{l=1}^n\frac{\Gamma(\frac32+\nu_l)}{\Gamma(1+\nu_l)}\,.
\end{equation}
As a check, we have evaluated the three-loop banana (four lines at $r_v=5$, heavy or light) and the four-loop banana (five heavy lines at $r_v=\frac{13}{2}$) for conformally coupled external scalars. The nonlocal signal fitted from the integral over $Z$ reproduces Eq.~\eqref{eq:bananaT0} to between $10^{-9}$ and $10^{-6}$, and away from the squeezed limit the integral agrees with a direct evaluation in position space to $4\times10^{-9}$ or better. The correlator would further require the cut weights of these diagrams, which we leave to future work.

\subsection{Special cases in closed form}\label{sec:close.closed}

The integral over $Z$ is the last one, and whether it can be done in closed form depends on the density. We find closed forms at conformal coupling for the bubble and the sunset, for the bubble of two massless lines, and for the coefficients of the background of the bubble on a lattice of points at any two masses. At generic orders, however, we find no reduction of the integral over $Z$, and we expect the one-integral form to be the most compact one there.

\paragraph{Conformal coupling.} With $\nu=\frac12$ on every line the density is a power of $Z^2-1$. The substitution $\sqrt{D}=kZ+\tau$ then turns the integral over $Z$ into logarithms and dilogarithms over four linear letters. For the bubble this reproduces the known closed form of the wavefunction coefficient~\cite{ref45,ref46,ref32}, which we recall over the whole shape in Appendix~\ref{app:ccsun}. More interestingly, the twice-subtracted sunset of Eq.~\eqref{eq:Zsun} has a closed form in the same alphabet at general kinematics. It is a rational function plus logarithms plus the weight-two group of the bubble, and no new letter appears (Eq.~\eqref{eq:ccsun} in Appendix~\ref{app:ccsun}). Both closed forms agree with the integral over $Z$, the bubble to $10^{-13}$ and the sunset to $10^{-17}$, at points covering every ordering of $k$, $x_1$, and $x_2$. For conformally coupled lines the sunset was previously known in closed form at $x_1=x_2=k$, both for the wavefunction coefficient~\cite{ref46} and for the in-in correlator~\cite{ref47}. The time-ordered sector of the in-in correlator is also known at general kinematics in dimensional regularization~\cite{ref40}.

\paragraph{Massless lines.} At the massless order $\nu=\frac32$ the density of two lines is $g_{3/2}^2=Z^2/(Z^2-1)^2+\partial_Z\big(\frac14ZL^2\big)$, with $L=\ln\frac{Z+1}{Z-1}$, so its integral from $Z$ to infinity is elementary. We integrate Eq.~\eqref{eq:psiren} by parts, with this integral as the antiderivative of the density. We then substitute $\sqrt{D}=kZ+\tau$, followed by $\tau=b-(b-t_0)w$ with $b=x_1x_2/k$ and $t_0=E-k$. With $u_v=(x_v-k)/(x_v+k)$, the logarithm becomes
\begin{equation}\label{eq:masslessL}
L(w)=\ln\frac{(1-u_1w)(1-u_2w)}{(1-w)(1-u_1u_2w)}\,,
\end{equation}
and for $k<x_v$, the physical region, we find
\begin{equation}\label{eq:masslessbubble}
\begin{aligned}
4\pi^2\psi_2^{\rm ren}&=\frac{1}{4E}\Big[1+\ln\frac{2E^2}{(x_1+k)(x_2+k)}\Big]+\frac{1}{4\Delta_x}\ln\frac{x_1+k}{x_2+k}-\frac{3\,I_1}{4k}+\frac{x_1x_2\,I_2}{4k^3}\\
&\quad+\frac{1}{k^2}\Big[E\,u_1u_2\,M(u_1u_2)-x_2u_1\,M(u_1)-x_1u_2\,M(u_2)\Big]\,,
\end{aligned}
\end{equation}
where $\Delta_x=x_1-x_2$, $I_n=\int_0^1L^n\,\dd w/w$, and $M(\alpha)=\int_0^1L\,\dd w/(1-\alpha w)$. Here $I_1=\zeta(2)+\Li_2(u_1u_2)-\Li_2(u_1)-\Li_2(u_2)$, and $M$ is a sum of dilogarithms. The only function of weight three is $I_2$, a sum of the functions $\sum_{m,n\ge1}\alpha^m\beta^n/[mn(m+n)]$ at the letters $\alpha,\beta\in\{1,u_1u_2,u_1,u_2\}$ (Appendix~\ref{app:ccsun}). Every term is real for $k<x_v$. Equation~\eqref{eq:masslessbubble} agrees with Eq.~\eqref{eq:psiren} to $5\times10^{-44}$ at seven points with $k/x_v$ between $0.017$ and $0.99$. It is the tensor bubble of Eq.~\eqref{eq:TTbubble} at $\nu=\frac32$ without its polarization weight $T_2(c)$. That weight depends on the angle between the two lines, so Eq.~\eqref{eq:Zform} does not apply to the tensor bubble itself.

\paragraph{The background coefficients on the lattice.} The background of Section~\ref{sec:bubble.trees} requires the constant $S_0(z)$ of Eq.~\eqref{eq:S0} at $z_l=r_2-\frac32+l$. For conformally coupled legs this is the half-integer lattice, on which $S_0$ takes a closed form at any two masses. The spectral weight of two lines of different masses is a very-well-poised series of the same type~\cite{MarolfMorrison2010a}, and no closed form of the equal-mass spectral function is known in three dimensions~\cite{ref7}. With $\omega_1=\nu_1^2-\frac14$, $\omega_2=\nu_2^2-\frac14$, and $\Psi_j=\gamma_E+\psi^{(0)}(\nu_j+\frac12)$, we find for $l\geq1$
\begin{equation}\label{eq:S0lat}
S_0(l+\tfrac12)=c_1^{(l)}\,\Psi_1+c_2^{(l)}\,\Psi_2+c_0^{(l)}+\tfrac12\gamma_E+\tfrac12\ln2-\tfrac14\,,
\end{equation}
where the coefficients are rational in $\omega_1$ and $\omega_2$,
\begin{equation}\label{eq:S0coef}
\begin{aligned}
l=1:&\quad c_1=-\frac{\omega_1}{2(\omega_1-\omega_2)}\,,\quad c_2=\frac{\omega_2}{2(\omega_1-\omega_2)}\,,\quad c_0=0\,,\\
l=2:&\quad c_1=-\frac{\omega_1(\omega_1-\omega_2-2)}{2\mathcal{D}}\,,\quad c_2=\frac{\omega_2(\omega_1-\omega_2+2)}{2\mathcal{D}}\,,\quad c_0=\frac{\omega_1+\omega_2-2}{2\mathcal{D}}\,,\\
&\quad\mathcal{D}=\omega_1^2-2\omega_1\omega_2-2\omega_1+\omega_2^2-2\omega_2\,.
\end{aligned}
\end{equation}
Higher $l$ follow from a recurrence that steps $z$ by two. With $w=(2z+1)(2z+3)$ and $c_K=\frac12\gamma_E+\frac12\ln2-\frac14$,
\begin{equation}\label{eq:S0rec}
p_1(z)\,S_0(z+2)+p_0(z)\,S_0(z)=q_1(z)\Psi_1+q_2(z)\Psi_2+q_0(z)+c_K\,[p_1(z)+p_0(z)]\,,
\end{equation}
where
\begin{equation}\label{eq:S0recc}
\begin{aligned}
&p_1=(z+\tfrac32)(z+\tfrac52)\prod_{\pm\pm}(z+\tfrac12\pm\nu_1\pm\nu_2)\,,\\
&p_0=-(z-\tfrac12)(z+\tfrac12)\prod_{\pm\pm}(z+\tfrac32\pm\nu_1\pm\nu_2)\,,\\
&q_1=\tfrac12\omega_1(z+1)(w-4\omega_1+4\omega_2)\,,\quad q_2=\tfrac12\omega_2(z+1)(w+4\omega_1-4\omega_2)\,,\\
&q_0=-\tfrac1{16}(z+1)\,w\,(w-4\omega_1-4\omega_2)\,.
\end{aligned}
\end{equation}
At equal masses the coefficients have poles, and their residues turn the difference of digamma functions into a trigamma function. For example,
\begin{equation}\label{eq:S0eq}
\begin{aligned}
S_0(\tfrac32;\nu,\nu)&=-\tfrac12\psi^{(0)}(\nu+\tfrac12)-\frac{\omega_1}{4\nu}\,\psi^{(1)}(\nu+\tfrac12)+\tfrac12\ln2-\tfrac14\,,\\
S_0(\tfrac52;\nu,\nu)&=-\tfrac12\psi^{(0)}(\nu+\tfrac12)-\frac{\omega_1-1}{4\omega_1}+\tfrac12\ln2-\tfrac14\,,
\end{aligned}
\end{equation}
and the trigamma function appears only at odd $l$.

Since $p_0(\frac12)=0$, the recurrence cannot reach the lowest point $z=\frac12$, which the bubble needs at $r_2=2$. Instead, $S(\nu_1)\equiv S_0(\frac12;\nu_1,\nu_2)$ obeys a first-order relation in the order of one line. With $x=\nu_1+\frac12$ and $P_b=\psi^{(0)}(\nu_2+\frac12)-\ln2+\frac12$,
\begin{equation}\label{eq:S0half}
\begin{aligned}
&(\nu_1^2-\nu_2^2)\,S(\nu_1+1)+\big((\nu_1+1)^2-\nu_2^2\big)\,S(\nu_1)=-F(\nu_1)\,,\\
&F=x^2\psi^{(0)}(x)+(\tfrac12-\ln2)\,x^2+\tfrac{x}{2}-\omega_2\,P_b\,,\\
&S_0(\tfrac12;\nu,\nu)=-\tfrac12\big[\psi^{(0)}(\nu+\tfrac12)+1-\ln2\big]\,,\\
&S_0(\tfrac12;\nu_1,\nu_2)=-(\nu_1^2-\nu_2^2)\sum_{m\geq0}\frac{(-1)^m\,F(\nu_1+m)}{[(\nu_1+m)^2-\nu_2^2]\,[(\nu_1+m+1)^2-\nu_2^2]}\,.
\end{aligned}
\end{equation}
The equal-mass value is in closed form. At unequal masses a single alternating sum remains, and it has a closed form again when $\nu_1-\nu_2$ is an integer. We derive all of these statements for real orders and reach the principal series by continuation in $\nu$. We have checked Eqs.~\eqref{eq:S0lat}--\eqref{eq:S0half} against direct summation of Eq.~\eqref{eq:S0} at heavy, light, and mixed orders, to $10^{-12}$ or better.

%% file: sec5.tex
\section{The triangle and the box}\label{sec:triangle}

With three and four vertices Mellin variables survive the loop integral, so the pieces no longer reduce to single sums. The triangle was treated in Ref.~\cite{ref26}. For conformally coupled legs, that reference gives the late-time logarithm of the wavefunction coefficient and shows that it cancels in the correlator. For inflaton legs, it derives a selection rule for the nonlocal signal and gives the pair-frequency signal to all orders in the squeezed limit, on the line $k_1=k_2$. In this section we extend that signal to every shape of the triangle. We then give the block of the box and its collapsed limit, which show where the reduction stops.

\begin{figure}[t]
\centering
\includegraphics{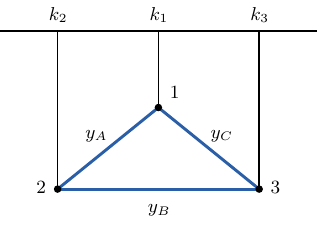}
\caption{The triangle. Vertex $v$ has energy $x_v=k_v$. Line $A$ joins vertices 1 and 2, $B$ joins 2 and 3, and $C$ joins 3 and 1, and line $l$ has energy $y_l$. Thus $A$ is opposite $k_3$.}
\label{fig:triangle}
\end{figure}

\subsection{The triangle at every shape}\label{sec:triangle.shape}

We take a $\dot\pi\sigma_i\sigma_j$ vertex at each corner of the triangle ($r_v=2$), with the soft leg $\mathbf{k}_3$ on the vertex joining lines $B$ and $C$ (Fig.~\ref{fig:triangle}). Non-analytic dependence on $k_3$ can then arise only from loop momenta $\ell\sim k_3$ with the soft-vertex time $t\sim1/k_3$. This is because the region $\ell\sim k_3$, $t\sim1$ is scaleless, while the region $\ell\sim k_1$ is analytic in $\mathbf{k}_3$. In the remaining region each soft line is $K_\nu(yt)$ at the soft vertex, and at its hard end it enters through its expansion at small argument. Section~\ref{sec:regions.split} applies the same argument to any one-loop diagram. In the wavefunction coefficient that end is an ordered kernel, $I_\nu(z)\to(z/2)^\nu/\Gamma(1+\nu)$, so the nonlocal signal oscillates only at $\mu_B+\mu_C$, with a single sign~\cite{ref26}. The cut and Feynman lines of the correlator supply both powers $(z/2)^{\pm\nu}$, and with them the other sign pairs, including the difference frequency $\mu_B-\mu_C$~\cite{ref26}. On $k_1=k_2$ the pair family is $k_1^{-3}(k_3/k_1)^{1+\nu_B+\nu_C}\sum_Nc_N(k_3/k_1)^{2N}$, where the $c_N$ are finite double sums of Gamma functions and moments of the hard line~\cite{ref26}.

The selection rule of Ref.~\cite{ref26} has a simple origin, and the same origin extends this series to every shape. Expanding the hard line around the soft region puts a numerator $e^{\bl\cdot\mathbf{J}}$ into the soft loop integral, which Eq.~\eqref{eq:C2J} of Appendix~\ref{app:box} evaluates as a series in $\mathbf{J}^2$. The term with $j$ powers of $\mathbf{J}^2$ reaches the time integral of the soft vertex with the factor $1/\Gamma(r_3-1-j)$, where $r_3$ is the power of that vertex. At $r_3=1$ no term survives, so the conformally coupled triangle has no pair family at any order, in agreement with Ref.~\cite{ref26}. This is a special case of the exclusion rule of Ref.~\cite{SilentHeavy2026}. At $r_3=2$ only $j=0$ survives, and in that term the two soft lines share the soft momentum collinearly, as $\lambda\mathbf{k}_3$ and $(1-\lambda)\mathbf{k}_3$. The pair family is then a single parameter integral over tree-level chains with three sites,
\begin{multline}\label{eq:pairshape}
\psi_3\supset\frac{1}{8\pi}\,\Gamma(2+\nu_{BC})\,\Gamma(-1-\nu_{BC})\;k_3\int_0^1\!\dd\lambda\,\big[\lambda(1-\lambda)\big]^{1/2}\int_0^\infty\!\dd t_1\,\dd t_2\;t_1^{\,r_1-1}\,t_2^{\,r_2-1}\,e^{-k_1t_1-k_2t_2}\\
\times I_{\nu_C}(\lambda k_3t_1)\;K_{\nu_A}(y_\lambda t_>)\,I_{\nu_A}(y_\lambda t_<)\;I_{\nu_B}\big((1-\lambda)k_3t_2\big)\,,
\end{multline}
where $y_\lambda=|\mathbf{k}_1+\lambda\mathbf{k}_3|$ is the energy of the hard line. Line $C$ ends on the vertex of $\mathbf{k}_1$ and line $B$ on that of $\mathbf{k}_2$. We expand in the variables
\begin{equation}\label{eq:shapevars}
R=\frac{2k_3}{k_1+k_2}\,,\qquad\delta=\frac{k_1-k_2}{k_3}\,,
\end{equation}
which a triangle restricts to $R\le2$ and $|\delta|\le1$. On $k_1=k_2$ we have $\delta=0$ and $R=k_3/k_1$. With $K=\frac12(k_1+k_2)$ the pair family at $r_3=2$ is
\begin{equation}\label{eq:scalarshape}
\psi_3\supset K^{\,1-r_1-r_2}\,R^{\,1+\nu_B+\nu_C}\sum_{N\ge0}A_N(\delta)\,R^N\,,\qquad A_{2N}(0)=\frac{c_N}{(2\pi)^3}\,,
\end{equation}
where $A_N$ is a polynomial in $\delta$ of degree at most $N$ and parity $(-1)^N$, given explicitly in Appendix~\ref{app:box}, and $c_N$ are the coefficients of Ref.~\cite{ref26}, which uses the bare loop measure. For equal soft lines and equal hard vertices the odd orders vanish. The zeros of $y_\lambda$ lie at $|R|\ge2$, so the series converges for $R<2$ at every $\delta$, up to the collinear point $k_3=k_1+k_2$. The series contains the moments
\begin{equation}\label{eq:hardmoment}
\mathcal{H}_{\nu}(\alpha_0,\alpha_1)=\int_0^\infty\!\dd t_0\,\dd t_1\;t_0^{\alpha_0}t_1^{\alpha_1}e^{-t_0-t_1}K_{\nu}(t_>)I_{\nu}(t_<)
\end{equation}
of the hard line at integer-shifted exponents, each a Gamma function times one ${}_3F_2$ at unit argument (Eq.~\eqref{eq:Hmoment} of Appendix~\ref{app:box}). They all follow from one of them, since the Bessel equation of the kernel gives
\begin{equation}\label{eq:Hrec}
(2\alpha_0+3)\,\mathcal{H}_\nu(\alpha_0+1,\alpha_1)=\big[(\alpha_0+1)^2-\nu^2\big]\,\mathcal{H}_\nu(\alpha_0,\alpha_1)+\frac{\Gamma(\alpha_0+\alpha_1+2)}{2^{\,\alpha_0+\alpha_1+2}}
\end{equation}
and the same relation in $\alpha_1$. The whole series therefore needs a single ${}_3F_2$ at unit argument.

At $\mu=(1,1.3,0.7)$ and $r_v=2$ the dependence on the shape is mild. On $k_1=k_2$ the modulus of the series exceeds that of its leading term by factors of $1.05$, $1.14$, and $1.78$ at $R=0.3$, $0.5$, and the equilateral shape $R=1$. At the folded shapes $k_1=2k_2=2k_3$ and $k_2=2k_1=2k_3$ the factors are $1.25$ and $1.31$. On $k_1=k_2$ the series reproduces $c_N/(2\pi)^3$, with the $c_N$ of Ref.~\cite{ref26}, to $10^{-30}$ through $N=12$, and it obeys the exchange of the hard legs order by order to $3\times10^{-27}$. Against the triangle integral of Ref.~\cite{ref26} at $\delta=\frac12$ and $\delta=1$, at 17 values of $R$ up to $0.5$ each, subtracting the series through $N=4$ leaves a polynomial in $R$ to $5\times10^{-14}$. Free fits then return $A_1$ and $A_2$ to $2\times10^{-4}$. Section~\ref{sec:compare} gives a closed approximation to the sum and compares it with the other loops.

\subsection{The box}\label{sec:triangle.box}

\begin{figure}[t]
\centering
\includegraphics{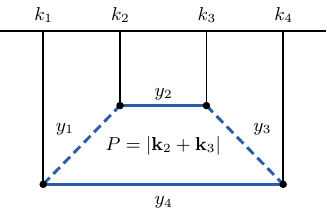}
\caption{The box. Line $i$ joins vertices $i$ and $i+1$ and has energy $y_i$. In the collapsed limit $P=|\mathbf{k}_2+\mathbf{k}_3|\to0$ the dashed lines 1 and 3 are soft, and lines 2 and 4 are hard.}
\label{fig:box}
\end{figure}

How far the reduction goes is set by the number of Mellin variables that survive every elementary integral. A one-loop polygon with $N\leq4$ lines retains one Mellin variable for each of the $N(N-1)/2-1$ ratios of its invariants. Thus the bubble keeps none, the triangle two, and the box five. The box block $C_4=\int\dd^3\ell\prod_i(y_i^2)^{-\delta_i}$ (Fig.~\ref{fig:box}) is then a five-fold Barnes integral (given in Appendix~\ref{app:box}), which agrees with an independent Feynman-parameter evaluation to $1.4\times10^{-6}$. Inserting it into Eq.~\eqref{eq:master} expresses the loop-integrated box as a thirteen-fold Mellin--Barnes integral on straight contours. For $\Sigma_4=\sum_i\delta_i$ the vertex conditions require $\frac32<\Re\Sigma_4<\min(3,\frac12\sum_vr_v)$, where the bound 3 comes from the Barnes form of the block. Since the decay of the integrand along its power-law direction requires only $\Re\Sigma_4<\frac72$, it does not narrow this strip further. Straight contours therefore exist for the box with conformally coupled legs ($r_v=1$) as well as with inflaton legs ($r_v=2$). In contrast to the bubble and the conformally coupled triangle, the box never crosses an ultraviolet pole of its block.

In the collapsed limit $P=|\mathbf{k}_2+\mathbf{k}_3|\to0$, lines 1 and 3 become soft, and they factor out of the time integrals. The soft momentum integral is then the two-line block of Eq.~\eqref{eq:C2}, and each hard line leaves one closed-form moment. With the hard-line energies $y_2=|\mathbf{k}_2|$ and $y_4=|\mathbf{k}_1|$ at $P=0$, we find
\begin{align}
\psi_\Box&\supset C_\Box\,P^{\,3+2\nu_1+2\nu_3}\,\big[1+O(P)\big]\,,\nonumber\\
C_\Box&=\frac{1}{8\pi^{3/2}\,4^{\,1+\nu_1+\nu_3}}\,\frac{\Gamma(\frac32+\nu_1)\,\Gamma(\frac32+\nu_3)\,\Gamma(-\frac32-\nu_1-\nu_3)}{\Gamma(1+\nu_1)\,\Gamma(1+\nu_3)\,\Gamma(3+\nu_1+\nu_3)}\nonumber\\
&\quad\times\frac{\mathcal{H}_{\nu_2}(r_2{-}1{+}\nu_1,\,r_3{-}1{+}\nu_3)}{y_2^{\,r_2+r_3+\nu_1+\nu_3}}\;\frac{\mathcal{H}_{\nu_4}(r_4{-}1{+}\nu_3,\,r_1{-}1{+}\nu_1)}{y_4^{\,r_4+r_1+\nu_1+\nu_3}}\,,\label{eq:Cbox}
\end{align}
where $\mathcal{H}_\nu$ is the hard-line moment of Eq.~\eqref{eq:hardmoment}, a single ${}_3F_2$ at unit argument (Eq.~\eqref{eq:Hmoment} of Appendix~\ref{app:box}). It has the factorized form that Ref.~\cite{ref9} derived for the leading nonlocal signal of the box at equal and unequal masses, a soft factor from the two soft lines times one hard moment on each side. We have not compared the normalizations. The factorization persists at every order in $P$, for scalar lines and for transverse helical lines, and the limit of one soft leg reduces in the same way to the soft factor of the triangle times a two-line tree (Appendix~\ref{app:box}). Away from such limits all five Barnes variables remain, and we find no further reduction.

%% file: secR.tex
\section{Squeezed limits and late-time templates}\label{sec:regions}

So far we have organized each diagram by its pieces and their weights. For phenomenology, however, the quantity of interest is usually the amplitude of the nonlocal signal in a squeezed or collapsed configuration. Most templates for loop signals estimate it at late times~\cite{ref4,ref78,ref17,ref20,ref23,ref24,AokiStrumia2026}. In this section we explain which part of such an estimate is exact and which part is not. The answer depends on whether the soft lines end on hard vertices only, and on whether the lines respect the de Sitter boosts. It also organizes the rest of the paper.

\subsection{Hard and soft regions}\label{sec:regions.split}

The argument of Section~\ref{sec:triangle.shape} applies to any one-loop diagram. Consider first a squeezed bispectrum, in which the soft momentum $k_3$ enters at one vertex, and let two lines join that vertex to the rest of the diagram. Non-analytic dependence on $k_3$ then arises only from loop momenta $\ell\sim k_3$ with the soft-vertex time $t\sim1/k_3$, since every other region is either scaleless or analytic in $\mathbf{k}_3$. In this region the hard vertices see the soft lines only through the small-argument expansion of their ends, so the leading coefficient factorizes into a hard part and a soft vertex,
\begin{equation}\label{eq:factorize}
\psi\supset\sum_\sigma\mathcal{H}_\sigma\,\mathcal{M}_\sigma\,,\qquad \mathcal{M}_\sigma\propto k_3^{\,3-r_2+\Sigma_\sigma}\,,
\end{equation}
where $\sigma$ labels the late-time branches of the two soft lines and $\Sigma_\sigma=\sigma_1\nu_1+\sigma_2\nu_2$. The hard part contains the hard vertices with the soft lines at their late-time ends $(y/2)^{\pm\nu}$, together with any hard line between them, that is, any internal line that carries a hard momentum. The soft vertex contains the two soft lines in full, together with the external mode at that vertex and the loop integral. For two scalar lines it is the moment $\mathcal{M}_\sigma$ of Eq.~\eqref{eq:twomoment}, a ratio of Gamma functions. The squeezed coefficient $\mathcal{M}_{\rm tri}$ of the triangle in Ref.~\cite{ref26} has the same structure, with the folded moment of the third line as its hard part. At higher orders in $k_3/k_1$ the hard part is expanded further, while the soft vertex remains a moment of the same kind (Sections~\ref{sec:triangle.shape} and~\ref{sec:dirac.shape}).

The late-time estimates of Refs.~\cite{ref4,ref78,ref17,AokiStrumia2026} treat the hard part in the same way (sometimes with further approximations of a hard line). What they replace is the soft vertex, where they keep only the late-time branches of the soft lines. In the bispectrum this replacement is not controlled. Reference~\cite{Song2026} reached the same conclusion for the fermion triangle, and Ref.~\cite{HanLeiQinYin2026} justified the late-time expansion at the hard vertex of the fermion bubble. Since the soft leg ends on the vertex at which the soft lines meet, the vertex time is $t\sim1/k_3$ and the soft lines are crossing the horizon there. In terms of the late-time series of the lines (in powers of $k_3/x_v$), the bispectrum sits on the boundary of convergence at $x_v=k_3$. Every order of that series, and the local signal as well, then lands on the same power of $k_3/k_1$ (as we will see for the helical vector in Section~\ref{sec:chem.bubble}). The leading late-time term is thus one contribution among many, rather than an approximation to the coefficient.

The collapsed limit of a four-point function is different. There the soft momentum $s=|\mathbf{k}_1+\mathbf{k}_2|$ flows through lines whose ends all sit at hard vertices, with energies $x_1,x_2\gg s$. Both ends of each soft line are then far outside the horizon, the late-time series converges, and its leading term is the coefficient. This is the factorization of Refs.~\cite{ref9,ref43}, in which the soft lines are taken to the boundary. The soft lines enter only through their late-time coefficients, and the loop integral is the massless block of Eq.~\eqref{eq:C2} at complex powers. Equation~\eqref{eq:suncol} for the sunset and Eq.~\eqref{eq:Cbox} for the box are examples. For the helical-vector bubble of Section~\ref{sec:chem.bubble}, whose late-time series converges for $s<\min(x_1,x_2)$, the collapsed coefficient is likewise the leading term of Eq.~\eqref{eq:vecleading}. In the bispectrum, by contrast, the same diagram exceeds that term by a factor of six to thirty.

We are thus led to three cases, which the remaining sections take in turn.
\begin{enumerate}[label=(\roman*),leftmargin=2em,itemsep=2pt]
\item If both ends of every soft line are hard, the late-time coefficient is exact. It is then a product of Gamma functions with the moments of the hard lines, as for the parity-odd box of Section~\ref{sec:chem.parity}.
\item If the soft lines end on the soft vertex and neither the lines nor the vertices break the de Sitter boosts, the late-time estimate misses by a universal factor. This factor depends only on the dimension of the pair and on the soft vertex (Section~\ref{sec:regions.universal}).
\item If a chemical potential breaks the boosts, no such factor exists, and the soft vertex must be computed. We do so for the helical vector, the charged scalar, and the Dirac pair and triangle in Sections~\ref{sec:chem} and~\ref{sec:dirac}.
\end{enumerate}

\subsection{A universal factor}\label{sec:regions.universal}

When the lines and vertices respect the boosts, the error of the late-time soft vertex can be corrected once and for all. For two scalar lines this follows directly from Eq.~\eqref{eq:twomoment}. We replace $K_\nu(yt)$ at the soft vertex by its late-time branch $\frac12\Gamma(-\nu)(yt/2)^{\nu}$ and integrate over the loop momentum with the block of Eq.~\eqref{eq:C2}. At equal orders the ratio of the exact moment to this late-time moment is then
\begin{equation}\label{eq:Univ}
\mathcal{U}(\Delta,r_2)=\frac{\sqrt{\pi}\;2^{\,1+\Delta-r_2}\,\Gamma(r_2-\Delta)}{\Gamma(r_2-1)\,\Gamma\big(\tfrac32-\Delta\big)}\,,\qquad \Delta=3+2\nu\,,
\end{equation}
where $\Delta$ is the dimension of the pair and $r_2$ is the power of the soft vertex. At $r_2=1$ (a non-derivative coupling of a conformally coupled leg) $\mathcal{U}$ vanishes through $1/\Gamma(r_2-1)$. The exact soft vertex then has no signal, although its late-time version does, in agreement with the exclusion rule of Ref.~\cite{SilentHeavy2026}. For the derivative couplings of the inflaton ($r_2=2$) the factor is large instead.

The same factor holds for lines of any spin contracted into a scalar vertex. To see this, we use the invariant-distance form of Section~\ref{sec:close.Z}. There the two soft lines depend only on $Z=(t^2+t'^2+\rho^2)/2tt'$, and at the hard vertex $t\to0$, so $Z\to(t'^2+\rho^2)/2tt'$. The leading term $c_\Delta Z^{-\Delta}$ of the pair density then becomes $c_\Delta(2t)^\Delta\,[t'/(t'^2+\rho^2)]^\Delta$. This is the bulk-to-boundary propagator of a scalar of dimension $\Delta$ in Euclidean anti-de~Sitter space, whose Fourier transform is $t'^{3/2}K_{\Delta-3/2}(kt')$ up to normalization. The late-time estimate keeps $c_\Delta$ but drops $t'^2$ against $\rho^2$, which leaves only the non-analytic leading power of this function. The ratio of the two soft-vertex integrals is Eq.~\eqref{eq:Univ}, which depends on the dimension of the pair and on the soft vertex but not on the spin of the lines. The subleading terms of the large-$Z$ expansion give the higher families. Each is a boundary operator of dimension $\Delta+2t$ in the late-time expansion of the pair, together with its descendants, and it is corrected by $\mathcal{U}$ at its own dimension, which we have checked to $10^{-30}$ for the first six families. At a fixed order in $k_3/k_1$, however, the families mix, so the ratio at that order is not a single $\mathcal{U}$. For the Higgs portal of Section~\ref{sec:SM} at $\mu=1$, it has modulus $14.4$ and $21.6$ at the orders $(k_3/k_{12})^2$ and $(k_3/k_{12})^4$, against $|\mathcal{U}|=58.5$ and $280$ at the dimensions of the corresponding families.

For heavy lines ($\nu=\ii\mu$) at $r_2=2$, the factor has modulus $10.1$, $11.6$, $14.8$, and $20.3$ at $\mu=\frac12$, $1$, $2$, and $4$, and it grows as $\mu^{1/2}$. Reference~\cite{ref17} found the same growth with the mass for a single exchange at tree level. For light lines the leading family has $\Delta=3-2\nu$, and $\mathcal{U}\to1$ as $\nu\to\frac32$. It is $1.03$ at $\nu=1.45$ and $1.65$ at $\nu=0.3$, but it diverges at $\nu=\frac12$ and is negative between $\nu=\frac12$ and $\frac34$. Exact bubbles of this kind have been computed at equal masses with the spectral decomposition, for derivatively coupled scalars and for massive vectors~\cite{ref5,ref14}. Equation~\eqref{eq:Univ} gives the size of the correction to the late-time templates in closed form, for any mass and spin. In terms of an operator basis, it covers every operator whose heavy bilinear is a scalar or a pseudoscalar under the de~Sitter isometries, such as $(\partial\phi)^2H^\dagger H$, $F_{\mu\nu}F^{\mu\nu}(\partial\phi)^2$, and $F_{\mu\nu}\tilde F^{\mu\nu}(\partial\phi)^2$. For the last of these the leading dimension is $4+2\nu$, and the soft moment summed over the nine helicity pairs follows $\mathcal{U}(4+2\nu,r_2)$ to $10^{-23}$.

A tensor bilinear $T_{\mu\nu}$ contracted with two gradients of the inflaton is not covered. The basis of Ref.~\cite{Craig2024} contains two such operators at the same order as $F_{\mu\nu}F^{\mu\nu}(\partial\phi)^2$, namely $(D_\mu H)^\dagger D_\nu H\,\partial^\mu\phi\,\partial^\nu\phi$ and $F_{\mu\rho}F_\nu{}^\rho\,\partial^\mu\phi\,\partial^\nu\phi$. With the hard vertex at late times, a boundary operator of dimension $\Delta$ then reaches the soft vertex through one of three functions that the isometries fix. They are the trace and the traceless part of $T_{\mu\nu}$ for a scalar operator, and the bulk-to-boundary function of a spin-2 operator. Each soft-vertex integral is again a moment of $K_{\Delta-3/2}$. For the late-time estimate we keep the leading late-time term of each massive mode function at the soft vertex. The ratios of the exact integrals to this estimate are then
\begin{equation}\label{eq:Uspin}
\mathcal{U}\,,\qquad-\,\mathcal{U}\,\frac{\Delta^2-3\Delta+2r_2^2-14r_2+20}{(\Delta+1)(\Delta+4r_2-8)}\,,\qquad\mathcal{U}\,\frac{\Delta-r_2-1}{2(\Delta-2)}\,,
\end{equation}
with $\mathcal{U}=\mathcal{U}(\Delta,r_2)$. At $r_2=2$ they become $\mathcal{U}$, $\mathcal{U}\,(3-\Delta)/(\Delta+1)$, and $\mathcal{U}\,(\Delta-3)/(2\Delta-4)$. The factor thus survives up to a rational function of $\Delta$ and $r_2$, which depends on the spin of the boundary operator and on the tensor structure of the vertex. We have checked these functions against the soft moments in momentum space, for two scalar lines to $10^{-29}$ and for two Proca lines with all polarizations to $5\times10^{-19}$. Two consequences follow at $r_2=2$. For $(D_\mu H)^\dagger D_\nu H$ the late-time soft vertex of the leading term vanishes, since it is a total derivative for a massless leg, while the exact one is $(\Delta-1)/(2\Delta+2)$ times that of the scalar vertex. For $F_{\mu\rho}F_\nu{}^\rho$ the leading term has an isotropic part with the ratio $2\Delta\,\mathcal{U}/(\Delta+1)$ and a part proportional to $P_2(\hat{\mathbf{k}}_1\cdot\hat{\mathbf{k}}_3)$ with the ratio $\nu\,\mathcal{U}/(1+2\nu)$, both at $\Delta=3+2\nu$. Their moduli are $19$ and $5.2$ at $\mu=1$, and $26$ and $7.2$ at $\mu=2$. At $\Delta=\frac32+\ii\mu$ the spin-2 integral is proportional to $(\mu^2+\frac94)/\cosh\pi\mu$, which is the mass dependence of the mixing of a massive spin-2 field with the inflaton at tree level~\cite{Kumar2019}.

A chemical potential falls outside Eq.~\eqref{eq:Univ}. It breaks the boosts, and a line with a chemical potential has no flat-propagator form (Section~\ref{sec:close.Z}), so there is no invariant-distance argument. The corresponding ratios are not universal, and we will find them to range from $0.41$ to $113$ (Sections~\ref{sec:chem} and~\ref{sec:dirac}).

\section{The Standard Model background}\label{sec:SM}

We now apply the results of Section~\ref{sec:regions} to the loops of the Standard Model background~\cite{ref4,ref78,ref22}, in which the Higgs, the electroweak gauge bosons, and the top quark couple to the inflaton through functions of its kinetic term $X$. For Standard Model signals when the Higgs has a large expectation value during inflation, see Refs.~\cite{ref21,ref76,ref77,ref79,ref81}. These couplings are derivative ($r_2=2$), so the exclusion rule of Ref.~\cite{SilentHeavy2026} does not apply, and the late-time templates miss by the factor $\mathcal{U}$. We quote signal sizes as $f_{\rm NL}^{\rm osc}$, the modulus of the coefficient of $(k_3/k_1)^{2+2\ii\mu}$ in the shape $(k_1k_2k_3)^2\langle\zeta^3\rangle'/(2\pi)^4P_\zeta^2$, with $P_\zeta=2.1\times10^{-9}$.

\subsection{The Higgs portal}\label{sec:SM.higgs}

\paragraph{The model.} The Higgs couples to the inflaton through $f_H(X)\,H^\dagger H$, and we use the vertex $-[\partial G\partial G+\mathbf{k}_1\cdot\mathbf{k}_2\,GG]$ of Eq.~(4.4) of Ref.~\cite{ref78}. We denote by $\nu_H$ the order of the Higgs line, with $\nu_H=\ii\mu_H$ for a heavy Higgs. For a light Higgs, $-\nu_H$ takes the place of $\nu_H$ in the leading family.

\begin{figure}[t]
\centering
\includegraphics{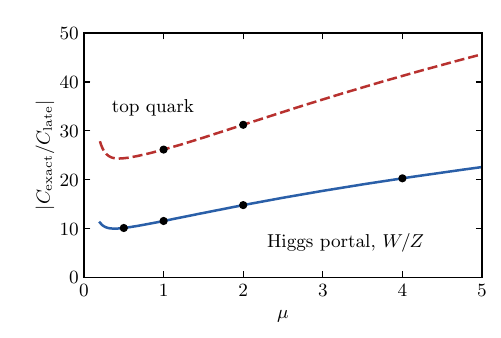}
\caption{The ratio of the exact squeezed coefficient to its late-time estimate, $|\mathcal{U}(\Delta,2)|$ of Eq.~\eqref{eq:Univ}, at the dimension of the leading pair operator. Solid, the Higgs-portal and $W/Z$ loops ($\Delta=3+2\ii\mu$). Dashed, the first term of the top-quark loop ($\Delta=4+2\ii\mu$). The points are the entries of Table~\ref{tab:summary}. For the light Higgs of non-Higgs inflation ($\nu_H=1.45$) the ratio is $1.03$.}
\label{fig:Umass}
\end{figure}

\paragraph{The leading coefficient.} Our late-time version of the bubble, with the soft lines on their same-branch late-time terms, reproduces the coefficient $C_H$ of Ref.~\cite{ref78} to $10^{-29}$ for heavy and light Higgs masses. The exact coefficient is therefore
\begin{equation}\label{eq:SMcorr}
C_H^{\rm exact}=\mathcal{U}(3+2\nu_H,2)\,C_H\,,\qquad \mathcal{U}(3+2\nu,2)=\frac{\sqrt{\pi}\;2^{\,2+2\nu}\,\Gamma(-1-2\nu)}{\Gamma(-\frac32-2\nu)}\,.
\end{equation}
For a Higgs heavier than about $1.5H$ the late-time template is too small by an order of magnitude (by a factor of $10$ to $20$ for $\mu$ between $\frac12$ and $4$), while for the light Higgs of non-Higgs inflation ($\nu\simeq1.4$ to $1.5$) it is accurate to a few percent. Figure~\ref{fig:Umass} shows the factor for the three loops of this section.

\paragraph{Beyond the squeezed limit.} The bispectrum follows at every shape in closed form. We use the normalization of Eq.~(4.4) of Ref.~\cite{ref78}, in which $\langle\delta\phi^3\rangle'=\mathcal{I}/(2k_1^3k_2^3k_3^3)$ at $f_H'=\dot\phi_0=H=1$. The soft vertex has $r_2=2$, and the hard vertex has the powers $r_1=J=2$, $3$, and $4$ with the coefficients $c_4=k_1^2k_2^2-(\mathbf{k}_1\cdot\mathbf{k}_2)\,k_1k_2$, $c_3=-(\mathbf{k}_1\cdot\mathbf{k}_2)\,k_{12}$, and $c_2=-\mathbf{k}_1\cdot\mathbf{k}_2$. At $x_1=k_{12}$ and $x_2=k=k_3$, the families of Section~\ref{sec:bubble.trees} then give, in $\overline{\rm MS}$ at $\mu_R=2$,
\begin{equation}\label{eq:higgsshape}
\mathcal{I}=-\frac{k_3^2}{2}\sum_{\epsilon=\pm}\sum_{J=2}^{4}c_J\bigg[\sum_{t\ge0}\frac{G_t\,\hat I_{\nu'_t}(J-\frac32,k_{12})}{\sqrt{2\pi k_3}\,\cos\pi\nu'_t}+\sum_{m\ge0}\frac{(-1)^m\zeta_m\,\hat I_{\zeta_m}(J-\frac32,k_{12})}{\sqrt{8\pi^3k_3}}\Big(S_0(\zeta_m)+\frac12\Big)\bigg]\,.
\end{equation}
The sum over $\epsilon$ combines the wavefunction coefficients at $\pm\nu_H$ into the correlator. Both lines have the order $\epsilon\nu_H$, so $\nu'_t=2\epsilon\nu_H+\frac32+2t$, the weights $G_t$ are those of Eq.~\eqref{eq:Gn}, and $\hat I$ is the moment of Eq.~\eqref{eq:Imoment} with line momentum $k_3$. The first sum is the nonlocal and local signal. At $x_2=k_3$ their two moments of the soft vertex combine into $\int_0^\infty\dd t\,t^{-1/2}e^{-k_3t}K_{\nu'_t}(k_3t)=\sqrt{\pi/2k_3}\,\pi/\cos\pi\nu'_t$, which gives the cosine. The second sum is the background, with the coefficients of Eq.~\eqref{eq:rung}. Since $r_2=2$, it sits on the half-integer lattice $\zeta_m=m+\frac12$, where $S_0(\zeta_m)$ at equal orders follows in closed form from Eqs.~\eqref{eq:S0rec}, \eqref{eq:S0eq}, and~\eqref{eq:S0half}. Reference~\cite{ref16} integrated this bispectrum numerically at general kinematics. For $\mu=1$ the signal exceeds its leading squeezed term by $4\%$, $9\%$, and $30\%$ at $R=0.4$, $0.6$, and $1$, with nearly the same values for $\mu$ between $\frac12$ and $4$ (Fig.~\ref{fig:smshape}, dash-dotted). As a complex amplitude the leading term is good to $10\%$ up to $R=0.62$ for the leading family at $\mu=1$, and up to $0.60$ with all families. For $\nu=1.4$ the two agree to $2\%$ up to $R=1$.

\paragraph{Checks.} The sums over trees and lattice points come from the Mellin--Barnes representation of Section~\ref{sec:bubble}. We have compared Eq.~\eqref{eq:higgsshape} with the integral over the invariant distance of Section~\ref{sec:close.Z}, which uses no Mellin variables. The two agree to $6\times10^{-14}$ at five shapes with $R$ between $0.1$ and $1.8$.

\paragraph{Size.} For the Higgs portal at $\mu_H=1$ the factor $\mathcal{U}$ raises $f_{\rm NL}^{\rm osc}$ from $9.0\times10^{-6}\,g_H^2$ (Eq.~(4.38) of Ref.~\cite{ref78}, with $g_H=f_H'\dot\phi_0$) to $1.0\times10^{-4}\,g_H^2$. This remains far below the reach of planned surveys, since the derivative expansion requires $g_H\lesssim1$.

\subsection{The W and Z bosons}\label{sec:SM.wz}

\paragraph{The model.} For the gauge bosons the vertex is $f_A(X)F_{\mu\nu}F^{\mu\nu}$, with all three polarizations of the Proca field. The pair enters only through the two-point function of $F_{\mu\nu}F^{\mu\nu}$, which depends on the invariant distance alone. The field strengths of a Proca line of order $\nu$ follow from the scalar line $g_\nu$ of Eq.~\eqref{eq:gnu}. With $h=g_\nu/4\pi^2$ and primes on $h$ for derivatives in $Z$,
\begin{equation}\label{eq:FFline}
\langle F_{\mu\nu}(x)F_{\rho'\sigma'}(x')\rangle=-4\,\Big[h'\,\partial_{[\mu}\partial_{[\rho'}Z+h''\,\partial_{[\mu}Z\,\partial_{[\rho'}Z\Big]\,\partial_{\nu]}\partial_{\sigma']}Z\,,
\end{equation}
where the brackets antisymmetrize each pair of indices with unit weight, and the mass enters through $\nu$ alone (the longitudinal mode included). Contracting two such lines, we find
\begin{equation}\label{eq:WZdensity}
\begin{aligned}
\langle F^2(x)\,F^2(x')\rangle&=\frac{24}{(4\pi^2)^2}\Big\{\big[2Zg_\nu'-M^2g_\nu\big]^2+4g_\nu'^{\,2}\Big\}=\frac{w_A(\Box)\,g_\nu^2}{(4\pi^2)^2}\,,\\
w_A(P)&=4P^2-(16M^2+24)P+24M^4+48M^2\,,
\end{aligned}
\end{equation}
with $M^2=\nu^2-\frac94$ and $\Box=(Z^2-1)\partial_Z^2+4Z\partial_Z$ the Laplacian on functions of $Z$. Each tree of the product identity of Eq.~\eqref{eq:product} is an eigenfunction of $\Box$ with the eigenvalue $P_t=\nu_t'^{\,2}-\frac94$. The $W/Z$ loop is therefore the scalar bubble of Section~\ref{sec:bubble} with the weight $w_A(P_t)$ under its sum over trees.

\paragraph{The leading coefficient.} The leading weight is $w_A(P_0)=\frac32(2\nu+1)^2(2\nu+3)^2$. It vanishes for a massless vector, as conformal invariance requires. The exact squeezed coefficient is thus $\mathcal{U}(3+2\nu,2)$ times its late-time version, in closed form. Our late-time coefficient differs from the coefficient $C_A$ of Ref.~\cite{ref78}. For the late-time two-point function of $F^2$ we find the factor $24(\frac12-\mu_1)^2$ in place of the $27(23-6\mu_1)^2H^4/M_A^4$ of their Eq.~(4.33), with $\mu_1=-\nu$. The two coefficients then differ by the factor $\frac89(\frac12-\mu_1)^2(\frac14-\mu_1^2)^2/(23-6\mu_1)^2$, whose modulus is $3.1\times10^{-3}$, $0.10$, and $3.5$ at $\mu=1$, $2$, and $4$. Our coefficient has a finite limit for a massless vector, whereas theirs grows as $M_A^{-4}$. The same construction reproduces their $C_H$, and their coefficient for $(D_\mu H)^\dagger D^\mu H$ (Eq.~(4.21) of Ref.~\cite{ref78}), to $5\times10^{-29}$.

\begin{figure}[t]
\centering
\includegraphics{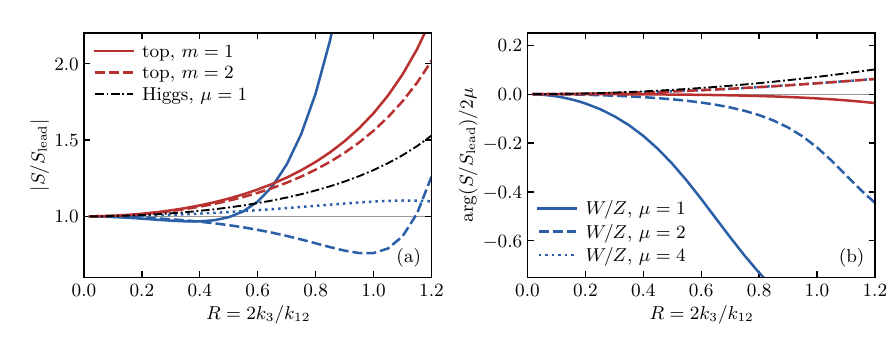}
\caption{The pair signal of the $W/Z$ and top-quark loops away from the squeezed limit, on $k_1=k_2$, with the Higgs portal at $\mu=1$ for comparison. (a) The modulus of the signal over its leading squeezed term. (b) The phase of the same ratio divided by $2\mu$. The $W/Z$ loop is shown at $\mu=1$, $2$, and $4$, and the top-quark loop at $m=1$ and $2$ with the vertex $(\partial_\mu\delta\phi)^2$.}
\label{fig:smshape}
\end{figure}

\paragraph{Beyond the squeezed limit.} The $W/Z$ loop follows at every shape in the same form as the Higgs portal. Its families are the sum over trees with the weights $w_A(P_t)$. Its background is the sum over the lattice points $\zeta_m=m+\frac12$ with the weights $w_A(\zeta_m^2-\frac94)$ and the same coefficients $S_0(m+\frac12)$. This renormalizes the composite $F^2$ by the $\overline{\rm MS}$ subtraction of the scalar pair, with $w_A(\Box)$ acting on the soft vertex. Any other subtraction adds a polynomial of degree two in $\zeta_m^2$ under the lattice sum, which amounts to three contact terms (against one for the Higgs portal). Away from the squeezed limit the higher trees matter sooner than for the Higgs portal, since their weights are larger, with $|w_A(P_1)/w_A(P_0)|=11.6$, $4.6$, and $1.7$ at $\mu=1$, $2$, and $4$. At $\mu=1$ the modulus of the signal stays within $4\%$ of its leading squeezed term up to $R=0.5$, but its phase has moved by $0.18$ and $0.57$~rad at $R=0.3$ and $0.5$, and at $R=1$ the signal is $3.9$ times the leading term (Fig.~\ref{fig:smshape}). At $\mu=2$ the ratio of the two has the modulus $0.98$, $0.94$, and $0.76$ at $R=0.3$, $0.5$, and $1$, and at $\mu=4$ it is $1.01$, $1.03$, and $1.10$. As a complex amplitude the leading term is good to $10\%$ up to $R=0.22$, $0.49$, and $0.52$ at $\mu=1$, $2$, and $4$ (Table~\ref{tab:shapedomain}).

\paragraph{Checks.} We have checked the sums against the integral over the invariant distance of Section~\ref{sec:close.Z}, with the density written through $g_\nu$ and $g_\nu'$ as in Eq.~\eqref{eq:WZdensity}. The third difference in $k^2$ at fixed vertex energies removes the contact terms and converges at coincidence without subtraction. The two agree to $5\times10^{-17}$ at five points with $k/x_1$ between $0.05$ and $0.9$, for $\mu=1$, $2$, and $4$. The closed form also reproduces our numerical soft moment to $10^{-23}$ at $\mu=1$ and $2$, at the leading and at the next order of the hard vertex. That moment is summed over the nine helicity pairs, with the electric and magnetic parts of $F_{\mu\nu}$ kept exactly at the soft vertex. The individual pairs, by contrast, have ratios between $1.4$ and $2\times10^4$ for $1\le\mu\le3$, so only their sum is universal.

\paragraph{Size.} In the normalization of Ref.~\cite{ref78} (their Eq.~(4.41)) and with our late-time coefficient, the factor $\mathcal{U}$ raises $f_{\rm NL}^{\rm osc}$ from $1.1\times10^{-4}\,g_A^2$ to $1.3\times10^{-3}\,g_A^2$ at $\mu=1$, with $g_A=f_A'H^4\dot\phi_0$. With their printed $C_A$ the late-time value would be $0.036\,g_A^2$. The derivative expansion requires $g_A<H^2/\dot\phi_0\simeq3\times10^{-4}$, so this signal stays far out of reach.

\subsection{The top quark}\label{sec:SM.top}

\paragraph{The model.} We first bring the operator $f_\Psi(X)\,\bar\Psi\,\ii\gamma^\mu D_\mu\Psi$ of Ref.~\cite{ref78} to a simpler form. By the equations of motion and an integration by parts it equals $m\,f_\Psi(X)\,\bar\Psi\Psi$ plus a coupling of $\partial_\mu f_\Psi$ to the vector current. The current term drops out of the loop. Its correlator with $\bar\Psi\Psi$ vanishes by charge conjugation, and its correlator with itself is removed by current conservation up to contact terms. The pair density of $\bar\Psi\Psi$ is again a function of the invariant distance. With $\nu=\ii\mu$ and the scalar lines $g_\pm=g_{\nu\pm\frac12}$ of the two chiral components (Appendix~\ref{app:diracline}),
\begin{equation}\label{eq:topdensity}
\langle\bar\Psi\Psi(x)\,\bar\Psi\Psi(x')\rangle=-\frac{4}{(4\pi^2)^2}\Big[g_+'^{\,2}+g_-'^{\,2}-2Z\,g_+'g_-'\Big]=\frac{2}{(4\pi^2)^2}\Big[\Box-\big(2\nu-\tfrac12\big)^2+\tfrac94\Big]\,g_{\nu+\frac12}^{\,2}\,.
\end{equation}

\paragraph{The leading coefficient.} The top-quark loop is thus the scalar bubble at the order $\nu+\frac12$, with the weight $2[\nu_t'^{\,2}-(2\nu-\frac12)^2]=4(2t+3)(2\nu+1+t)$ under the sum over trees and $\nu_t'=2\nu+\frac52+2t$. These weights are the spectral density of Ref.~\cite{ref25} (their Eq.~(3.37)) in our variables. The trees have the dimensions $4+2\nu+2t$, so the bubble starts at $(k_3/k_1)^{4+2\ii\mu}$, one power later than a boson loop. At late times this is the statement that $\bar\psi\psi$ has no component of dimension $3+2\nu$, because the two branches of each line sit in opposite eigenspaces of $\gamma^0$. The first term is $\mathcal{U}(4+2\nu,2)$ times its late-time estimate, with modulus $26$ at $m=1$ and $31$ at $m=2$. Every tree vanishes for a Yukawa coupling of the massless leg ($r_2=0$ or $1$) through $1/\Gamma(r_2-1)$, in agreement with Ref.~\cite{ref25}. The template of Ref.~\cite{ref78} for this loop (their Eq.~(4.40)) corresponds to a pair operator of dimension $2+2\ii\mu$, and their two-point function (their Eq.~(4.28)) falls as $(\tau\tau'/X^2)^{2+2\ii\mu}$. Since each fermion mode decays as $(-\tau)^{3/2\pm\ii\mu}$, no such term exists, and the signal starts two powers of $k_3/k_1$ later.

\paragraph{Beyond the squeezed limit.} The families carry the weights $4(2t+3)(2\nu+1+t)$ on the trees of the orders $\nu+\frac12$. The background is the lattice sum with the weights $2[\zeta_m^2-(2\nu-\frac12)^2]$ and the coefficients $S_0(m+\frac12)$ at those orders, up to two contact terms. The series contains only every second power, starting at dimension $4+2\nu$. With the vertex $(\partial_\mu\delta\phi)^2$ of Ref.~\cite{ref78}, the signal exceeds its leading term by $4\%$, $12\%$, and $67\%$ at $R=0.3$, $0.5$, and $1$ for $m=1$ (by $3\%$, $10\%$, and $56\%$ for $m=2$), and its phase moves by less than $0.04$~rad ($0.18$~rad). As a complex amplitude the leading term is good to $10\%$ up to $R=0.49$ at $m=1$ and $2$ (Fig.~\ref{fig:smshape} and Table~\ref{tab:shapedomain}).

\paragraph{Checks.} With the explicit propagator the Dirac equation holds to $10^{-41}$, and the trace of the vector current with $\bar\Psi\Psi$ vanishes to $10^{-41}$, so the two forms of the operator have the same pair density at separated points. In closed form the series reproduces our numerical coefficients of the orders one, three, five, and seven to $6\times10^{-29}$ at $m=1$ and $2$. The sums agree with the integral over the invariant distance of the density of Eq.~\eqref{eq:topdensity} to $2\times10^{-18}$ at the same five points as for the $W/Z$ loop, for $m=1$ and $2$.

\paragraph{Size.} In the same convention, the coefficient of the first term, $(k_3/k_1)^{3+2\ii m}$ in the shape, is $2.1\times10^{-3}\,g_\Psi^2$ at $m=H$ and $4.6\times10^{-5}\,g_\Psi^2$ at $m=2H$, with $g_\Psi=f_\Psi'H^2\dot\phi_0$. The derivative expansion again requires $g_\Psi<H^2/\dot\phi_0$, so this signal is also far out of reach.

%% file: sec7.tex
\section{Vectors and scalars with a chemical potential}\label{sec:chem}

A chemical potential is the line change with the largest phenomenological consequences, since it lifts the Boltzmann suppression of the collider signal~\cite{ref17,ref23,ref71}. In this section we show that, for the $F\tilde F$ chemical potential of a massive vector, it changes the weights of the pieces and leaves their exponents unchanged. We begin with a Barnes form of the helical line that keeps both time orderings in one kernel. We then give the pair-frequency signal of a loop of helical vectors in closed form at general masses, summed over all nine helicity pairs, and compare it with the late-time estimate of Ref.~\cite{ref17}. We restore the hard line of the model in the triangle, give the parity-odd collapsed trispectrum, and end with charged scalars, whose chemical potential sits on the vertex rather than on the lines.

\subsection{The helical line}\label{sec:chem.line}

A rolling background coupled to $F\tilde F$ gives a massive vector a chemical potential $\kappa$ that splits its transverse helicities $h=\pm1$~\cite{ref17}. We call such a field a helical vector. In conformal time $\tau$, the helicity-$h$ mode and its Bunch--Davies solution are
\begin{equation}\label{eq:helmode}
\frac{\dd^2A_h}{\dd\tau^2}+\Big(y^2+\frac{2h\kappa y}{\tau}+\frac{m^2}{\tau^2}\Big)A_h=0\,,\qquad
A_h=\frac{e^{\pi h\kappa/2}}{\sqrt{2y}}\,W_{K,\nu}(2\ii y\tau)\,,\qquad K=-\ii h\kappa\,,\quad\nu=\ii\mu\,,
\end{equation}
where $\mu^2=m^2-\frac14$ includes the vector offset of Eq.~\eqref{eq:mus}. Note that $\kappa$ leaves the order unshifted and enters only through the first index $K$. On the rotated contour the ordered kernel, normalized to unit jump (like $K_\nu(yt_>)I_\nu(yt_<)$), is
\begin{equation}\label{eq:helkernel}
\mathcal{K}_h(y;t,t')=\frac{\Gamma(a_+)}{\Gamma(1+2\nu)}\,\frac{W_{K,\nu}(2y\,t_>)\,M_{K,\nu}(2y\,t_<)}{2y\sqrt{t\,t'}}\,,\qquad
a_\pm=\tfrac12\pm\nu-K\,,
\end{equation}
which reduces to the scalar kernel at $\kappa=0$.

The Barnes form of Eq.~\eqref{eq:GBarnes} does not carry over directly to this line, but it does once we apply Kummer's transformation at one end. Inserting the Barnes form of $e^{z/2}M_{K,\nu}(z)$ at the end $t$ and that of Kummer's transformed series at the end $t'$, we find
\begin{equation}\label{eq:helBarnes}
\mathcal{K}_h(y;t,t')=e^{-y(t-t')}\iint\frac{\dd s\,\dd s'}{(2\pi\ii)^2}\;
\frac{\Gamma(-s)\,\Gamma(-s')\,\Gamma(a_++s)\,\Gamma(b)\,\Gamma(1-b)}{\Gamma(1{+}2\nu{+}s)\,\Gamma(1{+}2\nu{+}s')\,\Gamma(a_--s')}\,(2y)^{b-1}\,t^{\,\nu+s}\,t'^{\,\nu+s'}\,,
\end{equation}
with $b=1+2\nu+s+s'$, for all positive $t$ and $t'$ in either order. The straight contours lie in $\Re s<0$, $\Re s'<0$, $\Re(a_++s)>0$, and $0<\Re b<1$. When $\Re a_+\leq0$, as for two of the Dirac indices below, the contour of $s$ instead separates the poles of $\Gamma(-s)$ from those of $\Gamma(a_++s)$. To derive Eq.~\eqref{eq:helBarnes}, we close the contour of $s'$ to the right. On the poles of $\Gamma(-s')$ the residues sum to Kummer's series of $e^{-yt'}M_{K,\nu}(2yt')$, and the remaining integral is the Barnes form of $e^{yt}W_{K,\nu}(2yt)$. On the poles $b=1+k$ of $\Gamma(1-b)$ the relation $a_+-a_-=2\nu$ turns the two $\kappa$-dependent Gamma functions into a polynomial, so the integrand in $s$ becomes a rational function times $(t/t')^s$. On the principal series this function has no poles to the left of the contour, so these terms vanish for $t>t'$, while for $t<t'$ they supply the other time ordering. We have checked the identity numerically in both orderings, on the principal series to $3\times10^{-23}$, on the complementary series at $\nu=0.3$ to $5\times10^{-20}$, and at the four indices $K=\pm(\frac12\pm\ii\kappa)$ of the Dirac line of Section~\ref{sec:dirac} to $2\times10^{-22}$.

As in the scalar case, the pair $\Gamma(b)\Gamma(1-b)$ encodes the time ordering, while the chemical potential enters only through $a_+$ at one end and $a_-$ at the other. The price is the prefactor $e^{-y(t-t')}$, which shifts the two vertex energies by $\pm y$. One may ask whether this prefactor can be removed, so that $y$ enters only through a power. Within Gamma-function kernels it cannot, since the double Mellin transform of the ordered product obeys a three-term recurrence whose middle term is proportional to $K$. A Gamma-function kernel would require one of $\frac12\pm\nu\pm K$ to be a non-positive integer. Consequently, a helical line inside a loop enters the vertex energies as $x_v\pm y$, and the loop block does not apply to it directly. However, as we will see in Section~\ref{sec:chem.bubble}, the pair-frequency signal of a helical loop does not need it.

Using Eq.~\eqref{eq:helBarnes}, the vertex integrals of a helical single exchange become Euler integrals at the shifted energies $x_1{+}y$ and $x_2{-}y$. The exchange is then a two-fold Barnes integral that decays exponentially in every direction. It reduces to a product of two Gauss functions plus an ordering correction in one Kamp\'e de F\'eriet series~\cite{ref82,ref85} (Appendix~\ref{app:helical}), equivalent to the series of Refs.~\cite{ref67,ref53}. The ordering correction is not small (between $0.09$ and $0.67$ of the exchange at the points we checked).

The weights follow from the connection data of the line. At late times the Whittaker function splits into the two branches $M_{K,\pm\nu}$ with coefficients $\Gamma(\mp2\nu)/\Gamma(a_\mp)$, and the occupation of helicity $h$ is~\cite{TongXianyu2022,SouTongWang2021}
\begin{equation}\label{eq:helocc}
n_h=\frac{e^{2\pi h\kappa}+e^{-2\pi\mu}}{2\sinh2\pi\mu}\,,\qquad \frac{n_h}{1+n_{-h}}=e^{-2\pi(\mu-h\kappa)}\,,
\end{equation}
which reduces to the Bose--Einstein number at $\kappa=0$. Because the powers of the vertex time are free of $\kappa$, the exponents and the cut weights of Table~\ref{tab:weights} are those of the neutral vector line. The chemical potential changes only the moduli of the branch coefficients, and it turns the simple pole of the exchange at the partial-energy point $x_v=-y$ into a branch point with monodromy $e^{-2\pi h\kappa}$.

\subsection{The helical-vector bubble}\label{sec:chem.bubble}

\begin{figure}[t]
\centering
\includegraphics{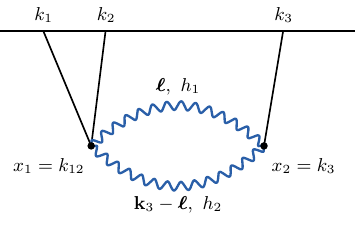}
\caption{The helical-vector bubble of Eq.~\eqref{eq:vecvertex}. Two vector lines of helicities $h_1,h_2$, with loop momenta $\bl$ and $\mathbf{k}_3-\bl$, join the quartic vertex at $x_1=k_{12}$ ($r_1=4$) to the cubic vertex at $x_2=k_3$ ($r_2=2$), on which the soft leg ends.}
\label{fig:vecbubble}
\end{figure}

\paragraph{The model.} We now turn to the model of Ref.~\cite{ref17}. A $U(1)$ gauge field is Higgsed by a scalar whose expectation value the rolling inflaton induces, so that $m_A=g\sigma_0$. The operator $-\phi F\tilde F/(4\Lambda_F)$ gives the chemical potential $\kappa=\dot\phi_0/\Lambda_F$. The leading one-loop bispectrum diagrams are a helical-vector triangle and a bubble dressed by inflaton--Higgs mixings. After the Higgs lines and the hard helical-vector line are contracted to local vertices (as in Ref.~\cite{ref17}), both diagrams reduce to one effective bubble with the vertices
\begin{equation}\label{eq:vecvertex}
\mathcal{H}_{\rm int}=\tfrac{\lambda_1}{4}\,\phi'^2\,\eta^{\mu\nu}A_\mu A_\nu+\tfrac{\lambda_2}{2}\,a\,\phi'\,\eta^{\mu\nu}A_\mu A_\nu\,,
\end{equation}
with $\lambda_1=-4\rho_4$ and $\lambda_2=-2\rho_3$ in terms of the effective couplings $\rho_4$ and $\rho_3$ of Ref.~\cite{ref17}. Section~\ref{sec:chem.hard} restores the hard line. The bubble is shown in Fig.~\ref{fig:vecbubble}. The contraction $\eta^{\mu\nu}A_\mu A_\nu$ couples all nine helicity pairs $(h_1,h_2)\in\{\pm1,0\}^2$, whereas Ref.~\cite{ref17} keeps the doubly enhanced transverse pair. At unit couplings we quote the shape
\begin{equation}\label{eq:vecshape}
S\equiv(k_1k_2k_3)^2\,\langle\delta\phi_{\mathbf{k}_1}\delta\phi_{\mathbf{k}_2}\delta\phi_{\mathbf{k}_3}\rangle'=\frac{k_1k_2k_3}{16}\,F(k_3;k_{12},k_3)\,,
\end{equation}
where $F(k;x_1,x_2)$ is the helicity-summed loop, normalized as in Appendix~\ref{app:helical}. For the enhanced helicity, this loop was computed numerically in Refs.~\cite{ref80,Niu2022gw,Niu2022tri}. Its pair-frequency signal for the enhanced transverse helicity and one quartic coupling was given as a series in Ref.~\cite{ref18}.

\paragraph{The pair-frequency pieces.} We first show that the time ordering drops out of the pair-frequency pieces. On the rotated contour the transverse mode $u$ and its conjugate decompose on the late-time branches,
\begin{equation}\label{eq:vecbranch}
u^*(\ii t)=B_+M_{-K,\nu}(2yt)+B_-M_{-K,-\nu}(2yt)\,,\qquad u(\ii t)=A_+M_{-K,\nu}(2yt)+A_-M_{-K,-\nu}(2yt)\,,
\end{equation}
with the coefficients of Appendix~\ref{app:helical}. The backward branch of the in-in contour follows by Schwarz reflection, $u(-\ii t)=\overline{u^*(\ii t)}$, since $\overline{W_{K,\nu}}=W_{-K,\nu}$ on the principal series. The cutting rule of Eq.~\eqref{eq:cutrule} therefore survives the chemical potential. The longitudinal line does not feel $\kappa$, since $F\tilde F$ vanishes on helicity zero. We now insert the branch decompositions into the four Schwinger--Keldysh assignments. A term in which a line sits on the same late-time branch at both vertices is a product $M_{\sigma\nu}(2yt_1)M_{\sigma\nu}(2yt_2)$, which is symmetric in $t_1\leftrightarrow t_2$. The two orderings then add to one, and the ordered kernel never appears. Hence the pair piece $(\sigma_1,\sigma_2)$, which scales as $k^{3+2\sigma_1\nu+2\sigma_2\nu+N}$ at order $N$ of the late-time expansion, is a sum of products of single-time integrals,
\begin{equation}\label{eq:vecpair}
F^{\sigma_1\sigma_2}_{\bl}=\sum_{a,b=\pm}w^{(1)}_{ab,\sigma_1}w^{(2)}_{ab,\sigma_2}\,T^{(a)}_{\sigma_1\sigma_2}(x_1;r_1)\,T^{(b)}_{\sigma_1\sigma_2}(x_2;r_2)\,,
\end{equation}
where
\begin{equation}\label{eq:vecT}
T^{(\pm)}_{\sigma_1\sigma_2}(x;r)=\int_0^\infty\!\dd t\,t^{\,r-2}e^{-xt}M_{\mp K_1,\sigma_1\nu}(2y_1t)\,M_{\mp K_2,\sigma_2\nu}(2y_2t)\,.
\end{equation}
The per-line weights $w_{ab,\sigma}$ have $\kappa$-independent phases, and the chemical potential appears only in their modulus,
\begin{equation}\label{eq:vecweight}
\big|2y\,\beta_+\big|=\frac{\sqrt{n_h(1+n_h)}}{2\mu}\,,\qquad \beta_\sigma\equiv w_{+-,\sigma}\,,
\end{equation}
with $n_h$ of Eq.~\eqref{eq:helocc}. The per-line amplitude of the pair piece is thus $\sqrt{n_h(1+n_h)}$, which is $\simeq e^{-\pi(\mu-h\kappa)}$ for $h\kappa<\mu$. This derives the rule, used in Refs.~\cite{ref17,ref23}, that the chemical potential replaces $\mu$ by $\mu-h\kappa$ in the Boltzmann factor of the enhanced helicity.

The remaining ingredients are elementary. The polarization contractions are Laurent polynomials in the opening cosine $c$ of the two lines. For two transverse lines, $|\bm{\epsilon}^{h_1}\cdot\bm{\epsilon}^{h_2}|^2=\frac14(1-h_1h_2c)^2$, so equal helicities have unit overlap at $c=-1$, as angular momentum conservation at a scalar vertex requires. The pairs with a longitudinal line are given in Appendix~\ref{app:helical}. As for the tensor bubble, every monomial $k^py_1^{d_1}y_2^{d_2}$ shifts the loop block by integers. The single-time integrals expand in the branch series $M_{K,\nu}(z)=z^{\frac12+\nu}\sum_ne_nz^n$, whose coefficients are terminating Gauss functions at argument two. Since $e_1=-K/(1+2\nu)$, the powers in this series advance in unit steps rather than in the even steps of the scalar bubble. Each piece is then a late-time series of Gamma functions (given in Appendix~\ref{app:helical}) that converges absolutely for $k<\min(x_1,x_2)$, and at $\kappa=0$ it reduces to the nonlocal signal of the scalar bubble. Its leading term is fully explicit,
\begin{equation}\label{eq:vecleading}
F^{++}_{N=0}=-\frac{\sinh^2\!\pi\mu}{\pi^2}\,\big(2y_1\beta_+^{(1)}\big)\big(2y_2\beta_+^{(2)}\big)\,2^{4\nu}\,\frac{\Gamma(r_1+2\nu)\Gamma(r_2+2\nu)}{x_1^{r_1+2\nu}x_2^{r_2+2\nu}}\,k^{\,3+4\nu}\,c_{\rm pol}(\nu)\,,
\end{equation}
where $c_{\rm pol}$ is the loop block summed over the polarization monomials. Before its vertex integrals, Eq.~\eqref{eq:vecleading} is the loop integral of the late-time propagators of Ref.~\cite{ref17}.

\paragraph{The leading coefficient.} In the bispectrum the soft leg ends on the cubic vertex, so that $x_2=k=k_3$ and both loop lines are soft. This configuration lies on the boundary of convergence of these series. With $k_1=k_2$ and $r\equiv k_3/k_1$ (which equals $R$ of Eq.~\eqref{eq:shapevars} on this line), every piece of the soft vertex (nonlocal or analytic at fixed $x_2$) lands on the same power of $r$. The squeezed signal then takes the form
\begin{equation}\label{eq:vectemplate}
S\supset2\Re\sum_{n\ge0}\mathcal{M}_n\,r^{\,2+2\ii\mu+n}\,,\qquad \mathcal{M}_A\equiv\mathcal{M}_0\,,
\end{equation}
in agreement with the exponent of Ref.~\cite{ref17}. The coefficients follow by taking only the hard-vertex residues and keeping the soft vertex whole. Each is a hard-vertex Gamma function times a soft moment. For every helicity pair and every monomial of the polarization sums, the soft moment is one Euler integral of elementary factors and a Gauss function with a unit parameter (Appendix~\ref{app:helical}, Eq.~\eqref{eq:vecKdF}). It is then a finite sum of Gamma functions, Gauss functions at argument $\frac12$, and Kamp\'e de F\'eriet functions of arguments $(1,\frac12)$, the class of function of the helical single exchange of Section~\ref{sec:chem.line}. For two transverse lines the first indices are equal or opposite, and the soft moments without inverse squares of the line momenta reduce further to the one-variable functions of Appendix~\ref{app:diracline}. We have found no such reduction for the pairs with one longitudinal line or for the inverse squares of the line momenta that the polarization sums contain. The reduction of Appendix~\ref{app:diracline} requires the first indices of the two lines to be equal or opposite up to an integer, which holds for a transverse and a longitudinal line only at $\kappa=0$.

We emphasize that, once the nine helicity pairs are summed, $\mathcal{M}_n$ are given by the factorized and doubly cut sectors alone. In the uncut sector every term of the ordered remainder contains the commutator of one line. Summed over the three polarizations of its line, this commutator reduces to the commutator of the Proca field, which is an entire function of the line momentum. Its loop integral is then analytic in $\mathbf{k}$, and its oscillatory contributions cancel among the nine pairs. For a single pair they do not cancel, and the ordered remainder oscillates at the single-line frequency at order $r^{4\pm2\ii\mu}$. Nor do the transverse pairs cancel on their own, so the cancellation requires the longitudinal polarization. Consequently, the statement of Ref.~\cite{ref9} that the nonlocal signal is the factorized part is exact for the vector summed over its polarizations.

\begin{figure}[t]
\centering
\includegraphics{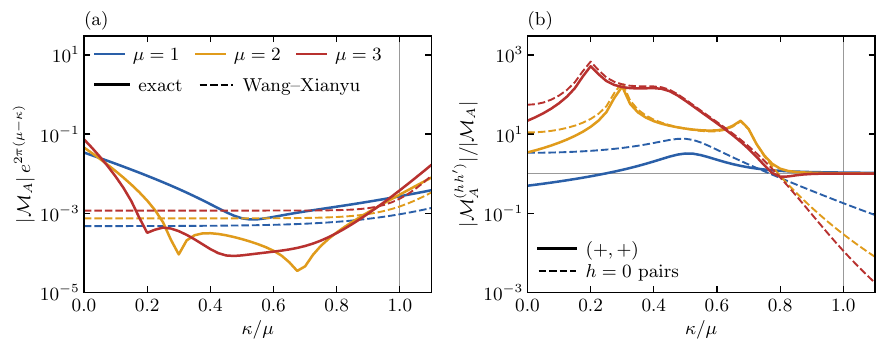}
\caption{The squeezed coefficient $\mathcal{M}_A$ against $\kappa/\mu$ for $\mu=1,2,3$. (a) $|\mathcal{M}_A|\,e^{2\pi(\mu-\kappa)}$ from the exact soft moment summed over the nine helicity pairs (solid) and from the closed form of Ref.~\cite{ref17} in the normalization of Eq.~\eqref{eq:vecshape} (dashed). (b) The modulus of the doubly enhanced transverse pair (solid) and the summed moduli of the five pairs with a longitudinal line (dashed), both relative to $|\mathcal{M}_A|$. Values above one signal destructive interference between pairs.}
\label{fig:vecM}
\end{figure}

Replacing the soft moment by its own leading residue gives the late-time coefficient $\mathcal{M}_{A,N=0}$. For the doubly enhanced pair it scales as $n_+(1+n_+)$, so the suppression $e^{-2\pi\mu}$ of the pair signal is lifted at $\kappa=\mu$. The exact coefficient, shown in Fig.~\ref{fig:vecM}, is much larger. At $\kappa=\mu$ it is $6.1$, $16.7$, and $30.0$ times its late-time leading term for $\mu=1$, 2, and 3. At $\kappa=0$ the ratio is between $11.6$ and $17.7$. The ratio is large because at $x_2=k_3$ every piece and every order of the late-time expansion at the soft vertex lands on the same power of $r$, and the soft moment collects all of them. The composition of the coefficient changes with $\kappa$. At small $\kappa$ the nine pairs interfere destructively, and the pairs with a longitudinal line dominate (their summed moduli are $3.4$ to $55$ times $|\mathcal{M}_A|$ at $\kappa=0$ for $1\le\mu\le3$). Near $\kappa/\mu\simeq0.5$ to $0.7$ the pairs interfere destructively in the exact coefficient, and $|\mathcal{M}_A|e^{2\pi(\mu-\kappa)}$ passes through a dip. From $\kappa/\mu\simeq0.9$ on, the doubly enhanced pair accounts for the modulus of the whole coefficient to $10\%$. At $\kappa=\mu$ the exact $|\mathcal{M}_A|$ is between $2.7\times10^{-3}$ and $3.8\times10^{-3}$ for $1\le\mu\le3$, and between $\mu=2$ and $4$ it grows as $\mu^{0.7}$.

\begin{figure}[t]
\centering
\includegraphics{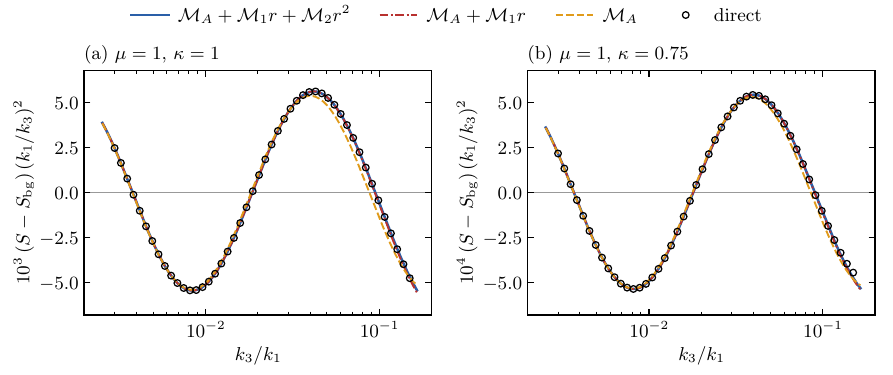}
\caption{The squeezed signal of the helical-vector bubble at $\mu=1$ and $k_1=k_2$, for (a) $\kappa=1$ and (b) $\kappa=0.75$. Circles show the full bispectrum of Eq.~\eqref{eq:vecshape} from direct numerical integration (nested over the vertex times and loop momentum), summed over all nine helicity pairs and four Schwinger--Keldysh sectors. After subtraction of a fitted power-and-logarithm background, the remainder is divided by $r^2$. Curves are $2\Re[(\mathcal{M}_A+\mathcal{M}_1r+\mathcal{M}_2r^2)\,r^{2\ii\mu}]$ (solid), the same without $\mathcal{M}_2$ (dash-dotted), and the leading term (dashed). A fit to the circles with free coefficients returns $\mathcal{M}_A$ in modulus and phase to $4\times10^{-4}$ in (a) and to $10^{-3}$ in (b).}
\label{fig:vecdirect}
\end{figure}

\paragraph{Comparison with the late-time estimate.} We can now make the comparison with Ref.~\cite{ref17} precise. Their $\tilde\nu$ is our $\mu$, their $\tilde\mu$ is our $\kappa$, and their enhanced helicity $h=-1$ is our $h=+1$. Apart from the loop factor, their closed form is the leading term of the late-time expansion for the doubly enhanced transverse pair. For loop momenta $\mathbf{q}$ and $\mathbf{p}=\mathbf{q}-\mathbf{k}_3$ they state the polarization weight $[1+\cos(\theta-\chi)]^2/4$. Since $\cos(\theta-\chi)=-c$, this is the same-helicity overlap $(1-c)^2/4$. Integrating that weight with the loop block gives one quarter of the loop function $L(a,b)$ of Ref.~\cite{ref18} at $a=b=2\ii\tilde\nu$. However, we find numerically that their printed loop factor (their Eq.~(42)) agrees instead with the loop integral of $(1+c)^2/4$. This is the overlap of opposite helicities, which the contraction of their polarization vectors produces when the second vector is not complex conjugated. Relative to the loop block their factor is
\begin{equation}\label{eq:WXratio}
\mathcal{R}_{\rm WX}(\tilde\nu)=\frac{R_0\,(1-e^{-4\pi\tilde\nu})^2}{1-R_0\operatorname{sech}2\pi\tilde\nu}\,,\qquad R_0=\frac{2(1+2\ii\tilde\nu)^4+1}{(1+4\ii\tilde\nu)(3+4\ii\tilde\nu)}\,,
\end{equation}
which is independent of $\kappa$, grows as $-2\tilde\nu^2$, and has modulus $2.4$ at $\tilde\nu=1$. With this factor, our late-time coefficient of the doubly enhanced pair reproduces the modulus of their closed form to $10^{-8}$ at $\mu=\kappa=3$, the precision of the stored value. The loop factor and the missing higher orders act in opposite directions. At $\kappa=\mu$ their amplitude is $0.35$, $0.50$, and $0.61$ of the exact one for $\mu=1$, 2, and 3. The compensation keeps it within a factor of three there, in line with the numerical study of Ref.~\cite{ref80}, which found amplitudes compatible with the estimates of the literature at $\kappa\simeq\mu$. Away from $\kappa=\mu$ the compensation fails, and at $\kappa=\mu/2$ and $\mu=3$ their amplitude is $13.5$ times the exact one.

\paragraph{Beyond the squeezed limit.} The first correction to the template is not small. Since $|\mathcal{M}_1/\mathcal{M}_A|$ is between 2 and 4 at $\kappa=\mu$, the leading term holds to $10\%$ as a complex amplitude only for $R\lesssim0.05$, $0.03$, and $0.02$ at our three benchmarks (Table~\ref{tab:shapedomain}). Multiplying it by $(1+R/2)^{-2\nu}$, as in the last column of Table~\ref{tab:shapedomain}, extends these ranges to $R\lesssim0.081$, $0.051$, and $0.050$. The bubble depends on the shape only through $k_{12}$ and $k_3$. We have not computed its analytic background, since the uncut sector contains the ordered remainder of Eq.~\eqref{eq:helkernel}. Its logarithmic divergence, however, has the same coefficient as for the neutral line, independent of $\kappa$ and of the mass, since in the ultraviolet each line takes its flat-space form.

\paragraph{Checks.} We have integrated the full bispectrum of Eq.~\eqref{eq:vecshape} directly, nested over the vertex times and the loop momentum, and summed over the nine helicity pairs and the four Schwinger--Keldysh sectors (Fig.~\ref{fig:vecdirect}). After subtraction of a fitted background of powers and logarithms, a fit with free coefficients returns $\mathcal{M}_A$ in modulus and phase to $4\times10^{-4}$ at $\mu=\kappa=1$ and to $10^{-3}$ at $(\mu,\kappa)=(1,0.75)$. The soft moments themselves can also be checked. For two transverse or two longitudinal lines and no inverse squares of the line momenta, the closed form of Appendix~\ref{app:diracline} reproduces our numerical moments to $3\times10^{-28}$ at the benchmarks of Fig.~\ref{fig:vecdirect} and at $(\mu,\kappa)=(2,2.5)$. Where two ladders of poles coincide at $\kappa=\mu=1$, the limit in $\kappa$ reproduces them to $8\times10^{-24}$. The remaining pairs and the inverse squares are covered by the Kamp\'e de F\'eriet form of Eq.~\eqref{eq:vecKdF}, which reproduces the numerical moments of every kind to $3\times10^{-26}$ or better at $(\mu,\kappa)=(1,1)$, $(2,2.5)$, $(3,3.75)$, $(1,0.75)$, and $(1,0)$.

\paragraph{Size.} In the convention of Ref.~\cite{ref17}, where $f_{\rm NL}^{\rm (osc)}$ multiplies $(2k_3/k_1)^{2+2\ii\mu}$ in the shape and is one quarter of our $f_{\rm NL}^{\rm osc}$, the contracted bubble with the exact $\mathcal{M}_A$ has $f_{\rm NL}^{\rm (osc)}=2.7$ at $(\mu,\kappa)=(3,4)$ and their $u=1$, against $2.6$ from their Eq.~(45). The size of the signal of the model is that of the triangle (Section~\ref{sec:chem.hard}).

\subsection{The helical-vector triangle}\label{sec:chem.hard}

\begin{figure}[t]
\centering
\includegraphics{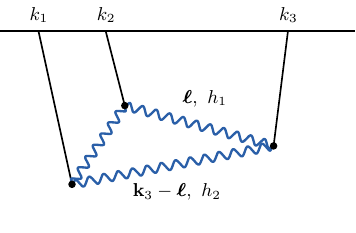}
\caption{The helical-vector triangle. The soft lines, with momenta $\boldsymbol{\ell}$ and $\mathbf{k}_3-\boldsymbol{\ell}$, meet at the vertex of the soft leg, and the hard line joins the vertices of $\mathbf{k}_1$ and $\mathbf{k}_2$. Shrinking the hard line to a point gives the bubble of Fig.~\ref{fig:vecbubble}.}
\label{fig:vectri}
\end{figure}

\paragraph{The model.} The bubble above is a contracted diagram. In the model of Ref.~\cite{ref17} the two soft helical-vector lines end on a third, hard line (Fig.~\ref{fig:vectri}), and the bubble is the triangle with that line shrunk to a point. The contraction is exact for a hard line much heavier than the soft ones, but not at equal masses, since a line at horizon crossing does not localize.

\paragraph{The leading coefficient.} To quantify the effect of the contraction, we have computed the leading squeezed coefficient with the hard line kept as the full helical propagator, including all three hard helicities and all orderings of the three vertices. We quote it as the ratio $\mathcal{R}_{\rm hard}=\mathcal{M}_A^{\triangle}/\mathcal{M}_A$, where $\mathcal{M}_A$ is the coefficient of the contracted bubble in the normalization of Ref.~\cite{ref17} (whose quartic coupling includes the factor $e^{2\pi(\kappa-\mu)}$ for $\kappa>\mu$) and $\mathcal{M}_A^{\triangle}$ is the coefficient of the same power of $r$ in the triangle. We also define $\mathcal{R}_{\rm hard,0}=e^{2\pi(\kappa-\mu)}\mathcal{R}_{\rm hard}$.

\begin{figure}[t]
\centering
\includegraphics{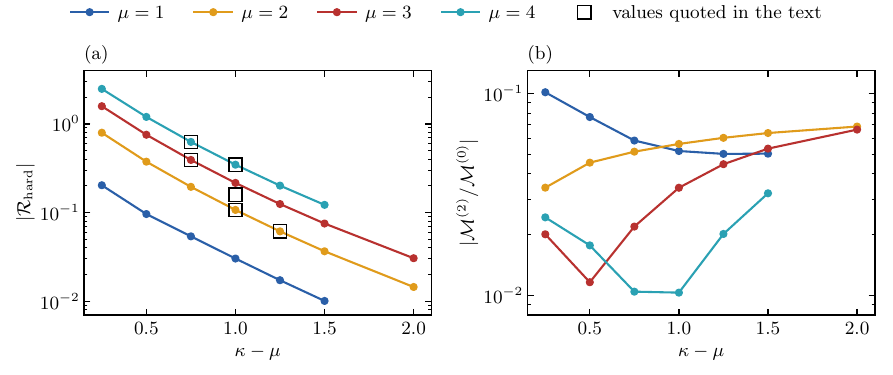}
\caption{The squeezed coefficient with the hard line kept, for the soft helicity pair $(+,+)$ and a hard line of the same mass and chemical potential as the soft lines. (a) The modulus of the ratio $\mathcal{R}_{\rm hard}$ against $\kappa-\mu$. (b) The quadrupole over the monopole.}
\label{fig:hard}
\end{figure}

The structure of the calculation fixes four properties. First, the monopole of the ratio is the same for every pair of soft helicities, so the hard line rescales the coefficient without changing its helicity content. Second, the dependence on the angle between the hard and soft momenta is a monopole plus a quadrupole (with no dipole), because the parity-odd soft tensor is antisymmetric under exchange of the two soft lines. Third, when the hard line alone is made heavy, $\mathcal{R}_{\rm hard,0}\to2$, with corrections of order $\mu_{\rm hard}^{-2}$. The limit is 2 rather than 1 because the symmetry factor of the triangle is one and that of the bubble is one half. Fourth, the $A_0$ component of the hard line first contributes one power of $r$ later. Numerically, the monopole agrees between all nine soft pairs to $10^{-15}$, the dipole vanishes to $10^{-11}$, and $|\mathcal{R}_{\rm hard,0}|$ approaches 2 as the hard mass grows ($2.02$ at $\mu_{\rm hard}=10$).

Figure~\ref{fig:hard} shows the ratio for a hard line of the same mass and chemical potential as the soft ones. At $\mu=2$ and $3$, with $\kappa-\mu$ between $0.75$ and $1.25$, $|\mathcal{R}_{\rm hard}|$ lies between $0.06$ and $0.4$, and the quadrupole is between two and six percent of the monopole for the $(+,+)$ soft pair. With the bubble normalized as in Ref.~\cite{ref17}, the contracted diagram thus overestimates the squeezed amplitude by a factor of $2.6$ to $16$ there (or $5$ to $33$ if the bubble is normalized as the heavy-hard-line limit of the triangle, which is twice the normalization of Ref.~\cite{ref17}). The growth with the chemical potential is exponential only asymptotically. In units of the rate $2\pi$ of Ref.~\cite{ref17}, $\dd\ln|\mathcal{R}_{\rm hard,0}|/\dd(\kappa-\mu)$ rises from about $0.5$ at small $\kappa-\mu$ to about $0.7$ at $\kappa-\mu\simeq2$. At $\kappa=\mu$, on the other hand, the coefficient with the hard line kept is $0.8$, $2.1$, and $3.9$ times that of the contracted diagram for $\mu=1$, 2, and 3.

\paragraph{The hard line in closed form.} The moments of the hard line, which we evaluated numerically above, have closed forms at every order. At leading order the hard line carries $k_1$, and both of its vertices have energy $|\mathbf{k}_1|$, so the line sits at the doubly folded point $x_1=x_2=y$. In the Barnes form of Eq.~\eqref{eq:helBarnes}, the prefactor $e^{-y(t-t')}$ then gives the $M$ end of the kernel zero energy. That end's time integral sets $s'$ by a residue, as for a vertex without legs, and the remaining integral is of the Barnes type $\int\Gamma(a{+}s)\Gamma(b{+}s)\Gamma(c{+}s)\Gamma(d{-}s)\Gamma({-}s)/\Gamma(e{+}s)$. An Euler integral over the beta function $\Gamma(c{+}s)\Gamma(d{-}s)$ turns it into a ${}_3F_2$, and we find
\begin{equation}\label{eq:helmoment}
\begin{split}
\int_0^\infty\!\!\dd t_0\,\dd t_1\,t_0^{\alpha_0}t_1^{\alpha_1}e^{-t_0-t_1}\mathcal{K}_h(1;t_0,t_1)={}&\frac{\Gamma(\lambda)\Gamma(\alpha_0{+}1{+}\nu)\Gamma(\alpha_1{+}1{+}\nu)\Gamma(a_+)}{2^{\lambda}\,\Gamma(1+2\nu)\,\Gamma(\lambda+a_+)}\\
&\times{}_3F_2\!\left[\begin{matrix}a_+,\ \alpha_0{+}1{+}\nu,\ \alpha_1{+}1{+}\nu\\ \lambda+a_+,\ 1+2\nu\end{matrix};1\right],
\end{split}
\end{equation}
with $\lambda=\alpha_0+\alpha_1+2$ and $a_+=\frac12+\nu-K$. The chemical potential enters only through $a_+$, and at $\kappa=0$ ($a_+=\frac12+\nu$) Eq.~\eqref{eq:helmoment} is the scalar moment $\mathcal{H}_\nu$ of Appendix~\ref{app:box}. The excess of the ${}_3F_2$ is one for all exponents, so the series always converges. The branch decomposition of the modes (Appendix~\ref{app:helical}) then gives all four Schwinger--Keldysh moments of a transverse hard line. The two factorized ones are products of single-time Whittaker moments, which are Gamma functions. The two ordered ones are linear combinations of Eq.~\eqref{eq:helmoment} at the orders $\pm\nu$, with the branch coefficients of Appendix~\ref{app:helical} (and $K\to-K$ for the conjugate branch). They reproduce the quadrature at $(\mu_h,\kappa_h)=(1,0.75)$ and $(2,2.5)$ to $10^{-11}$, for both transverse helicities and all four moments. The longitudinal line does not feel $\kappa$. Its derivative ends reduce to the same moments, as we show in Section~\ref{sec:chem.parity}. At higher orders in $r$ the exponents shift and $y$ moves off the vertex energies. By scaling, however, each $y$ derivative at the folded point is a finite combination of Eq.~\eqref{eq:helmoment} at shifted exponents, so the hard side of every order of the series below has a closed form. What remains at higher orders is the soft side. Expanding $|\bl+\mathbf{k}_1|$ inserts powers of $\bl\cdot\hat{\mathbf{k}}_1$ into the soft moments. After the average over the azimuth of $\bl$ these are again monomials in the two line energies, and each is one more soft moment of the kind of Eq.~\eqref{eq:vecKdF} (Appendix~\ref{app:helical}). No line energy enters with an inverse power higher than two, so both sides of every order of the series reduce to named functions. The hard side gives ${}_3F_2$'s at unit argument, and the soft side gives Kamp\'e de F\'eriet functions.

\paragraph{Beyond the squeezed limit.} The calculation extends to every order in the soft momentum and to every shape, with the steps of the scalar pair family in Section~\ref{sec:triangle.shape}. At the hard vertices each soft line is expanded in its series at small argument, Eq.~\eqref{eq:en}, and the hard line is kept in full. Its energy $|\mathbf{k}_1+\bl|$ and the energy $k_2$ of the second hard leg move off $k_1$ at higher orders, and both are restored by scaling (Appendix~\ref{app:helical}). Since $e_1=-K/(1+2\nu)$ does not vanish for $\kappa\neq0$, the series advances in unit steps, as for the bubble. In the variables of Eq.~\eqref{eq:shapevars} we find
\begin{equation}\label{eq:vecDseries}
S\supset2\Re\sum_{N\ge0}\mathcal{M}^{\triangle}_N(\delta)\,R^{\,2+2\ii\mu+N}\,,\qquad
\mathcal{M}^{\triangle}_0(\delta)=\mathcal{M}_{\rm mono}+\tfrac12\big(3\delta^2-1\big)\,\mathcal{M}_{\rm quad}\,,
\end{equation}
where each $\mathcal{M}^{\triangle}_N$ is a polynomial in $\delta^2$, so the series is symmetric under $\mathbf{k}_1\leftrightarrow\mathbf{k}_2$ term by term. Since $\hat{\mathbf{k}}_1\cdot\hat{\mathbf{k}}_3\to-\delta$ in the squeezed limit, the leading coefficient is the monopole and quadrupole found above. From the second order on, the temporal component of the hard line also enters through a contact term (Appendix~\ref{app:helical}).

The corrections are large. For a hard line of the same mass and chemical potential as the soft lines, we have computed the series through $N=8$ at $(\mu,\kappa)=(1,1)$ and $(2,2.5)$ and through $N=12$ at $(3,3.75)$. On $k_1=k_2$ its modulus is $1.14$, $1.58$, and $1.80$ times that of its leading term at $R=0.3$, and $2.09$, $3.27$, and $3.70$ times at the equilateral shape. At the folded shape $k_1=2k_2=2k_3$ the factors are $1.19$, $2.80$, and $3.78$. The leading term alone is thus accurate to $10\%$ only for $R\le0.078$, $0.042$, and $0.029$ (Table~\ref{tab:shapedomain}). The contracted bubble has larger corrections still, so the ratio of the two diagrams depends on the shape. With the bubble normalized as in Ref.~\cite{ref17}, the contracted diagram exceeds the triangle on $k_1=k_2$ by factors of $1.3$, $2.6$, and $2.5$ at small $R$, and of $2.5$, $5.8$, and $4.7$ at the equilateral shape.

\paragraph{Checks.} We have checked the series in four ways. The symmetry under $\mathbf{k}_1\leftrightarrow\mathbf{k}_2$, which the expansion does not make manifest, holds to $5\times10^{-27}$ through $N=8$. When the hard line alone is made heavy, each order through $N=4$ tends to twice the contracted bubble, with corrections of order $\mu_{\rm hard}^{-2}$. The odd orders vanish at $\kappa=0$. Finally, we have integrated the triangle directly at fixed shape, as for the bubble in Fig.~\ref{fig:vecdirect}, and fitted an analytic background together with the series. At $\mu=\kappa=1$ the first three coefficients agree with the series to $2\times10^{-5}$, $4\times10^{-4}$, and $0.5\%$ for collinear momenta, and to $2\times10^{-5}$, $7\times10^{-4}$, and $1.1\%$ at $\hat{\mathbf{k}}_1\cdot\hat{\mathbf{k}}_3=\frac12$. At $(2,2.5)$ they agree to $4\times10^{-6}$, $8\times10^{-5}$, and $1.5\times10^{-3}$. The orders $N\ge4$ rest on the first three checks. We have not computed the part without oscillations, and the Higgs lines of the model remain contracted.

\paragraph{Size.} The hard line also sets the size of the signal. At $(\mu,\kappa)=(3,4)$ and $u=1$ it lowers $f_{\rm NL}^{\rm (osc)}$ of the triangle on $k_1=k_2$ from $2.7$ to $0.59$, in the convention of Ref.~\cite{ref17}. At eight points from $(\mu,\kappa)=(2,3)$ to $(20,20.3)$, where their Eq.~(45) gives $f_{\rm NL}^{\rm (osc)}$ between $0.1$ and $4.3$, the exact triangle is $0.12$ to $11.5$ times that estimate. It is smaller for $\mu\lesssim5$ and larger for heavier vectors (by $4.3$ at $\mu=10$ and $11.5$ at $\mu=20$). Since the exact amplitude grows as about $e^{5\pi(\kappa-\mu)}$ at these points, a shift of $\kappa$ by at most $0.16$ compensates any of these factors, and the signal remains within reach of planned surveys.

\subsection{The parity-odd trispectrum}\label{sec:chem.parity}

\begin{figure}[t]
\centering
\includegraphics{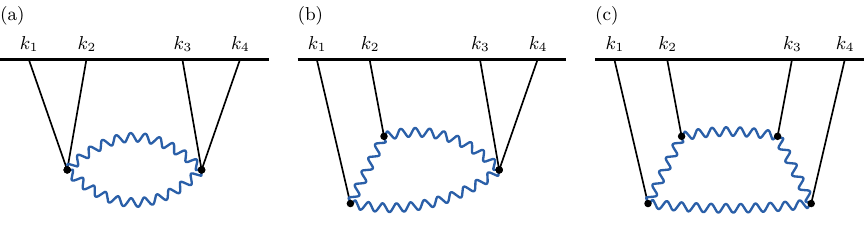}
\caption{The diagrams of the collapsed trispectrum. Two soft lines carry $s=|\mathbf{k}_1+\mathbf{k}_2|$ between the legs $(12)$ and $(34)$. On each side the soft lines meet either at the quartic contact vertex or at two cubic vertices joined by a hard line. This gives (a) two contact sides, (b) one contact side and one hard pair, and (c) two hard pairs, the box.}
\label{fig:collapsed}
\end{figure}

\paragraph{The leading coefficient.} The same hard line determines the collapsed trispectrum of the model, in which the soft momentum $s=|\mathbf{k}_1+\mathbf{k}_2|$ flows between the pairs of legs $(12)$ and $(34)$ (Fig.~\ref{fig:collapsed}). It is the momentum $P$ of Section~\ref{sec:triangle.box} and Appendix~\ref{app:box}, where the legs are labeled cyclically around the box, and the configuration is case (i) of Section~\ref{sec:regions.split}. Both ends of each soft line are hard, so the soft lines enter only through their late-time coefficients. On each side the two soft lines end either on the quartic vertex of Eq.~\eqref{eq:vecvertex} or on two cubic vertices joined by a hard line. With these vertices, the Wick contractions give
\begin{equation}\label{eq:triweights}
16\,k_1k_2k_3k_4\,\langle\delta\phi^4\rangle'\supset\tfrac12\lambda_1^2\,I_{\rm cc}+\lambda_1\lambda_2^2\,\big(I_{\rm cp}+I_{\rm pc}\big)+\lambda_2^4\,\big(I_\square+I_\square^{\times}\big)\,,
\end{equation}
where each $I$ is the loop integral with unit vertices. The first term is the bubble between two quartic vertices, the second the diagrams with one quartic vertex and one hard pair, and the third the box. The soft lines can join the two hard pairs in two ways, so the crossed box $I_\square^{\times}$ is the first with $\mathbf{k}_3\leftrightarrow\mathbf{k}_4$. Every term has a closed form. The hard pairs reduce to the moments of Eq.~\eqref{eq:helmoment} and to Gamma functions, so the collapsed coefficient of the box is a finite sum of Gamma functions and ${}_3F_2$'s at unit argument, with no integral left (Appendix~\ref{app:helical}, Eq.~\eqref{eq:Mbox}). It agrees with direct quadrature to $10^{-11}$ in every harmonic.

The contact vertices carry no direction. All dependence on the dihedral angle $\beta$ between the planes $(\mathbf{s},\mathbf{k}_1)$ and $(\mathbf{s},\mathbf{k}_3)$ therefore comes from diagrams with a hard pair, and at leading order the non-analytic signals of the box contain only the harmonics $e^{\ii n\beta}$ with $|n|\le2$. Parity acts as $\beta\to-\beta$, and the part odd in $\beta$ vanishes at $\kappa=0$. Reversing $\kappa$ exchanges the harmonics $c_n$ and $c_{-n}$. At $\kappa\neq0$ the dihedral quadrupole is almost entirely of one helicity, with $|c_{-2}/c_{+2}|=160$ at $\kappa=0.5$ and $\mu=1$ for $\hat{\mathbf{k}}_1$ and $\hat{\mathbf{k}}_3$ perpendicular to $\mathbf{s}$. The quartic coupling of Section~\ref{sec:chem.bubble} is the hard line of Ref.~\cite{ref17} contracted to a point, which the hard pair already contains. The quartic vertex that accompanies the hard pair is instead the one at which two Higgs lines end. By Eq.~(37) of Ref.~\cite{ref17} it is smaller by $2e^{2\pi(\kappa-\mu)}$ for $\kappa>\mu$ (and by $2$ otherwise), which gives $\lambda_2^2/\lambda_1=-2m_A^2$. The parity-odd part divided by the monopole, $|c_2-c_{-2}|/2|c_0|$, is then $0.2$ to $2.3$ for $0.25\le\kappa\le\mu$ at $\mu=1$, and $0.13$ to $3.5$ at $\mu=2$. It stays of order one for $\kappa>\mu$. In the collapsed limit, therefore, the parity-odd part of the pair-frequency signal is comparable to its parity-even part. It is carried by the dihedral angle as $\sin2\beta$, and away from the perpendicular configuration also as $\sin\beta\,\sin2\alpha_1\sin2\alpha_3$, with $\alpha_{1,3}$ the angles of $\hat{\mathbf{k}}_{1,3}$ to $\hat{\mathbf{s}}$.

This makes the parity-odd trispectrum a natural observable for this model. It is the lowest-order scalar correlator that can violate parity, and at tree level, with scale invariance and the Bunch--Davies vacuum, it vanishes for scalar exchanges and is factorized in general~\cite{CabassParity2023,StefanyszynTongZhu2024}. A parity-odd signal thus requires spinning fields with a chemical potential, a departure from these assumptions, or loops~\cite{LiuTongWangXianyu2020,CabassParity2023,ref72,StefanyszynTongZhu2024}. Loops of tachyonic helical gauge fields in axion inflation have been studied numerically away from the collapsed limit. There the parity-odd part was found one to two orders of magnitude below the parity-even part~\cite{Niu2022tri}, or about ten percent of it at equilateral momenta and comparable nearby~\cite{Fujita2024} (see also Refs.~\cite{Reinhard2025,OrtolaSlepian2026}). These models and configurations differ from ours. The parity-odd trispectrum has been searched for in the four-point function of BOSS galaxies~\cite{HouSlepianCahn2022,Philcox2022,CabassGhosts2023} and in the cosmic microwave background~\cite{Philcox2023cmb}. The early hints in BOSS are now attributed to the covariance of the mock catalogs~\cite{PhilcoxEreza2024,Krolewski2024}, and the first DESI measurement is consistent with zero~\cite{HouDESI2025}. Searches with more data will need templates of this kind.

\paragraph{Further signals.} For helical lines the pair frequency is not the only non-analytic signal of the collapsed limit. A line on opposite late-time branches at its two ends has a helicity-odd part that is not analytic in its momentum. It gives signals that scale as $s^{3\pm2\ii\mu}$, at half the pair frequency, and terms $s^{3+N}$ without oscillation (Appendix~\ref{app:helical}). These signals vanish at $\kappa=0$, but at $\mu=\kappa=1$ the monopole of the single-frequency signal is $13$ times that of the pair signal. The statements above about the parity-odd fraction refer to the pair-frequency signal.

\paragraph{Beyond the collapsed limit.} The collapsed coefficient is the first term of a series in $s$ over the hard momenta, and every order has a closed form of the same kind (Appendix~\ref{app:helical}). The corrections are not small. In the criterion of Table~\ref{tab:shapedomain}, the leading term reproduces the dominant harmonic $n=-2$ to $10\%$ as a complex amplitude only for $s\le0.028\,k$ at $(\mu,\kappa)=(1,1)$, $0.052\,k$ at $(1,0.5)$, and $0.019\,k$ at $(2,2)$. Most of the change is a phase, and the parity-odd fraction changes little (Fig.~\ref{fig:parityseries}a). At $\mu=\kappa=1$ it is $2.27$ at leading order and $2.51$ at $s=0.3\,k$, where the series through $N=6$ is still good to one percent.

\paragraph{Checks.} The leading term of the series reproduces Eq.~\eqref{eq:Mbox} to $2\times10^{-23}$, and the series obeys the exchange of the two legs of a side to $5\times10^{-24}$ through $N=6$. It also has the expected limit when the hard line is made heavy. Direct integration of the bubble between two quartic vertices, and of the diagram with one quartic vertex and one hard pair, confirms the orders $N\ge1$ to $2\times10^{-3}$ and $2\times10^{-4}$ (Appendix~\ref{app:helical}). Direct integration of the box at $\mu=\kappa=1$ and equal external momenta confirms the same orders to $5\times10^{-5}$ in the monopole and to $3\times10^{-5}$ in the dominant harmonic $n=-2$.

\paragraph{Size.} With the Higgs lines contracted as in Ref.~\cite{ref17}, $\lambda_2=-2m_A^2/(u\dot\phi_0)$ and $\lambda_1=-2m_A^2/(u^2\dot\phi_0^2)$ in units of $H$, where $u$ is their parameter for the Higgs expectation value. For $\zeta=\delta\phi/\dot\phi_0$ and equal external momenta $k$, with $r=s/k$, Eq.~\eqref{eq:triweights} then gives
\begin{equation}\label{eq:tauloop}
\frac{\langle\zeta^4\rangle'}{4P_\zeta(k)^2P_\zeta(s)}=2\pi^2\Delta_\zeta^2\,\frac{m_A^4}{u^4}\;r^6\,\mathcal{A}(r,\beta)\,,
\end{equation}
where $\mathcal{A}$ is the dimensionless function multiplying $\lambda_1^2s^3k^{-8}$ (for the pair signal, twice the real part in Eq.~\eqref{eq:parityseries}). The left-hand side is $\tau_{\rm NL}$ for the local shape, and it is the quantity that the direction-dependent templates of Ref.~\cite{PhilcoxCMB1} parametrize. At leading order the parity-odd part of the pair signal is $\ii\,\hat{\mathbf{s}}\cdot(\hat{\mathbf{k}}_1\times\hat{\mathbf{k}}_3)$ times a combination of $\hat{\mathbf{k}}_1\cdot\hat{\mathbf{k}}_3$ and $(\hat{\mathbf{k}}_1\cdot\hat{\mathbf{s}})(\hat{\mathbf{k}}_3\cdot\hat{\mathbf{s}})$, so it lies in the channels $(n_1,n_3,n)=(2,2,1)$ and $(2,2,3)$ of that basis. The corresponding amplitudes are $|\tau_{\rm NL}^{221}|\simeq6\times10^{-8}\,u^{-4}r^6$ at $\mu=\kappa=1$ and $6\times10^{-6}\,u^{-4}r^6$ at $\mu=\kappa=2$ for the pair signal, and about the same or twice that for the single-frequency signal. The Planck bound on the scale-invariant template is $\tau_{\rm NL}^{221}=(-3.7\pm4.8)\times10^4$~\cite{PhilcoxCMB3}, and the collapsed templates for a heavy spin-one particle are bounded at the level of $3\times10^7$. For $u$ of order one and masses of a few $H$ the loop is thus ten or more orders of magnitude below these bounds, even at $s\simeq k$, and the factor $r^6$ lowers the signals further in the collapsed configurations that these templates weight. Only the corner $u\sim10^{-3}$, near the lower limit $u\gtrsim4\times10^{-4}$ that the equilateral bispectrum of the same model sets at $\kappa=\mu$ (Eq.~(56) of Ref.~\cite{ref17}), or much larger masses could approach them. The part of the box that is analytic in $\mathbf{s}$ also violates parity, and it falls more slowly. We have integrated the box without a loop cut at $\mu=\kappa=1$ and equal external momenta for $0.05\le r\le0.6$, where its parity-odd part converges. At small $r$ the analytic part of the coefficient of $\ii\sin2\beta$ is linear in $r$. In the normalization of Eq.~\eqref{eq:tauloop} it gives about $2\times10^{-9}\,u^{-4}r^4$, against an amplitude of $4.0\times10^{-9}\,u^{-4}r^6$ for the pair signal. It exceeds the oscillating signals for $s\lesssim0.3\,k$, so in this model the parity-odd loop falls as $r^4$ relative to the local shape, not as $r^6$. This does not change the comparison with the bounds above. The dependence of the analytic part on the shape, and with it a template for searches, remains open.

\subsection{Charged scalars}\label{sec:chem.vertex}

\begin{figure}[t]
\centering
\includegraphics{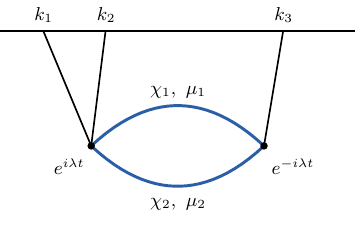}
\caption{The charged-scalar bubble. The lines $\chi_1$ and $\chi_2$ have masses $\mu_1$ and $\mu_2$, and the two vertices carry the phases $e^{\pm\ii\lambda t}$ of the rolling inflaton.}
\label{fig:charged}
\end{figure}

The chemical potential of Refs.~\cite{ref19,ref73} is of a different kind, since it sits on a vertex rather than on the lines.

\paragraph{The model.} In the model of Refs.~\cite{ref19,ref73}, a pair of charged scalars couples to the inflaton through a vertex that breaks their $U(1)$ (Fig.~\ref{fig:charged}). After a field redefinition the rolling inflaton appears only as a phase of that vertex, which shifts its energy, while the lines remain Hankel functions of the masses $\mu_1$ and $\mu_2$. In the present representation this phase is a complex shift of the vertex powers, $r_1\to r_1-\ii\lambda$ at one vertex and $r_2\to r_2+\ii\lambda$ at the other. The families of Sections~\ref{sec:block} and~\ref{sec:bubble} then apply unchanged, with the vertex phase in the arguments of their sines, and the bispectrum is a bubble whose two vertices carry $t^{\mp\ii\lambda}$. The background of a single diagram, however, needs the derivative form of the vertices, as we show below.

\paragraph{The leading coefficient.} The squeezed signal follows from Eq.~\eqref{eq:factorize}, with a hard-vertex Gamma function and the soft moment of Eq.~\eqref{eq:twomoment} at complex powers. We quote it in the normalization of Ref.~\cite{ref19}, for their $F$ at $|\alpha|^2/\Lambda^2=1$ and $k_1=k_2$. With $p=(k_1+k_2)/k_3$ and $b=\ii(\mu_1+\mu_2-\lambda)$ we find
\begin{equation}\label{eq:chargedA}
\begin{aligned}
F&\supset A\,p^{-3-b}+\text{c.c.}\,,\\
A&=\frac{5\lambda}{24\pi^5}\,\Gamma(-\ii\mu_1)\,\Gamma(-\ii\mu_2)\,\sinh\frac{\pi(\lambda+\mu_1+\mu_2)}{2}\,\sin\frac{\pi b}{2}\\
&\quad\times\Big[\Gamma(b)+\Gamma(1+b)+\tfrac14\Gamma(2+b)\Big]\Big[\mathcal{M}_{++}(\ii\lambda)+\mathcal{M}_{++}(1+\ii\lambda)\Big]\,,
\end{aligned}
\end{equation}
where $\mathcal{M}_{++}(r_2)$ is the moment of Eq.~\eqref{eq:twomoment} with $\sigma_1=\sigma_2=+1$ at $k_3=1$. The first bracket is the hard vertex. The second is the soft vertex at its two powers, which come from the two terms of the external mode. This is a closed form at any masses and any $\lambda$. We have checked the late-time coefficients and the rotated prefactors against direct evaluation, to $10^{-13}$ and $10^{-25}$, and the reality of the bispectrum holds exactly.

\paragraph{Comparison with the template.} This gives a check of the stationary-phase and threshold expressions of Ref.~\cite{ref19}, which hold at large $\lambda$ and large masses. For $\mu_1=\mu_2=\mu$ the moduli agree to $3\times10^{-4}$ at $(\lambda,\mu)=(40,8)$ and $(50,10)$. At their benchmark $(40,4)$ the ratio is $0.98$, and it is $0.93$ at $(20,2)$ and $0.83$ at $(10,1)$, with phase offsets between $0.2$ and $0.65$ rad. This is an independent check of the normalization of both calculations. Close to the threshold $\lambda=\mu_1+\mu_2$, however, the asymptotic expressions fail. The ratio is $0.46$ at $\lambda=10$ and $\mu=5.1$, just below threshold. Further below, the exact signal is suppressed as $e^{-\pi(\mu_1+\mu_2-\lambda)}$ (it is $1.7\times10^{-4}$ against $0.074$ at $\lambda=10$ and $\mu=6$), whereas the asymptotic one is not.

At the resonance $\lambda=\mu_1+\mu_2$ the hard factor stays finite, since its sum over residues gives $\Gamma(b)\sin(\pi b/2)$ with $b=\ii(\mu_1+\mu_2-\lambda)$. The pole sits instead in the soft moment, through $\Gamma(-3-b)$ and $\Gamma(-2-b)$, and it comes from loop momenta between $k_3$ and $k_{12}$. The pole is simple and its residue is continuous across the resonance. Near the resonance the family appears as $c_{\rm res}(p^{-b}-1)/b$ with $p=k_{12}/k_3$, so the $1/(\lambda-\mu_1-\mu_2)$ of its coefficient is cancelled by the local term and the signal stays bounded by $|c_{\rm res}|\ln p$. On resonance it becomes $c_{\rm res}\ln(k_{12}/k_3)$, with $|c_{\rm res}|$ between a quarter and a half of the amplitude at $\lambda=\mu_1+\mu_2+1$ for $1\le\mu\le8$. At unequal masses a second line appears at $\lambda-|\mu_1-\mu_2|$. Its relative amplitude is $5\times10^{-3}$ at $\lambda=10$ and $(\mu_1,\mu_2)=(1,3)$, and it falls with the smaller mass roughly as $e^{-2\pi\mu_1}$, rather than with $\lambda$ (it is $10^{-10}$ at $\lambda=40$ and $(\mu_1,\mu_2)=(4,8)$).

\paragraph{Beyond the squeezed limit.} The representation of Section~\ref{sec:bubble} also gives this bispectrum at every shape. With the vertex phases, the weights of Eq.~\eqref{eq:patterns} become
\begin{equation}\label{eq:chargedweights}
w_\epsilon=\epsilon_1\epsilon_2\,\frac{\sinh[\frac\pi2(M_\epsilon-\lambda)]\,\sinh[\frac\pi2(M_\epsilon+\lambda)]}{\sinh\pi\mu_1\,\sinh\pi\mu_2}\,,\qquad M_\epsilon=\epsilon_1\mu_1+\epsilon_2\mu_2\,,
\end{equation}
and the bispectrum is the imaginary part of the weighted sum over the four sign patterns. The logarithmic divergence of the loop is real, so it drops out and the result does not depend on the scheme. The background needs one more step. The vertices of Ref.~\cite{ref19} are non-derivative, with $r_1=j-\ii\lambda$ and $r_2=j'+\ii\lambda$ for $j+j'\le3$, and the background of a single diagram then has late-time poles. Their residue is proportional to $k_1^3+k_2^3-k_3^3$, and they cancel in the sum of all diagrams, since the interaction depends only on differences of $\delta\phi$. We remove them diagram by diagram with the contiguous relation of the single-time moments, which trades each non-derivative vertex for the derivative vertices of Ref.~\cite{ref19}. On the locus $x_2=k_3$ the families are then one sum over trees, and the background is one sum over the points $\zeta_m=\frac12+\ii\lambda+m$.

The oscillating signal leaves its leading squeezed term much sooner than the background does (Fig.~\ref{fig:shapes}). At $(\lambda,\mu)=(40,4)$ its modulus stays within $10\%$ of the leading term up to $R=1.14$, but its phase shifts by about $(\lambda-\mu_1-\mu_2)^2R^2/[16(\mu_1+\mu_2)]$, which is $0.33$ at $R=0.2$. As a complex amplitude the leading term is therefore accurate to $10\%$ only for $R<0.11$, and for $R<0.15$ and $0.23$ at $(20,2)$ and $(10,1)$ (Table~\ref{tab:shapedomain}).

\paragraph{Checks.} This result agrees with a numerical spectral integral over tree exchanges to $1.5\times10^{-13}$, at five shapes with $R$ between $0.1$ and $1.8$. At the benchmark $(\lambda,\mu)=(40,4)$ of Ref.~\cite{ref19}, their smooth background agrees with ours to $0.3\%$ for $R\le1$, although it fails near threshold.

\paragraph{Size.} At the benchmark $(\lambda,\mu)=(40,4)$ with the largest coupling that Ref.~\cite{ref19} allows, the exact coefficient changes their amplitude from $0.0163$ to $0.0160$ (in their normalization of $|F_{\rm osc}|$ at $p=2$). Their conclusion that only 21-cm surveys could reach this signal is unchanged.

%% file: sec8.tex
\section{Fermions with a chemical potential}\label{sec:dirac}

The last change of line is statistics. Without a chemical potential, a fermion loop falls under the second case of Section~\ref{sec:regions.split}, and the late-time error is the universal factor of Eq.~\eqref{eq:Univ}. For the top quark the leading term is absent, and the first term, of dimension $4+2\nu$, is $|\mathcal{U}(4+2\nu,2)|=26$ times its late-time estimate at $m=H$ (Section~\ref{sec:SM}). A chemical potential lifts the Boltzmann suppression of fermion loops and removes this universality. Earlier estimates of these signals~\cite{ref20,ref21,ref23} have recently been revised down by orders of magnitude~\cite{AokiStrumia2026}, and the squeezed coefficients of the bubble and the triangle have just been computed exactly~\cite{HanLeiQinYin2026,Song2026}. In this section we obtain both independently, with the soft vertex in position space, and treat each loop in turn. The late-time soft vertex misses the bubble by factors of $0.41$ to $13$ and the triangle by factors of $2.4$ to $113$. Away from the squeezed limit the leading term of the triangle soon fails, and at a large chemical potential much of the error is a phase set by the total energy.

\begin{figure}[t]
\centering
\includegraphics{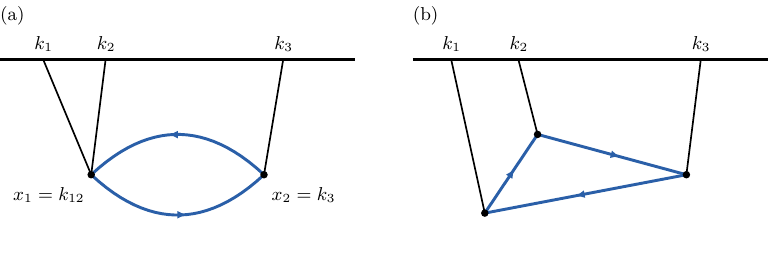}
\caption{Loops of a Dirac fermion with an axial chemical potential $\kappa$, with arrows for the fermion flow. (a) The bubble, with a quartic vertex for the legs $\mathbf{k}_1,\mathbf{k}_2$ and a cubic vertex for the soft leg. (b) The triangle, whose three vertices are $\gamma^\mu\gamma^5\partial_\mu\delta\tau$. The other orientation of the triangle contributes equally.}
\label{fig:fermions}
\end{figure}

\subsection{The Dirac line}\label{sec:dirac.line}

After the spinor is rescaled by $a^{-3/2}$, the two chiral amplitudes of a Dirac fermion of mass $m$ obey a first-order system. Without a chemical potential their sums and differences are Bessel functions of the orders $\frac12\mp\ii\mu$, with $\mu=m$~\cite{ref20,AdsheadSfakianakis2015}. The Dirac line therefore enters the representation with the exponents of a scalar line shifted by one half. This half unit is the source of the statistics sign. It makes the cut weights real and gives the time-ordered kernel the Fermi--Dirac weights $(1-n_F,\,n_F)$, with $n_F=(e^{2\pi\mu}+1)^{-1}$, in place of the $(1+\bar n,\,-\bar n)$ of Eq.~\eqref{eq:GF} (Appendix~\ref{app:diracline}).

An axial chemical potential $\kappa$, induced by the coupling $\partial_\mu\phi\,\bar\psi\gamma^\mu\gamma^5\psi/\Lambda$ with $\kappa=\dot\phi_0/\Lambda$, shifts the momentum term of the system by $h\kappa/\tau$ for helicity $h$. The chiral amplitudes then become Whittaker functions,
\begin{equation}\label{eq:diracmodes}
\alpha\propto\tau^{-1/2}\,W_{K_\alpha,\,\ii\bar\mu}(2\ii k\tau)\,,\qquad \beta\propto\tau^{-1/2}\,W_{K_\beta,\,\ii\bar\mu}(2\ii k\tau)\,,
\end{equation}
where $K_{\alpha,\beta}=\pm h\big(\tfrac12\mp\ii\kappa\big)$ and $\bar\mu^2=\mu^2+\kappa^2$~\cite{AdsheadSfakianakis2015,ref20,SouTongWang2021}, and the half-unit of statistics now appears in the first index. The occupation and the per-line pair amplitude are
\begin{equation}\label{eq:diracocc}
n_h=\frac{e^{2\pi h\kappa}-e^{-2\pi\bar\mu}}{2\sinh2\pi\bar\mu}\,,\qquad \text{per-line amplitude}=\frac{m}{2\bar\mu}\sqrt{n_h(1-n_h)}\,,
\end{equation}
where the first relation is the occupation of Ref.~\cite{AdsheadSfakianakis2015} and reduces to $n_F$ at $\kappa=0$. Since Eq.~\eqref{eq:helBarnes} holds at the Dirac indices, the Dirac line with a chemical potential has the same two-fold Barnes kernel as the helical vector.

In position space the soft vertex of both loops reduces to a single radial integral. With the projectors traced, the loop integrand is a sum of monomials $y_1^Ay_2^B$ (with $y_{1,2}$ the soft-line momenta in units of $k_3$), multiplied by products of Whittaker functions of $2y_1t$ and $2y_2t$ and by the external mode at the vertex. We use the convolution theorem for the loop integral and the Laplace transform~\cite{GradshteynRyzhik}
\begin{multline}\label{eq:WLaplace}
\int_0^\infty\!\dd z\,e^{-sz}z^{p-1}W_{K,\ii\bar\mu}(z)\\
=\frac{\Gamma(p+\frac12+\ii\bar\mu)\,\Gamma(p+\frac12-\ii\bar\mu)}{\Gamma(p+1-K)}\;{}_2F_1\!\Big(p+\tfrac12+\ii\bar\mu,\,p+\tfrac12-\ii\bar\mu;\,p+1-K;\,\tfrac12-s\Big)
\end{multline}
for the soft lines. The time integral at the vertex then gives a Gamma function, and each monomial becomes
\begin{equation}\label{eq:Jpos}
J=\frac{\Gamma(Q)}{(2\pi)^6}\int\!\dd^3w\;\phi_A(w)\,\phi_B(w)\,\big(1+\ii\,\mathbf{w}\cdot\hat{\mathbf{k}}_3\big)^{-Q}\,,
\end{equation}
where $Q$ is fixed by the powers of the vertex and the monomial, and $\phi_A(w)$ is the difference of two ${}_2F_1$'s of Eq.~\eqref{eq:WLaplace} at $s=\frac12\mp\frac{\ii w}{2}$. The angular integral is elementary, so a single radial integral of Gauss functions remains, with no loop-momentum grid and no time grid. The ultraviolet region of the loop is continued analytically by $\Gamma(Q)$. One point needs care. At large $w$ the integrand falls only as $w^{-2}$ (times $w^{\pm2\ii\bar\mu}$), and its coefficients are of size $e^{\pi\bar\mu}$, since $\arg(1\mp\ii w)\to\mp\frac\pi2$ in the powers of Eq.~\eqref{eq:Jpos} and in the Gauss functions. We therefore add the tail beyond the last node from its asymptotic expansion, whose powers are known. Cutting the integral off at $w=e^{12}$ instead leaves an error of order $10^4$ at $\kappa=10$, against a coefficient of order $10$. At large $\kappa$ the radial integrand is much larger than the result (for the triangle of Section~\ref{sec:dirac.triangle} it reaches $5\times10^{11}$ at $\kappa=10$ and $6\times10^{26}$ at $\kappa=20$, against coefficients of modulus $17.6$ and $231$), so the coefficients, the quadrature nodes, and the sums must be carried to thirty to sixty digits. The evaluation takes a few minutes on one core.

The radial integral can also be done. The first indices of the two soft lines of a Dirac loop are equal or opposite up to an integer, and each soft moment is then a finite sum of Gamma functions, hypergeometric functions at argument $\pm1$, and Meijer $G$ functions at argument $-1$ (Appendix~\ref{app:diracline}). This is the kind of closed form that Ref.~\cite{Song2026} found for the leading coefficient of the triangle, and it holds for every monomial. We quote numbers from the radial integral and use the closed form as a check.

\subsection{The fermion bubble}\label{sec:dirac.bubble}

\paragraph{The model.} The bubble of the scalar bilinear (Fig.~\ref{fig:fermions}a) has been computed for a Weyl fermion with a Majorana mass~\cite{HanLeiQinYin2026}, including its squeezed bispectrum for the vertices $\delta\phi'\,\psi\psi$ and $\delta\phi'^2\,\psi\psi$ (plus conjugates), which are the time-derivative parts of the couplings of Ref.~\cite{ref20}. A Dirac fermion with an axial chemical potential is two such copies. At these vertices $r_1=4$ and $r_2=2$, with $x_2=k_3$. Without a chemical potential the nonlocal and local signals of the bubble are known exactly~\cite{ref25}, and in our variables its cut term is also a single integral over the invariant distance that needs no subtraction, with a density built from two Gauss functions per line.

\paragraph{The leading coefficient.} The squeezed coefficient is a hard-vertex residue times one soft moment, which has the closed form of Section~\ref{sec:dirac.line} since $K_1=\pm K_2$ for the two lines. The bispectrum sums two pair branches over the four Schwinger--Keldysh sectors. At order $n$ in $k_3/k_{12}$ its coefficient is
\begin{equation}\label{eq:bubblebranches}
\mathcal{B}_n=\big[1-(-1)^ne^{-2\pi\bar\mu}\big]\,\Gamma(4+2\nu+n)\,S_n^++\big[1-(-1)^ne^{2\pi\bar\mu}\big]\,\big[\Gamma(4-2\nu+n)\,S_n^-\big]^*\,,
\end{equation}
where $S_n^\pm$ are the soft moments of the two branches and $\nu=\ii\bar\mu$. The second branch is smaller than the first by $10^{-5}$ to $10^{-8}$, but its weight $e^{2\pi\bar\mu}$ makes up for this. At $n=0$ Eq.~\eqref{eq:bubblebranches} is the combination of Eq.~(5.41) of Ref.~\cite{HanLeiQinYin2026}. Inserted into their Eqs.~(5.39)--(5.43), our soft moments reproduce the amplitude and the phase of their Fig.~8 to the accuracy of the figure, at all $21$ points we compared (masses from $1$ to $7$ and chemical potentials from $1$ to $20$, in units of $H$). We compare the exact coefficient with the late-time pair term, in which both soft lines keep their pair branch at the soft vertex. At late times the two branches are complex conjugates, so this estimate carries the factor $2-2\cosh2\pi\bar\mu$. The exact coefficient is $13$, $8.9$, $2.5$, and $0.41$ times the estimate at $(\mu,\kappa)=(0.5,0.25)$, $(1,0.5)$, $(1,1)$, and $(2,2)$. At $\kappa=0$ the pair term vanishes, as noted in Section~\ref{sec:SM}. For the covariant quartic vertex $(\partial_\mu\phi)^2\bar\psi\psi$ of Ref.~\cite{ref20}, the gradient term multiplies the leading coefficients by $1+(3\pm\ii\bar\mu)/(1\pm\ii\bar\mu)$, independently of the angle between $\mathbf{k}_1$ and $\mathbf{k}_2$.

\paragraph{Beyond the squeezed limit.} The same soft moments give the higher orders $\mathcal{B}_n$ of Eq.~\eqref{eq:bubblebranches}, whose powers of $k_3/k_{12}$ advance in steps of one. The second branch is $0.10$ and $0.44$ of the first in $\mathcal{B}_0$ and $\mathcal{B}_1$ at $(\mu,\kappa)=(1,1)$, and $0.85$ and $0.94$ at $(2,2)$, so it cannot be dropped. The pair signal exceeds its leading term by $2\%$ at $R=2k_3/k_{12}=0.2$ and by a factor of $2.2$ at $R=0.6$ for $(1,1)$, whereas at $(2,2)$ it lies $11\%$ and $9\%$ below it. Its phase moves sooner, so that as a complex amplitude the leading term is good to $10\%$ only for $R\le0.054$ and $0.034$. At the benchmarks of the triangle, $(\mu,\kappa)=(1,1.3)$, $(1,5)$, and $(1,10)$, the same test gives $R=0.055$, $0.019$, and $0.010$ (Table~\ref{tab:shapedomain}). For $\kappa\gg\mu$ the second branch is small, $2.3\%$ of the first at $\kappa=5$ and $0.8\%$ at $\kappa=10$ for $\mu=1$. The series of the first branch then follows in closed form, Eq.~\eqref{eq:bubbleshape} of Section~\ref{sec:compare.phase}. At $\kappa=5$ it reproduces that series to $10\%$ as a complex amplitude up to at least $R=0.38$, where the computed orders end, and the factor $(1+R/2)^{-2\ii\bar\mu}$ alone holds up to $R=0.26$. Table~\ref{tab:shapedomain} quotes only $R>0.04$, because its stricter test trusts the series only while its last two orders change the sum by less than $0.2\%$.

\paragraph{Checks.} We have checked the series of the bubble in three ways. First, its leading coefficient reproduces Fig.~8 of Ref.~\cite{HanLeiQinYin2026}, as described above. Second, we have integrated the bubble directly in the time domain, with the exact Schwinger--Keldysh functions of the pair, both time orderings, and a cut on the loop momentum, which changes only terms analytic in $k_3$. We then fit an analytic background of $12$ to $16$ powers of $k_3$, with and without $\ln k_3$, together with the series. At $(\mu,\kappa)=(1,1)$, for $2\times10^{-4}\le R\le0.6$, the series with no free parameter leaves a residual below $1.3\times10^{-5}$ of its envelope, and the leading term alone leaves one at least $1200$ times larger. A free factor on the sum of the corrections returns one to $7\times10^{-4}$. Fitted one at a time, the first two corrections are confirmed to $7\times10^{-4}$ and $2.5\%$, and with $16$ background terms the third is confirmed to $0.8\%$. At $(2,2)$, for $0.02\le R\le0.6$, the same factor returns one to $4\times10^{-4}$, and the first four corrections are confirmed to $3\times10^{-4}$, $0.13\%$, $0.6\%$, and $2\%$. At $(1,0.5)$, where the signal has one period per $2.8$ e-folds of $k_3$, the factor returns one to $4\times10^{-4}$ with $14$ or more background terms and to $3\%$ with $12$. Third, the doubly cut and the time-ordered sectors contain different combinations of the two branches, and each agrees with its own direct integral to $2\times10^{-5}$ at $(1,1)$ and $(2,2)$. With the first branch alone the factor on the corrections is off by $0.34$ at $(1,1)$ and by $0.43$ to $0.48$ at $(2,2)$. The orders $n\ge4$ at $(1,1)$ and $n\ge5$ at $(2,2)$ are not tested one by one.

\paragraph{Size.} Neither Ref.~\cite{ref20} nor Ref.~\cite{HanLeiQinYin2026} quotes a signal size for this bubble. We take the operator $(\partial_\mu\phi)^2NN/\Lambda^3$ of Ref.~\cite{ref20} at both vertices, with the same cutoff as the coupling that sets the chemical potential, $\kappa=\dot\phi_0/\Lambda$. Equation~(2.48) of Ref.~\cite{HanLeiQinYin2026} then gives $f_{\rm NL}^{\rm osc}=1.5\times10^{-7}$ at $(m,\kappa)=(1,10)$ and $1.0\times10^{-4}$ at $(1,20)$ for one Weyl fermion, before the factor of about two from the gradient term. This estimate uses their normalization, which we have not rederived, with the amplitude of their Fig.~8, which our soft moment reproduces. The triangle of the same fermion has $7.3\times10^{-7}$ and $8.6\times10^{-6}$ at these points (from the closed form of Ref.~\cite{Song2026}). The bubble of this dimension-seven operator is therefore not negligible next to the triangle for $\kappa\gtrsim10$, although both remain small.

\subsection{The fermion triangle}\label{sec:dirac.triangle}

\paragraph{The model.} In the model of Ref.~\cite{AokiStrumia2026}, the inflaton is a modulus $\tau$, the CP phases of the Yukawa couplings evolve during inflation, and the Standard Model fermions acquire chemical potentials in a Higgs condensate. We keep the axial one, $\kappa$. The vector one drops out of the correlators of $\delta\tau$, since the Dirac mass conserves fermion number and a rotation of the fermion phase removes the coupling to the vector current. The leading signal is the fermion triangle (Fig.~\ref{fig:fermions}b), with two vertices on the hard legs $\mathbf{k}_1,\mathbf{k}_2$ and one on the soft leg $\mathbf{k}_3$. In the squeezed limit we write $\langle\delta\tau^3\rangle'\supset C\,k_1^{-6}(k_3/k_1)^{2\ii\bar\mu}$, so that $S=-4\pi^2P_\zeta\,C$.

\paragraph{The leading coefficient.} At this order $C$ does not depend on the angle between $\mathbf{k}_1$ and $\mathbf{k}_3$, and the corrections in $k_3/k_1$ follow below. As in Section~\ref{sec:regions.split}, $C$ splits into a hard part and a soft vertex. The hard part contains the two hard vertices with the soft lines at their late-time ends. It is a nested double time integral for the time-ordered branches and a product of single integrals for the cut ones, and both have closed forms. The cut ones are products of Gamma functions, and the time-ordered ones are sums of ${}_4F_3$ functions at argument $-1$ (Eq.~\eqref{eq:fermimoment} of Appendix~\ref{app:box}), so no contour needs to be chosen. The soft vertex contains the two soft lines at the vertex time, and Ref.~\cite{AokiStrumia2026} replaces them by their late-time branches. Here we keep them exactly.

Our coefficient agrees with the closed form of Ref.~\cite{Song2026} for a Majorana fermion. A Dirac fermion with an axial chemical potential is two Majorana copies, and the couplings $1/\Lambda$ of that reference become $\kappa$. Twice that closed form then reproduces our $C$ (Table~\ref{tab:fermiC}), in modulus and phase, to better than $10^{-23}$ at $\kappa=1.3$, $5$, and $10$, and to $10^{-18}$ at $\kappa=20$. With the soft moments in closed form the agreement is $3\times10^{-29}$, $2\times10^{-40}$, $2\times10^{-44}$, and $9\times10^{-50}$ at these four points, at working precisions of $30$, $45$, $60$, and $90$ digits. A direct evaluation of the soft vertex on a grid in loop momentum and vertex time, with the same hard part, also agrees at $\mu=1$, to $2\times10^{-6}$ at $\kappa=1.3$ and to $3\times10^{-4}$ at $\kappa=5$. The closed form of Ref.~\cite{Song2026} needs about $80$ digits at $\kappa=20$, as does our radial integral. The closed soft moments lose only about five digits there, so $30$ digits give $C$ to $6\times10^{-25}$.

\begin{table}[t]
\centering
\begin{tabular}{@{}cccc@{}}
\toprule
$(\kappa,\mu)$ & $C$ & $|S|/P_\zeta$ & $|C/C_{\rm late}|$\\
\midrule
$(1.3,1)$ & $(-6.104+1.511\,\ii)\times10^{-4}$ & $0.0248$ & $2.38$\\
$(5,1)$ & $-1.0638+0.1083\,\ii$ & $42.2$ & $36.8$\\
$(10,1)$ & $-17.578+0.4925\,\ii$ & $694$ & $113$\\
$(20,2)$ & $-230.86-6.511\,\ii$ & $9.12\times10^{3}$ & $66.9$\\
\bottomrule
\end{tabular}
\caption{The squeezed coefficient $C$ of the fermion triangle, $\langle\delta\tau^3\rangle'\supset C\,k_1^{-6}(k_3/k_1)^{2\ii\bar\mu}$, with $|S|/P_\zeta=4\pi^2|C|$. $C_{\rm late}$ is the same coefficient with the late-time soft lines of Ref.~\cite{AokiStrumia2026} at the soft vertex.}
\label{tab:fermiC}
\end{table}

The triangle also differs strongly from its late-time estimate. For $\mu=1$ we find $|S|/P_\zeta=0.025$, $42$, and $694$ at $\kappa=1.3$, $5$, and $10$ (Table~\ref{tab:fermiC}). Replacing the soft lines at the soft vertex by the late-time branches of Ref.~\cite{AokiStrumia2026} (with the hard part unchanged) lowers these by factors of $2.4$, $37$, and $113$, and the closed estimate of that reference (their Eq.~(60)) is lower again by a factor of about two. At $\kappa=20$ and $\mu=2$ we find $|S|/P_\zeta=9.1\times10^3$, which is $67$ times the late-time value and $116$ times the estimate of their Eq.~(60). As explained in Section~\ref{sec:regions.split}, the soft vertex sits at horizon crossing of $k_3$, where the soft lines are not yet at late times. With a chemical potential, the two late-time branches of each soft line carry weights that differ by factors of order $e^{\pi\kappa}$, through the Gamma functions of the connection formula. If all branch assignments at the soft vertex are kept at leading late-time order, the result at $\kappa=10$ is $7.4\times10^3$ times the single assignment of Ref.~\cite{AokiStrumia2026} and $66$ times the exact soft vertex. Neither truncation is accurate, and the exact soft vertex lies between them. In the squeezed limit the leading power $(k_3/k_1)^{2\ii\bar\mu}$ and its hard-vertex factor are unaffected, so the signal has the form of Ref.~\cite{AokiStrumia2026}, with a different size and phase. Away from that limit the shape changes as well, as we show below.

The growth with the chemical potential also changes. Between $\kappa=5$ and $10$ the exact coefficient grows by a factor of $16$ (roughly as $\kappa^4$). At large $\kappa$ the closed form of Ref.~\cite{Song2026} behaves as $\kappa^3e^{-11\pi\mu^2/4\kappa}$ (their Eqs.~(20) and~(24)), whose logarithmic slope $3+11\pi\mu^2/4\kappa$ is close to four in this range. The exact $|C|$ exceeds this limiting form by $21\%$, $4.8\%$, and $2.1\%$ at $(\kappa,\mu)=(5,1)$, $(10,1)$, and $(20,2)$. For comparison, Eq.~(60) of Ref.~\cite{AokiStrumia2026} grows by a factor of five, and the earlier estimates collected in their Eq.~(16) by a factor of $44$. In other words, the exact soft vertex restores part of the growth with $\kappa$ that the late-time soft loop integral suppresses.

\paragraph{Beyond the squeezed limit.}\label{sec:dirac.shape} The region argument of Section~\ref{sec:regions.split} holds at every order, as for the scalar triangle of Section~\ref{sec:triangle.shape}. The pieces it leaves out are proportional to the anticommutator of the field, which is analytic in the momentum of the line once the helicities are summed. Three expansions enter. The hard ends of the soft lines follow the branch series of Eq.~\eqref{eq:en} at the Dirac indices, and since these indices do not vanish, the powers of $k_3$ advance in unit steps rather than in steps of two. The hard line is expanded around the folded point, where its moments are Gamma functions and the ${}_4F_3$ series of Eq.~\eqref{eq:fermimoment}. Its energy, and that of the second hard leg, follow by scaling as in Eq.~\eqref{eq:Dfold}. Finally, the spin traces are polynomials in the loop momentum, and each monomial is one soft moment of the form of Eq.~\eqref{eq:Jpos}. Each of these moments has the closed form of Appendix~\ref{app:diracline}, since each soft line contributes at most one inverse power of its momentum. With the closed form the coefficients agree with their position-space values to $6\times10^{-28}$ through second order at $\kappa=1.3$, and to $2\times10^{-32}$ through first order at $\kappa=5$.

At a fixed angle between $\mathbf{k}_1$ and $\mathbf{k}_3$ the series in $k_3/k_1$ converges only for $k_3<k_1$, and its coefficients grow. The signal is symmetric under the exchange of $\mathbf{k}_1$ and $\mathbf{k}_2$, however, and in variables that respect this symmetry the coefficients fall. In the variables $R$ and $\delta$ of Eq.~\eqref{eq:shapevars}, the pair-frequency signal is
\begin{equation}\label{eq:fermishape}
\langle\delta\tau^3\rangle'\supset\Big(\frac{2}{k_1+k_2}\Big)^{6}\,R^{\,2\ii\bar\mu}\sum_{N\ge0}A_N(\delta)\,R^N+\text{c.c.}\,,\qquad A_0=C\,,
\end{equation}
where each $A_N$ is an even polynomial in $\delta$ of degree at most $N+1$. The exchange of the hard legs maps $\delta$ to $-\delta$, so the series respects it term by term. Equation~\eqref{eq:fermishape} takes $\mathbf{k}_3$ as the soft leg, and a template symmetric in the three legs is the sum over the three choices.

The leading coefficient $A_0$ does not depend on $\delta$, nor on the angle between $\mathbf{k}_1$ and $\mathbf{k}_3$. Ref.~\cite{ref24} found this with late-time soft lines, and it holds with the exact soft vertex as well. At leading order $\hat{\mathbf{k}}_1$ enters only through the hard part, which is a spin matrix built from $1$ and $\bm{\sigma}\cdot\hat{\mathbf{k}}_1$, so $C$ is at most linear in $\hat{\mathbf{k}}_1\cdot\hat{\mathbf{k}}_3$. The exchange of $\mathbf{k}_1$ and $\mathbf{k}_2$ reverses this cosine at leading order, so the linear term vanishes (numerically it is below $10^{-29}$ of $C$). The collinear coefficient above, and that of Ref.~\cite{Song2026}, is therefore the leading coefficient of every squeezed triangle. (For the helical vector of Section~\ref{sec:chem.hard} the polarization sum is quadratic in $\hat{\mathbf{k}}_1$, and a quadrupole survives.)

\begin{table}[t]
\centering
\begin{tabular}{@{}llcccccc@{}}
\toprule
$\kappa$ & & $N=1$ & $2$ & $3$ & $4$ & $5$ & $6$\\
\midrule
$1.3$ & $|A_N/A_0|$ & 2.40 & 2.40 & 0.60 & 0.55 & 0.33 & 0.057\\
$5$ & $|A_N/A_0|$ & 5.14 & 13.1 & 21.6 & 25.5 & 23.7 & 20.0\\
 & $|G_N|$ & 0.50 & 0.28 & 0.47 & 0.59 & 0.62 & 0.57\\
$10$ & $|A_N/A_0|$ & 10.0 & 49.8 & 164 & 405 & 796 & 1301\\
 & $|G_N|$ & 0.66 & 0.29 & 0.20 & 0.23 & 0.63 & 0.79\\
\bottomrule
\end{tabular}
\caption{Coefficients of Eqs.~\eqref{eq:fermishape} and~\eqref{eq:fermiphase} on $k_1=k_2$, for $\mu=1$. The $G_N$ are the coefficients with the factor $(1+R/2)^{-2\ii\bar\mu}$ removed. The complex $A_N/A_0$ are listed in Appendix~\ref{app:fermicoef}.}
\label{tab:fermishape}
\end{table}

\begin{figure}[t]
\centering
\includegraphics{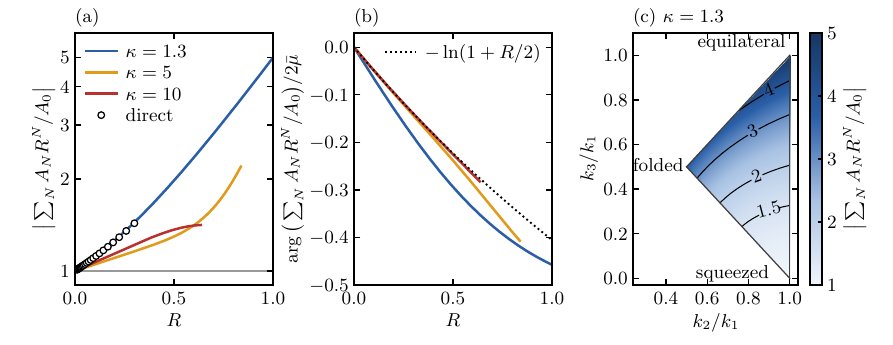}
\caption{The pair-frequency signal of the fermion triangle away from the squeezed limit, for $\mu=1$. (a) Modulus of the series of Eq.~\eqref{eq:fermishape} over that of its leading term, on $k_1=k_2$. Each curve ends where one of its last two terms reaches $2\%$ of the sum. Circles show the direct integration at $\kappa=1.3$, with $A_0$, $A_1$, and $A_2$ fitted and the higher orders taken from the series. (b) The phase of the same ratio divided by $2\bar\mu$. The dotted line is the phase of the factor $(1+R/2)^{-2\ii\bar\mu}$ of Eq.~\eqref{eq:fermiphase}. (c) The modulus ratio at $\kappa=1.3$ over the triangles in which $k_3$ is the shortest side.}
\label{fig:fermishape}
\end{figure}

The corrections are large. At $\kappa=1.3$ and $\mu=1$ the series converges up to the equilateral shape, and we have computed it through $N=10$ (Table~\ref{tab:fermishape}). On $k_1=k_2$ its modulus exceeds that of the leading term by factors of $1.44$, $1.98$, and $4.98$ at $R=0.3$, $0.5$, and $1$ (Fig.~\ref{fig:fermishape}). At the folded shape $k_1=2k_2=2k_3$ the factor is $3.78$. The last two orders change these values by at most $0.4\%$. The leading term alone reproduces the signal to $10\%$ as a complex amplitude only for $R\le0.043$ (Table~\ref{tab:shapedomain}), and to $1\%$ only for $R\le0.004$.

The corrections grow with $\kappa$. For $\kappa\gg\mu$ we find $|A_N/A_0|\simeq\bar\mu^N/N!$ on $k_1=k_2$, to within $20\%$ through $N=6$ at $\kappa=5$ and $10$. The first coefficient is $A_1/A_0=0.50-5.11\,\ii$ at $\kappa=5$ and $0.66-9.99\,\ii$ at $\kappa=10$. The leading term alone is then good to $10\%$ only for $R\lesssim0.02$ and $R\lesssim0.01$. Most of this growth is a phase. We write
\begin{equation}\label{eq:fermiphase}
R^{\,2\ii\bar\mu}\sum_{N\ge0}A_N(\delta)\,R^N=C\,\Big(\frac{2k_3}{k_T}\Big)^{2\ii\bar\mu}\sum_{N\ge0}G_N(\delta)\,R^N\,,\qquad k_T=k_1+k_2+k_3\,,
\end{equation}
with $G_0=1$. On $k_1=k_2$ the $G_N$ are of order one through the orders computed ($N\le10$ at $\kappa=5$ and $N\le8$ at $\kappa=10$). At a large chemical potential the signal therefore oscillates in $k_3/k_T$ rather than in $k_3/k_1$ (Fig.~\ref{fig:fermishape}b), and with this factor the leading term is good to $10\%$ up to $R=0.21$ at $\kappa=5$, and at least up to $R=0.14$, where the series ends, at $\kappa=10$. The modulus changes less, by factors of $1.16$ and $1.22$ at $R=0.3$.

For $\kappa\gg\mu$ the factor can be derived. A chiral rotation of the fermion by the fluctuations of the two hard legs removes their axial vertices and moves the coupling to the mass term, $m\,\bar\psi\,e^{2\ii\kappa\gamma^5\delta\tau_h/\dot\tau_0}\psi$~\cite{ref21,AokiStrumia2026}. Here $\delta\tau_h$ is the part of $\delta\tau$ that holds the two hard legs, and $\dot\tau_0$ is the background velocity. The term of second order is a quartic vertex, and with the axial vertex of the soft leg it forms a bubble of the kind of Fig.~\ref{fig:fermions}a. The term of first order, taken twice, forms a triangle with two pseudoscalar vertices. The triangle of Fig.~\ref{fig:fermions}b is exactly the sum of these two diagrams. We have verified this on the coefficients $A_N$ with $N\le2$, to $10^{-16}$ at $\kappa=5$, $10$, and $20$. The first diagram has a local hard vertex, so the argument of Section~\ref{sec:compare.phase} applies, and for $\kappa\gg\mu$ its series is
\begin{equation}\label{eq:fermicontact}
\Big(\frac{2k_3}{k_T}\Big)^{2\ii\bar\mu}\Big(1+\frac R2\Big)^{2\ii(\bar\mu-\kappa)}\Big(1-\frac R2\Big)^{-2}\Big(1-\frac{R^2\delta^2}{4}\Big)^{-2}
\end{equation}
times its leading term, with corrections of order $1/\kappa$. This is the factor of Eq.~\eqref{eq:fermiphase}, with the exponent $2\ii\kappa$ on the total energy as for the bubble in Eq.~\eqref{eq:bubbleshape}. With that factor removed, the first coefficient of the first diagram on $k_1=k_2$ is $1+\ii(\bar\mu-\kappa)$ at this order. Its exact values are $0.97+0.13\,\ii$, $1.00+0.06\,\ii$, and $1.01+0.03\,\ii$ at $\kappa=5$, $10$, and $20$, which agree to $0.04$, $0.014$, and $0.006$. The second diagram is smaller by a factor of order $\mu^2/\kappa$, and we have not derived its shape. At $R\to0$ it is $-0.35+0.02\,\ii$, $-0.21+0.05\,\ii$, and $-0.11+0.05\,\ii$ of the first at these three values of $\kappa$, and $-0.39+0.13\,\ii$ at $\kappa=20$ and $\mu=2$. It varies on the scale $R\sim\kappa^{-1/2}$, since the hard line is then close to the turning point of its enhanced mode. Numerically it lowers $G_1$ from the value of the first diagram to $0.50-0.02\,\ii$, $0.66+0.06\,\ii$, and $0.75+0.05\,\ii$, a shift of $-1.1\,\mu^2/\sqrt\kappa$ in the real part to within $0.02$. With this shift as a factor $e^{-1.1\mu^2R/\sqrt\kappa}$, Eq.~\eqref{eq:fermicontact} reproduces the series on $k_1=k_2$ to $10\%$ as a complex amplitude up to $R=0.35$ at $\kappa=5$, and up to at least $R=0.20$ at $\kappa=10$, where the computed orders end.

Equation~\eqref{eq:fermiphase} thus describes the first diagram, and it requires $\mu^2\ll\sqrt\kappa$. It fails at $\mu=\kappa=5$, where we find $A_1/A_0=6.5-15.7\,\ii$, whereas the phase alone would give $-7.1\,\ii$. The $G_N$ also grow with the mass, to $|G_N|=0.29$, $1.2$, $1.6$, and $4.9$ for $N=1$ to $4$ at $\kappa=20$ and $\mu=2$, where $\mu^2/\sqrt\kappa=0.9$. At $\kappa=10$ they grow with $N$ away from $k_1=k_2$, and for collinear momenta they reach $7.7$ at $N=8$. The exact coefficients of the first diagram fall there, from $1.26$ at $N=2$ to $0.22$ at $N=8$, so this growth belongs to the second diagram.

With the couplings divided out, the coefficients are odd in $\kappa$, as parity requires for three axial vertices. The pair signal of the triangle therefore vanishes without a chemical potential at every order. At small $\kappa$ the first correction falls more slowly than the leading term. Between $\kappa=0.01$ and $0.1$ we find $A_0/\kappa^3\propto\kappa^{2.99}$, $A_1/\kappa^3\propto\kappa^{0.97}$, and $A_2/\kappa^3\propto\kappa^{2.98}$. The first correction then dominates unless $R<|A_0/A_1|$, which is $0.034$ at $\kappa=0.1$.

\paragraph{Checks.} We have checked the series in three ways. First, the symmetry under $\mathbf{k}_1\leftrightarrow\mathbf{k}_2$ is not built into the expansion, which treats the two hard legs differently. It holds order by order to $3\times10^{-27}$ at $\kappa=1.3$ and $7\times10^{-27}$ at $\kappa=5$ (through $N=10$), and to $7\times10^{-34}$ at $\kappa=10$ (through $N=8$). The other orientation of the loop is the first one with the hard legs exchanged, so the same symmetry makes the two orientations equal, as charge conjugation requires. Second, the leading coefficient agrees with twice the closed form of Ref.~\cite{Song2026} to better than $10^{-23}$ at these three values of $\kappa$. Third, we have integrated the triangle directly in the time domain at fixed shape. This uses the exact Schwinger--Keldysh functions of all three lines, all time orderings, and a cut on the loop momentum (which changes only terms analytic in $k_3$). We then fit an analytic background of $12$ to $16$ powers of $k_3$, with and without $\ln k_3$, together with the series. At $\kappa=1.3$ on $k_1=k_2$, for $0.01\le R\le0.30$, the series with no free parameter leaves a residual below $2\times10^{-6}$ of its envelope, and the leading term alone leaves one at least $1500$ times larger. A free factor on the sum of the corrections returns one to $4\times10^{-4}$. Fitted one at a time, $A_1$, $A_2$, and $A_3$ are confirmed to $4\times10^{-4}$, $3\times10^{-3}$, and $9\%$. For collinear momenta (up to $R=0.29$) and at $\delta=-\frac12$ (up to $R=0.28$) the corrections are confirmed to $3\%$ and $0.8\%$. These fits are less precise, because the background of the cut integral has singularities closer to the fitted range. At $\kappa=3$ the same comparison for collinear momenta returns $A_0$, $A_1$, and $A_2$ to $4\times10^{-5}$, $6\times10^{-4}$, and $2\%$. At $\kappa=5$, where the corrections reach $1.4$ times the leading term at $R=0.29$, it returns $A_0$ to $4\times10^{-4}$, $A_1$ to $0.2\%$, $A_2$ to $0.5\%$, and $A_3$ to $3\%$. This tests Eq.~\eqref{eq:fermiphase} through $N=3$. The orders $N\ge4$, and all orders at $\kappa=10$, rest on the exchange symmetry alone.

\paragraph{Size.} In the minimal scenario of Ref.~\cite{AokiStrumia2026}, where the couplings are tied to $P_\zeta$, the signal nevertheless remains small. With three colors and $P_\zeta=2.1\times10^{-9}$, $|f_{\rm NL}^{\rm osc}|=|S|$ is $4.4\times10^{-6}$ at $(\kappa,\mu)=(10,1)$ and $5.7\times10^{-5}$ at $(20,2)$, far below the reach of planned surveys. In models where the couplings are not tied to $P_\zeta$, such as the right-handed neutrino triangle of Refs.~\cite{ref24,Song2026} (see also Ref.~\cite{CuiXianyu2022}), the signal can be larger. For a cutoff $\Lambda=60H$, at the edge of the derivative expansion ($\Lambda^2>\dot\phi_0$), Ref.~\cite{Song2026} finds $f_{\rm NL}=5\times10^{-4}$ at a mass of $4H$ (in a normalization that is $10/9$ of ours). Their largest value, $1.4\times10^{-2}$, requires $\Lambda=30H$, where $\dot\phi_0^2/\Lambda^4\simeq15$.

%% file: sec9cmp.tex
\section{Signals across models}\label{sec:compare}

The coefficients of the previous sections multiply the leading power of $k_3/k_1$, and templates use this term at all shapes. In this section we compare the loops away from the squeezed limit, ask in which variable their signals oscillate, and end with their reach in surveys. The region argument of Section~\ref{sec:regions.split} holds at every order in $k_3$, so the pair signal of each loop is a series in the variables $R$ and $\delta$ of Eq.~\eqref{eq:shapevars}, as in Eq.~\eqref{eq:scalarshape}. We have given these series for the scalar triangle (Section~\ref{sec:triangle.shape}), the Higgs-portal and charged-scalar bubbles (Sections~\ref{sec:SM} and~\ref{sec:chem.vertex}), the helical-vector bubble and triangle (Sections~\ref{sec:chem.bubble} and~\ref{sec:chem.hard}), and the fermion bubble and triangle (Sections~\ref{sec:dirac.bubble} and~\ref{sec:dirac.triangle}). Figure~\ref{fig:shapes} and Table~\ref{tab:shapedomain} compare them on $k_1=k_2$.

\subsection{Where the leading term holds}\label{sec:compare.range}

\begin{figure}[t]
\centering
\includegraphics{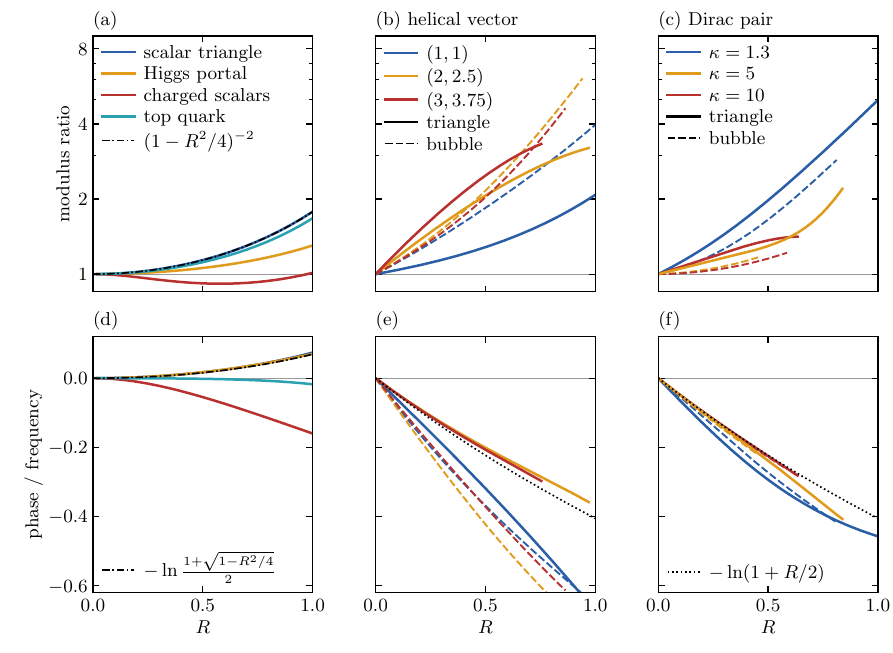}
\caption{The pair signal away from the squeezed limit, on $k_1=k_2$. Top row, the modulus of the signal over its leading squeezed term. Bottom row, the phase of the same ratio divided by the frequency of the signal. (a, d) Loops whose lines carry no chemical potential, with the large-mass forms of Eq.~\eqref{eq:scalarrapidity} (dash-dotted). The charged scalars have a chemical potential on the vertex. (b, e) The helical-vector triangle (solid) and contracted bubble (dashed) at $(\mu,\kappa)=(1,1)$, $(2,2.5)$, and $(3,3.75)$. (c, f) The fermion triangle (solid) and bubble (dashed) at $\mu=1$, with the phase of $(1+R/2)^{-2\nu}$ (dotted). Series end where one of their last two terms reaches $2\%$ of the sum.}
\label{fig:shapes}
\end{figure}

Without a chemical potential on the lines the corrections start at $R^2$ and are mild. The leading term is good to $10\%$ up to $R=0.44$ for the scalar triangle and up to $R=0.62$ for the Higgs portal (Table~\ref{tab:shapedomain}). The $W/Z$ loop is good to $10\%$ only up to $R=0.22$ at $\mu=1$, where its higher families are large, and up to about $R=0.5$ at $\mu=2$ and $4$. The top-quark loop is good up to $R=0.49$. The charged scalars are the exception, since the phase on their vertices shifts the frequency of the signal. Their modulus stays within $10\%$ of the leading term beyond $R=1$, but their phase drifts quadratically in $R$ (Section~\ref{sec:chem.vertex}). A chemical potential on the lines changes this. The corrections then start at the first order in $R$, and their coefficients grow with $\kappa$. At our benchmarks the leading term is good to $10\%$ only for $R\lesssim0.08$ for the helical vector and $R\lesssim0.055$ for the Dirac pair, and less at larger $\kappa$. Much of this growth is a phase (Fig.~\ref{fig:shapes}, bottom row).

\begin{table}[t]
\centering
\footnotesize
\setlength{\tabcolsep}{4pt}
\begin{tabular}{@{}l>{\raggedright\arraybackslash}p{3.5cm}>{\raggedright\arraybackslash}p{3.1cm}>{\raggedright\arraybackslash}p{2.5cm}@{}}
\toprule
Loop & Benchmarks & Leading term & With phase factor\\
\midrule
Scalar triangle & $\mu=(1,1.3,0.7)$ & $0.44$ & $>1.4$\\
Higgs portal & $\mu_H=\frac12$, $1$, $2$, $4$; $\nu_H=1.45$ & $0.66$, $0.62$, $0.53$, $0.42$; $1.46$ & --\\
$W/Z$ & $\mu=1$, $2$, $4$ & $0.22$, $0.49$, $0.52$ & --\\
Top quark & $m=1$, $2$ & $0.49$, $0.49$ & --\\
Charged scalars & $(\lambda,\mu)=(40,4)$, $(20,2)$, $(10,1)$ & $0.11$, $0.15$, $0.23$ & --\\
\midrule
Helical-vector bubble & $(\mu,\kappa)=(1,1)$, $(2,2.5)$, $(3,3.75)$ & $0.051$, $0.027$, $0.021$ & $0.081$, $0.051$, $0.050$\\
Helical-vector triangle & $(\mu,\kappa)=(1,1)$, $(2,2.5)$, $(3,3.75)$ & $0.078$, $0.042$, $0.029$ & $0.21$, $0.065$, $0.050$\\
Fermion bubble & $(\mu,\kappa)=(1,1.3)$, $(1,5)$, $(1,10)$ & $0.055$, $0.019$, $0.010$ & $0.16$, $>0.04$, $>0.02$\\
Fermion triangle & $(\mu,\kappa)=(1,1.3)$, $(1,5)$, $(1,10)$ & $0.043$, $0.020$, $0.010$ & $0.087$, $0.21$, $>0.14$\\
\bottomrule
\end{tabular}
\caption{The largest $R=2k_3/(k_1+k_2)$ on $k_1=k_2$ up to which the leading squeezed term, with its exact coefficient, reproduces the pair signal to $10\%$ as a complex amplitude, $|S_{\rm lead}/S-1|\le0.1$. The ratio $R$ has no units. It vanishes in the squeezed limit and equals one at the equilateral shape. The last column repeats the test after multiplying the leading term by the phase factor of Section~\ref{sec:compare.phase}, Eq.~\eqref{eq:scalarrapidity} for the scalar triangle and $(1+R/2)^{-2\nu}$ for the lines with a chemical potential. That factor is not derived for the other loops (--). An entry $>R$ means that the criterion still holds where the series stops being reliable. Masses and chemical potentials are in units of $H$. The Higgs portal keeps its leading family, while the $W/Z$ and top-quark loops keep all families and the vertex $(\partial_\mu\delta\phi)^2$.}
\label{tab:shapedomain}
\end{table}

\subsection{The phase of the signal}\label{sec:compare.phase}

\paragraph{The rapidity of the pair.} For heavy lines the phase follows from a stationary point of the time integrals at the hard vertices. Consider the scalar triangle in the form of Eq.~\eqref{eq:pairshape}, with $|\nu_B|$ and $|\nu_C|$ large. The soft lines are then waves, $I_\nu(\nu\zeta)\propto(1+\zeta^2)^{-1/4}e^{\nu\eta(\zeta)}$ with $\eta'(\zeta)=\sqrt{1+\zeta^2}/\zeta$. The exponent is stationary at $\lambda=\nu_C/\nu_{BC}$, where the two soft lines have the same $\zeta$ and propagate as a single wave of order $\nu_{BC}$ and momentum $k_3$. Stationarity in the time then gives
\begin{equation}\label{eq:scalarrapidity}
R^{\,\nu_{BC}}\sum_{N\ge0}\frac{A_N}{A_0}\,R^N\;\to\;\Big(1-\frac{R^2}{4}\Big)^{-2}\,\bigg[\frac{4k_3}{k_{12}+\sqrt{k_{12}^2-k_3^2}}\bigg]^{\nu_{BC}}\,.
\end{equation}
The signal thus oscillates in the rapidity $\operatorname{arccosh}(k_{12}/k_3)$ of the pair, as the tree-level exchange of a single particle of mass $\mu_B+\mu_C$ does at large mass. This differs from $\ln(k_{12}/k_3)$ only at order $R^2$. For the tree-level exchange of a single particle, a saddle point of the time integral at the hard vertex gives the same dependence on $\operatorname{arccosh}(k_{12}/k_3)$, with the prefactor $(k_{12}^2-k_3^2)^{-3/4}$~\cite{QinRenauxPetelTongWerthZhu2025}.
The prefactor comes from the hard line and from the amplitudes of the waves, and at large masses it depends on the shape through $R$ alone. Equation~\eqref{eq:scalarrapidity} also holds well outside this regime. On $k_1=k_2$ it reproduces the series to about $1\%$ up to $R=1$, for equal masses between $\frac12$ and $10$ and for $\mu=(1,1.3,0.7)$. It is less accurate when the hard line is much heavier than the pair, with an error of $14\%$ at $R=1$ for $\mu=(10,2,2)$.

\paragraph{The total energy.} A chemical potential on the lines changes the wave. The hard end of a soft line is then the branch function $M_{K,\nu}(2yt)$ of Eq.~\eqref{eq:en}, and at $K=\frac12+\nu$ it is a pure exponential, $M_{\frac12+\nu,\nu}(2yt)=(2yt)^{\frac12+\nu}e^{-yt}$. Such a line acts on its hard vertex only through a power of the time and the shift $k_i\to k_i+y_i$ of the vertex energy. The hard part then becomes a power of $k_1+k_2+y_1+y_2$. If the soft vertex also holds the loop momentum on the segment $y_1+y_2=k_3$, the signal oscillates in $k_3/k_T$, with $k_T=k_1+k_2+k_3$. A Dirac line has the first indices $\pm\frac12\pm\ii\kappa$ and $\nu=\ii\bar\mu$ (Eq.~\eqref{eq:diracmodes}), so on one helicity it approaches this limit for $\kappa\gg\mu$, where the fermion is effectively massless. This accounts for the factor $(2k_3/k_T)^{2\ii\bar\mu}$ that we find for the fermion triangle in Section~\ref{sec:dirac.shape}. At $\kappa=10$ the assignment with the two hard vertices on one branch and the soft vertex on the other gives $97\%$ of its leading coefficient. The first correction of this assignment is $A_1/A_0=0.32-10.00\,\ii$, and the soft vertex holds $(y_1+y_2)/k_3$ within $0.03$ to $0.10$ of one. For the imaginary part the factor $(2k_3/k_T)^{2\ii\bar\mu}$ gives $-10.05\,\ii$, and the exponent $2\ii\kappa$ of the total energy derived below gives $-10\,\ii$. For the fermion bubble the factor reduces the imaginary parts of the first four coefficients of $\ln\sum_N(A_N/A_0)R^N$ from $10.2$ to at most $0.15$ at $\kappa=10$. A scalar line has $K=0$ and never reaches this limit, and neither does a charged scalar, whose chemical potential enters only through the vertices. The helical vector, with $K=\mp\ii\kappa$ and $\nu=\ii\mu$, follows the pattern only in part. For the triangle the imaginary part of $A_1/A_0$ is $-1.24$, $-1.90$, and $-2.87$ at our three benchmarks, close to the value $-\mu$ of the total energy, but its real part grows to $0.37$, $1.63$, and $2.13$. Both helicities contribute, and for $\kappa>m$ the enhanced mode has a turning point, so a single stationary point need not dominate.

For $\kappa\gg\mu$ these statements can be derived for a loop with two vertices. The two time integrals are dominated by stationary points, at which the energy of the vertex equals the sum of the frequencies of the two lines. These are the resonant saddles that Ref.~\cite{QinRenauxPetelTongWerthZhu2025} identified for tree-level exchange. With the lines of the previous paragraph they give the powers $(k_1+k_2+y_1+y_2)^{-2\nu}$ and $(k_3+y_1+y_2)^{2\nu}$, up to integer powers set by the vertices. The phase of their product is monotonic in $y_1+y_2$, and the triangle inequality bounds $y_1+y_2$ from below by $k_3$. The loop integral is therefore dominated by its end point $y_1+y_2=k_3$, within a width of order $k_3/\kappa$. The segment is thus an end point of the loop integral rather than a stationary point. For the fermion bubble this gives
\begin{equation}\label{eq:bubbleshape}
\sum_{N\ge0}\frac{A_N}{A_0}\,R^N\;\to\;\Big(1+\frac R2\Big)^{-2\ii\kappa}\Big(1-\frac{R^2}{4}\Big)^{-2}\bigg[1+\frac{4\rho}{c}+\frac{6\rho^2}{c\,(c+1)}\bigg]\,,
\end{equation}
with $\rho=2R/(2-R)$ and $c=2+2\ii\bar\mu$. The exponent of the total energy is $2\ii\kappa$, whereas the squeezed frequency remains $2\bar\mu$. The second factor arises because the soft vertex cannot produce a pair of equal helicities with parallel momenta, so the integrand vanishes linearly at the end point. The bracket is the complete series in $1/\bar\mu$ for lines with $K=\frac12+\nu$. Equation~\eqref{eq:bubbleshape} holds to first order in $\bar\mu-\kappa$, up to a factor that differs from one by $\ii(\bar\mu-\kappa)R^2/12$ at leading order. It describes the first of the two pair branches of Eq.~\eqref{eq:bubblebranches}. For $\mu=1$ it reproduces the first four coefficients of $\ln\sum_N(A_N/A_0)R^N$ for that branch to $0.03$ at $\kappa=5$ and to $0.006$ at $\kappa=10$, and the remainder falls as $1/\kappa^2$. The second branch is $2.3\%$ and $0.8\%$ of the first at these two values of $\kappa$, and with it the agreement is $0.03$ and $0.008$. Section~\ref{sec:dirac.shape} gives the corresponding statements for the fermion triangle. For its leading coefficient, Ref.~\cite{Song2026} showed that one assignment dominates, with the loop momentum on the collinear boundary $y_1+y_2=k_3$.

For the exchange of a single field with a chemical potential at tree level, Ref.~\cite{ref18} finds the signal as a function of $2r/(1+r)$ with $r=k_3/k_{12}$, which is $2k_3/k_T$. The same reference noted that the phase of loop signals departs from its squeezed form for $R\gtrsim0.02$ with a chemical potential and for $R\gtrsim0.2$ without one. Reference~\cite{ref80} found a similar departure in a numerical evaluation of the helical loop. For tree-level exchange of a helical field with a reduced sound speed of the inflaton, Ref.~\cite{QinRenauxPetelTongWerthZhu2025} finds a further, transient signal that oscillates linearly in the ratio of the long to the short momenta.

\subsection{Reach in surveys}\label{sec:compare.summary}

In terms of $f_{\rm NL}^{\rm osc}$, and against the most recent template for each model, none of the corrections of this paper moves a signal into or out of the reach of current or planned surveys. The fermion triangle had already left it~\cite{AokiStrumia2026,Song2026}. The helical-vector loop of Ref.~\cite{ref17} remains the only loop studied here that can reach $f_{\rm NL}^{\rm osc}$ of order one, and a shift of the chemical potential by less than $0.2$ absorbs the effect of the hard line on its size. The Standard Model loops and the fermion loops stay far below that level within the derivative expansion. No search has yet used the template of a loop signal, with its power $(k_3/k_1)^2$ and its frequency $2\mu$ (or $\mu_1+\mu_2$).

%% file: sec9.tex
\section{Conclusions}\label{sec:concl}

We have computed the one-loop correlators of massive fields beyond their late-time limits. For every diagram and every kind of line we determined the pieces, their weights in the correlator, and where their coefficients have a closed form. For diagrams with two vertices the answer is complete. The bubble and the two-loop sunset split into exactly three pieces for all masses, each with a coefficient given by one sum over tree diagrams. On the locus of the bispectrum each piece of the bubble is itself a single sum, which gives the bispectra of the Higgs-portal and charged-scalar loops at every shape. Their background is fixed by renormalization without a free constant. The integral over the invariant distance then has a closed form at conformal coupling, for two massless lines, and, for the constants of the background, on a lattice of points. The same integral gives banana diagrams at any loop order, as we showed for massive lines at three and four loops. With three and four vertices the loop integral retains Mellin variables. There we find closed forms for the pair-frequency signal of the triangle at every shape and for the collapsed limit of the box.

Changing the line affects the answer in three distinct ways. Spin shifts the exponents by integers and, in the tensor bubble, adds pieces with no scalar counterpart. A chemical potential leaves the exponents unchanged and modifies only the weights (through the connection data of the Whittaker line). Statistics displaces the pieces by half a unit and turns the weights into Fermi--Dirac ones. In each case the squeezed signal is a set of frequencies fixed by the exponents, with amplitudes set by the weights and by the loop integral. As the helical-vector and fermion bubbles illustrate, the loop integral can change these amplitudes by large factors relative to late-time estimates.

The amplitudes of the signals follow a simple rule in both limits. The leading coefficient is the product of a hard factor with the soft vertex, and only the first is treated correctly at late times. In the collapsed limit every soft line ends on hard vertices, so the late-time estimate already gives the full coefficient. In the bispectrum, if neither the lines nor the vertices break the de Sitter isometries, the soft vertex differs from its late-time form by the universal factor of Eq.~\eqref{eq:Univ}. This factor is between $10$ and $20$ for the heavy Higgs and gauge-boson loops and close to one for the light Higgs, and it applies to each family at its own dimension. For the top quark, whose first family has dimension $4+2\nu$, it is about $26$ at $m=H$. Once a chemical potential is present there is no universal factor, and the exact soft vertex then changes the coefficient by factors between $0.41$ and $113$ in the examples of Table~\ref{tab:summary}. For Dirac loops this soft vertex has a closed form in Gamma functions and Meijer $G$ functions at argument $-1$, and for the helical vector it reduces to Kamp\'e de F\'eriet functions of two variables. For the fermion triangle of Ref.~\cite{AokiStrumia2026} it also restores part of the growth with the chemical potential that the late-time soft vertex suppresses.

Once $k_3/k_1$ is no longer small, the leading term itself is a poor guide. If no chemical potential is present, it holds to ten percent up to $k_3$ between $0.22k_1$ and $0.66k_1$ (Table~\ref{tab:shapedomain}). When the chemical potential sits on the vertex it holds up to $k_3$ between $0.11k_1$ and $0.23k_1$, and with one on the lines only for $k_3$ below a few percent of $k_1$. In the last case the signal at the equilateral shape is two to five times larger. A large part of this error is a shift of phase. For scalar lines the oscillation follows $\operatorname{arccosh}(k_{12}/k_3)$, and for fermions whose chemical potential is large it follows $\ln(k_T/k_3)$. In all of these cases the late-time templates have the right leading power but not the right amplitude or phase, so they should be corrected before they are compared with data~\cite{Sohn2024,Cabass2025,Kumar2026planck,Kumar2026scalars,Philcox2026dissect}.

Loops of helical vectors also give a parity-odd collapsed trispectrum, and in the model of Ref.~\cite{ref17} the parity-odd part of its pair signal is comparable to the parity-even part (Section~\ref{sec:chem.parity}). At $\mu=\kappa=1$ the part of the box that is analytic in $\mathbf{s}$ also violates parity and exceeds the oscillating signals for $s\lesssim0.3\,k$, so relative to the local shape the loop falls as $(s/k)^4$ rather than $(s/k)^6$. For $u$ of order one both parts lie ten or more orders of magnitude below the Planck bound~\cite{PhilcoxCMB3}, out of reach of the present parity-odd searches~\cite{Philcox2022,HouSlepianCahn2022,Philcox2023cmb,HouDESI2025}.

Several questions remain open. We have not obtained the analytic background of the helical-vector bubble in closed form, since the uncut sector retains the ordered remainder of the helical kernel. Beyond the limits treated here, the triangle and the box remain multiple Barnes integrals, and at generic orders we find no reduction of the integral over the invariant distance. For banana diagrams beyond the sunset we have computed the wavefunction coefficient, while the correlator will require the corresponding cut weights. The hard line changes the squeezed coefficient of the helical-vector model at order one. At larger $k_3/k_1$ the corrections are of the same size, and templates will also need the part of the triangle without oscillations and the Higgs lines of the model, which we have not computed. For the fermion triangle the part of the bispectrum without oscillations remains open, and so does the shape of the smaller of its two diagrams at a large chemical potential. For the trispectrum we have the analytic part of the box only at equal external momenta, and its dependence on the shape, which a template for searches would need, remains open. Finally, all of our external legs are weakly mixed with the inflaton. The representation of Ref.~\cite{ref26} also covers strong mixing, which has recently been studied at tree level~\cite{An2018nonpert,Iyer2018strong,Wang2026strong,Huenupi2026,Huenupi2026linear,Huenupi2026sound,Pinol2026,Pinol2026strong,Belrhali2026strong}. A heavy field that cannot mix with the inflaton (for instance because it is odd under a $\mathbb{Z}_2$) then enters only through loops whose legs are strongly mixed, and we leave this case to future work.

\paragraph{Note added.} Refs.~\cite{SilentHeavy2026,HanLeiQinYin2026,Song2026,Bassett2026} appeared while this paper was being completed. Our results were obtained independently, and we note in the text where they overlap.

\acknowledgments
This work was carried out under the Anthropic STEM Fellows Program. Code development, calculations, and manuscript preparation were assisted by Claude Fable 5.1, Claude Fable 5.2, and Claude Opus 5.5. The author verified the results and takes full responsibility for them. The views expressed are the author's own and are not endorsed by Anthropic.

%% file: appccsun.tex
\section{Closed forms of the bubble and the sunset}\label{app:ccsun}

In this appendix we collect the conformally coupled bubble and sunset of Section~\ref{sec:close.closed}, and the functions of the massless bubble of Eq.~\eqref{eq:masslessbubble}. For the bubble, Eq.~\eqref{eq:Zbub} with density $(Z^2-1)^{-2}$ reproduces the known wavefunction coefficient~\cite{ref45,ref46,ref32}. In terms of $\Delta_x=x_1-x_2$, $t_0=E-k$, and $b=x_1x_2/k$, it reads
\begin{equation}\label{eq:ccbubble}
\begin{aligned}
4\pi^2\big[\psi_2(k)-\psi_2(0)\big]&=\frac{1}{4E}\ln\frac{x_1x_2}{(x_1+k)(x_2+k)}+\frac{1}{4\Delta_x}\ln\frac{(x_1+k)\,x_2}{(x_2+k)\,x_1}-\frac{J_k-J_0}{4}\,,\\
J_0&=\frac{2\,[E\ln E-x_1\ln x_1-x_2\ln x_2]}{x_1x_2}\,,\qquad J_k=\frac1k\,\Re\sum_a\sigma_a\,j_a\,,\\
&\hspace{-2em}(a,\sigma_a)\in\{(k+\Delta_x,+),(k-\Delta_x,+),(E-k,-),(-k-E,-)\}\,,
\end{aligned}
\end{equation}
where the four terms follow from the substitution $\sqrt{D}=kZ+\tau$ of Section~\ref{sec:close.closed},
\begin{equation}\label{eq:ja}
j_a=\begin{cases}
\ln^2(b-a)+\ln(t_0-a)\ln\frac{b-t_0}{b-a}+\Li_2\frac{t_0-a}{b-a}-\frac{\pi^2}{6}\,, & b>t_0,\ a\leq t_0\,,\\[2pt]
\ln(a-b)\ln(b-t_0)-\Li_2\big({-}\frac{b-t_0}{a-b}\big)\,, & b>t_0,\ a>t_0\,,\\[2pt]
\ln(b-a)\ln(t_0-b)-\Li_2\big({-}\frac{t_0-b}{b-a}\big)\,, & b<t_0,\ a<b\,,\\[2pt]
\ln(a-b)\ln(t_0-b)-\Li_2\frac{t_0-b}{a-b}\,, & b<t_0,\ a>b\,,
\end{cases}
\end{equation}
with principal branches (at $a=t_0$ only the first and last terms of the first line survive). This covers every ordering of $k$, $x_1$, and $x_2$, with the surfaces $k=x_v$ following as limits. For $k<x_v$, every letter satisfies $a\leq t_0<b$, so the constants $\pi^2/6$ cancel and every term is real. On the diagonal $x_1=x_2=x$, with $w=k/x$, this reduces to
\begin{equation}\label{eq:ccdiag}
2\pi^2k\,\big[\psi_2(k)-\psi_2(0)\big]=\frac18\Big[-\frac{w^2}{1+w}+w\ln\frac{16}{1+w}+\ln^2\frac{1+w}{2}+2\,\Li_2\frac{1-w}{2}-\frac{\pi^2}{6}\Big]\qquad(w\leq1)\,,
\end{equation}
and at the doubly folded point $x_1=x_2=k=x$ the value of $4\pi^2[\psi_2(k)-\psi_2(0)]$ is $(3\ln2-\frac12-\frac{\pi^2}{6})/(4x)$.

For the sunset, with three lines at $\nu=\frac12$, $r_v=\frac72$, $c=2x_1x_2$, and the $J_k$ of Eq.~\eqref{eq:ccbubble}, the twice-subtracted sunset of Eq.~\eqref{eq:Zsun} at $x_1\neq x_2$ and $k\neq x_v$ is
\begin{equation}\label{eq:ccsun}
\begin{aligned}
\psi_{\rm sub}&=-\frac{1}{16\pi^4}\Big[P_k(b)-\mathrm{FP}\,P_k(t_0)-\tfrac{3}{16}J_k+F_0-k^2F_2\Big]\,,\\
P_k(\tau)&=\sum_{i=1}^4\Big[-\frac{A_i}{\tau-a_i}+B_i\ln|\tau-a_i|\Big]\,,\\
a&=(E-k,\,-E-k,\,k+\Delta_x,\,k-\Delta_x)\,,\\
A&=\Big(-\tfrac{(x_1-k)(x_2-k)}{16E^2},\,-\tfrac{(x_1+k)(x_2+k)}{16E^2},\,\tfrac{(x_2-k)(x_1+k)}{16\Delta_x^2},\,\tfrac{(x_1-k)(x_2+k)}{16\Delta_x^2}\Big)\,,\\
B&=(B_E,-B_E,B_\Delta,-B_\Delta)\,,\\
B_E&=\frac{k^2+3x_1^2+7x_1x_2+3x_2^2}{16E^3}\,,\qquad B_\Delta=\frac{k^2+3x_1^2-7x_1x_2+3x_2^2}{16\Delta_x^3}\,,\\
\mathrm{FP}\,P_k(t_0)&=\frac{A_1d_2}{d_1^2}+B_1\ln|d_1|+\sum_{i=2}^4\Big[-\frac{A_i}{t_0-a_i}+B_i\ln|t_0-a_i|\Big],\\
d_1&=\frac{(x_1-k)(x_2-k)}{E}\,,\qquad d_2=\frac{k^2}{2E}-\frac{(c+2k^2)^2}{8E^3}\,.
\end{aligned}
\end{equation}
The momentum-independent pieces are
\begin{equation}\label{eq:ccsunF}
\begin{aligned}
F_0&=-\frac{\alpha_E}{E}+\beta_E\lambda-\alpha_\Delta\Big(\frac{1}{E-\Delta_x}+\frac{1}{E+\Delta_x}\Big)-\beta_\Delta\ell_x-\frac{3}{8c}\big(E\ell_c+\Delta_x\ell_x\big)\,,\\
F_2&=\frac{3E}{8c^2}-\frac{\gamma_0}{E}+\beta'_E\lambda-\beta'_\Delta\ell_x\\
&\quad-\frac{3}{16c^3}\Big[\tfrac{E^3}{3}\ell_c+\tfrac{4cE}{3}+\tfrac{\Delta_x^3}{3}\ell_x-(E^2+\Delta_x^2)(E\ell_c+\Delta_x\ell_x)-E^2\Delta_x^2\Big(\frac{\ell_c}{E}+\frac{\ell_x}{\Delta_x}\Big)\Big]\,,
\end{aligned}
\end{equation}
with $\ell_c=\ln\frac{c}{2E^2}$, $\ell_x=\ln\frac{x_1}{x_2}$, $\lambda=\ln\frac{c}{4E^2}$, and
\begin{equation}\label{eq:ccsunc}
\begin{aligned}
&\alpha_E=-\frac{x_1x_2}{16E^2},\quad \beta_E=\frac{3x_1^2+7x_1x_2+3x_2^2}{16E^3},\quad \beta'_E=-\frac{1}{16E^3},\\
&\alpha_\Delta=\frac{x_1x_2}{16\Delta_x^2},\quad \beta_\Delta=\frac{3x_1^2-7x_1x_2+3x_2^2}{16\Delta_x^3},\quad\beta'_\Delta=-\frac{1}{16\Delta_x^3},\\
&\gamma_0=-\frac{(x_1^2+x_2^2)(3x_1^4-14x_1^2x_2^2+3x_2^4)}{32x_1^2x_2^2\Delta_x^2E^2}\,.
\end{aligned}
\end{equation}
The sunset thus combines a rational function, logarithms, and the weight-two group of the bubble. This form holds at every ordering of $k$ and the vertex energies with $x_1\neq x_2$ and $k\neq x_v$, and the surfaces follow as limits. Both forms agree with the integral over $Z$ (the bubble to $10^{-13}$ and the sunset to $10^{-17}$ or better) at points covering every ordering.

For the bubble of two massless lines, the steps behind Eq.~\eqref{eq:masslessbubble} are as follows. With $H(Z)$ the right-hand side of Eq.~\eqref{eq:masslessH} below, the integration by parts gives $2\pi^2\psi_2^{\rm ren}=1/(8E)+\mathrm{FP}\int_1^\infty\dd Z\,H\,\partial_ZK$. The substitution $\sqrt D=kZ+\tau$ turns $\partial_ZK\,\dd Z$ into $\dd\tau/[2k(b-\tau)]$, and $\tau=b-(b-t_0)w$ maps $1<Z<\infty$ onto $1>w>0$ with
\begin{equation}\label{eq:masslessZw}
Z-1=\lambda\,\frac{(1-w)(1-u_1u_2w)}{w}\,,\qquad Z+1=\lambda\,\frac{(1-u_1w)(1-u_2w)}{w}\,,
\end{equation}
where $\lambda=(x_1+k)(x_2+k)/2k^2$, so that $2\pi^2\psi_2^{\rm ren}=1/(8E)+(2k)^{-1}\,\mathrm{FP}\int_0^1H\,\dd w/w$. The rational part of $H$ gives the logarithms of Eq.~\eqref{eq:masslessbubble}, its term $L/4$ gives $I_1$, and its term $-ZL^2/4$ gives $I_2$ and, after one integration by parts in $w$, the functions $M$. Near $w=1$ the cut $Z-1=\epsilon$ is $1-w=\epsilon/[\lambda(1-u_1u_2)]$, which fixes the logarithm in the first term. We then write the logarithm of Eq.~\eqref{eq:masslessL} as $L=\sum_is_i\ln(1-\beta_iw)$, with the letters $\beta=(1,u_1u_2,u_1,u_2)$ and the signs $s=(-1,-1,1,1)$. The functions of Eq.~\eqref{eq:masslessbubble} are then $M(\alpha)=\sum_is_iB(\alpha,\beta_i)$ and $I_2=\sum_{i,j}s_is_j\Lambda(\beta_i,\beta_j)$, with
\begin{equation}\label{eq:masslessB}
\begin{aligned}
B(\alpha,\beta)&=\int_0^1\frac{\ln(1-\beta w)}{1-\alpha w}\,\dd w=\frac1\alpha\Big[\ln\frac\alpha\beta\,\ln(1-\alpha)-\tfrac12\ln^2(1-\alpha)+\Li_2\rho-\Li_2\frac{\rho}{1-\alpha}\Big]\,,\\
\Lambda(\alpha,\beta)&=\int_0^1\ln(1-\alpha w)\,\ln(1-\beta w)\,\frac{\dd w}{w}=\sum_{m,n\ge1}\frac{\alpha^m\beta^n}{mn(m+n)}\,,
\end{aligned}
\end{equation}
where $\rho=1-\alpha/\beta$, for $0<\alpha<1$ and $0<\beta\le1$. Three cases of $\Lambda$ are classical, namely $\Lambda(1,1)=2\zeta(3)$, $\Lambda(1,\beta)=\Li_3\beta+S_{1,2}(\beta)$, and $\Lambda(\alpha,\alpha)=2S_{1,2}(\alpha)$, with the Nielsen polylogarithm $S_{1,2}(x)=\frac12\int_0^x\ln^2(1-t)\,\dd t/t$. The remaining three, at the pairs $(u_1,u_2)$, $(u_1u_2,u_1)$, and $(u_1u_2,u_2)$, depend on both letters. At $x_1=x_2$ the second term of Eq.~\eqref{eq:masslessbubble} tends to $1/[4(x_1+k)]$. The integration by parts uses
\begin{equation}\label{eq:masslessH}
\int_Z^\infty g_{3/2}^2\,\dd Z'=\frac{Z}{2(Z^2-1)}+\frac L4-\frac{ZL^2}{4}\,,
\end{equation}
and its boundary term at coincidence is $1/(8E)$. We have checked Eqs.~\eqref{eq:masslessB} and~\eqref{eq:masslessH} against quadrature to $10^{-60}$.

%% file: appB.tex
\section{Details of the triangle and the box}\label{app:box}

In this appendix we give the box block, the collapsed limit of the box at every order for scalar, helical, and Dirac lines, its limit with one soft leg, and the coefficients of the scalar triangle at general shape (Sections~\ref{sec:triangle.shape} and~\ref{sec:triangle.box}). Line $i$ enters through $(y_i^2)^{-\delta_i}$, with $y_i=|\bl-\mathbf{q}_i|$ and dual points $\mathbf{q}_i=\mathbf{k}_1+\dots+\mathbf{k}_i$. We write $M_{ij}=(\mathbf{q}_i-\mathbf{q}_j)^2$, $\Sigma_4=\sum_i\delta_i$, and $\lambda_4=\Sigma_4-\frac32$. A Feynman-parameter splitting against the $M_{13}$ channel (the three-dimensional massless case of the $N$-point representation of Refs.~\cite{ref83,ref84}) leaves one Barnes variable $z_{ij}$ for each ratio $x_{ij}=M_{ij}/M_{13}$,
\begin{align}
C_4&=\frac{\pi^{3/2}\,M_{13}^{-\lambda_4}}{\prod_i\Gamma(\delta_i)\,\Gamma(3-\Sigma_4)}\biggl[\prod_{(ij)}\int\frac{\dd z_{ij}}{2\pi\ii}\,\Gamma(-z_{ij})\,x_{ij}^{z_{ij}}\biggr]\,\Gamma(\lambda_4+Z)\,\textstyle\prod_{i=1}^4\Gamma(a_i)\,,\nonumber\\
Z&=\textstyle\sum_{(ij)}z_{ij}\,,\quad a_1=\delta_1-\lambda_4-z_{23}-z_{24}-z_{34}\,,\quad a_3=\delta_3-\lambda_4-z_{12}-z_{14}-z_{24}\,,\nonumber\\
a_2&=\delta_2+z_{12}+z_{23}+z_{24}\,,\quad a_4=\delta_4+z_{14}+z_{34}+z_{24}\,,\label{eq:W4}
\end{align}
where $(ij)$ runs over $\{12,14,23,24,34\}$. All five contours can be taken straight when the $\delta_i$ are real with $0<\delta_i<\frac32$ and $\frac32<\Sigma_4<3$. The upper bound is necessary, since $a_1+a_2+a_3+a_4=3-\Sigma_4$. We have evaluated Eq.~\eqref{eq:W4} on such contours and compared it with a sector-decomposed Feynman-parameter integral, finding agreement to $1.4\times10^{-6}$ with exponential convergence in the step size. At $\Sigma_4=3$ only a two-fold integral survives, which is Symanzik's star formula~\cite{ref86}. On this slice, inversion about $\mathbf{q}_4$ maps the box block onto the triangle block, $C_4=\prod_{i=1}^3M_{i4}^{-\delta_i}\,C_3$ at the momenta $k_1^2=M_{13}/(M_{14}M_{34})$, $k_2^2=M_{12}/(M_{14}M_{24})$, and $k_3^2=M_{23}/(M_{24}M_{34})$, which we have checked to $2.5\times10^{-14}$.

\paragraph{The correlator.} In the cutting rule of Eq.~\eqref{eq:cutrule}, cutting $c\geq1$ of the four lines of the box leaves $c$ components, joined in a ring by the cut lines. Adjacent components must lie on opposite branches, which is possible only for even $c$. For $c=1$ the cut line joins its component to itself, and for $c=3$ the three components form an odd cycle. Eight cut sets therefore contribute, namely the uncut box, the four pairs of adjacent lines, the two pairs of opposite lines, and the fully cut box. Each of them has exactly two branch assignments, which are complex conjugates of each other when the vertex phases are real, as they are for inflaton legs ($r_v=2$). The ordered kernel has the Barnes representation of Eq.~\eqref{eq:GBarnes}. The Feynman kernel of Eq.~\eqref{eq:GF} adds a multiple of $K_\nu(yt)K_\nu(yt')$, which is also the cut kernel, and this product is a product of two single Barnes integrals. In each representation the line energy enters only through a power $(y_i^2)^{-\delta_i}$. Every cut set therefore leads to the same momentum integral, the block $C_4$ of Eq.~\eqref{eq:W4}, at shifted $\delta_i$.

\paragraph{The collapsed limit.} When $P=|\mathbf{k}_2+\mathbf{k}_3|\to0$ the invariant $M_{13}=P^2$ is soft, and lines 1 and 3 are soft together. At small argument, the ordered kernel has a non-analytic part made of powers of the two times,
\begin{equation}\label{eq:KIsmall}
K_\nu(yt_>)\,I_\nu(yt_<)\ \supset\ -\frac{\pi}{2\sin\pi\nu}\,\frac{(y/2)^{2\nu}\,(t\,t')^{\nu}}{\Gamma(1+\nu)^2}\,,
\end{equation}
so the soft lines factor out of the time integrals. The soft momentum integral is the two-line block of Eq.~\eqref{eq:C2}, and each hard line leaves one two-time moment, which gives Eq.~\eqref{eq:Cbox} of Section~\ref{sec:triangle.box}. In that equation $\mathcal{H}_\nu(\alpha_0,\alpha_1)=\int\dd t_0\,\dd t_1\,t_0^{\alpha_0}t_1^{\alpha_1}e^{-t_0-t_1}K_\nu(t_>)I_\nu(t_<)$ is the moment of a hard line at unit energies. Summing its two time orderings term by term gives a very-well-poised ${}_6F_5$ at argument $-1$, which Whipple's transformation~\cite{ref57} brings to
\begin{equation}\label{eq:Hmoment}
\mathcal{H}_\nu(\alpha_0,\alpha_1)=\frac{\sqrt{\pi}\,\Gamma(L)\,\Gamma(\alpha_0{+}1{+}\nu)\,\Gamma(\alpha_1{+}1{+}\nu)}{2^{L+2\nu}\,\Gamma(1{+}\nu)\,\Gamma(L{+}\nu{+}\frac12)}\;{}_3F_2\!\left[\begin{matrix}\nu{+}\frac12,\ \alpha_0{+}1{+}\nu,\ \alpha_1{+}1{+}\nu\\ L{+}\nu{+}\frac12,\ 2\nu{+}1\end{matrix};1\right],
\end{equation}
with $L=\alpha_0+\alpha_1+2$.
This series has parameter excess one and converges at every mass and vertex exponent. After a Thomae transformation it coincides with the hypergeometric term of the doubly folded seed integral of Ref.~\cite{ref53}. The coefficient $C_\Box$ is therefore a closed expression in Gamma functions and two values of this series.

\paragraph{All orders in $P$.} The factorization of Eq.~\eqref{eq:Cbox} holds at every order of the collapsed limit. We shift $\bl$ so that $y_1=|\bl|$ and $y_3=|\bl-\mathbf{P}|$, with $\mathbf{P}=\mathbf{k}_2+\mathbf{k}_3$. The hard lines then carry $y_2=|\bl-\mathbf{k}_2|$ and $y_4=|\bl+\mathbf{k}_1|$, and no vertex time is soft. In the soft region each soft line contributes the non-analytic piece of Eq.~\eqref{eq:KIsmall} with the full series of both Bessel functions, $-\frac{\pi}{2\sin\pi\nu}I_\nu(yt)I_\nu(yt')$, since a piece analytic in $y_1$ or $y_3$ gives a scaleless integral. The wavefunction coefficient therefore has a single nonlocal family,
\begin{equation}\label{eq:boxseries}
\psi_\Box\supset P^{\,3+2\nu_1+2\nu_3}\sum_{N\ge0}c_N\,P^N\,,\qquad c_0=C_\Box\,,
\end{equation}
where the odd orders enter through the angles of $\mathbf{P}$ with $\mathbf{k}_1$ and $\mathbf{k}_2$. The higher orders expand the hard lines around the folded point. A hard line enters through its moment $D_\nu(\alpha_0,\alpha_1;x_0,x_1,y)$, the integral of $t_0^{\alpha_0}t_1^{\alpha_1}e^{-x_0t_0-x_1t_1}$ against $K_\nu(yt_>)I_\nu(yt_<)$. Since the kernel depends on $y$ only through $yt$ and $yt'$, rescaling the times gives
\begin{equation}\label{eq:Dfold}
D_\nu=y^{-\alpha_0-\alpha_1-2}\sum_{p_0,p_1\ge0}\frac{(1-x_0/y)^{p_0}\,(1-x_1/y)^{p_1}}{p_0!\,p_1!}\;\mathcal{H}_\nu(\alpha_0+p_0,\alpha_1+p_1)\,,
\end{equation}
so each derivative at the folded point is a finite sum of the moments of Eq.~\eqref{eq:Hmoment}. The line energies $y_2,y_4$ and the vertex energies $x_3=|\mathbf{P}-\mathbf{k}_2|$ and $x_4=|\mathbf{k}_1+\mathbf{P}|$ then expand in the components of $\bl$ and $\mathbf{P}$. The soft integral of each monomial is the two-line block with a polynomial numerator, and the Feynman-parameter form of Eq.~\eqref{eq:C2} gives all of them through
\begin{multline}\label{eq:C2J}
\int\!\dd^3\ell\;e^{\bl\cdot\mathbf{J}}\,y_1^{-2\delta_1}y_3^{-2\delta_3}=\frac{\pi^{3/2}}{\Gamma(\delta_1)\Gamma(\delta_3)}\sum_{j,m\ge0}\frac{(\mathbf{J}^2)^j\,(\mathbf{P}\cdot\mathbf{J})^m}{4^j\,j!\,m!}\\
\times\frac{\Gamma(\mathcal{S}-\frac32-j)\,\Gamma(\frac32-\delta_1+j+m)\,\Gamma(\frac32-\delta_3+j)}{\Gamma(3-\mathcal{S}+2j+m)}\,P^{\,3-2\mathcal{S}+2j}
\end{multline}
(with $\mathcal{S}=\delta_1+\delta_3$). Each $c_N$ is thus a finite sum of Gamma functions times values of Eq.~\eqref{eq:Hmoment}. At the collinear point of the check below, the ratio of successive coefficients tends to $1/|\mathbf{k}_2|$ (we computed them through $N=12$), so the series converges for $P$ below the smaller hard momentum. We have checked Eq.~\eqref{eq:boxseries} against a direct evaluation of the box at 26 collapsed points, for collinear momenta with $|\mathbf{k}_1|=1$, $|\mathbf{k}_2|=0.8$, and $0.003\le P\le0.08$, and for real orders $\nu=(0.13,0.4,0.12,0.3)$ at $r_v=2$. At fixed loop momentum the four time integrals are a sum over the 24 time orderings, each a nested integral of functions of one time, and the loop integral is split around the four points where a line energy vanishes (with the region of the two soft points scaled with $P$). A polynomial fit of the direct values (with $\psi_\Box\simeq0.03$) leaves a residual of $1\times10^{-12}$. Subtracting Eq.~\eqref{eq:boxseries} through $N=0$, $1$, and $2$ lowers it to $1\times10^{-14}$, $3\times10^{-15}$, and $1\times10^{-16}$, where it levels off. With the coefficients left free, the fit returns $c_0$ to $5\times10^{-4}$, $c_1$ to $2\times10^{-3}$, and $c_2$ to $3\times10^{-2}$.

Since the ring trace of Eq.~\eqref{eq:patterns} holds at fixed loop momentum, the correlator follows from Eq.~\eqref{eq:boxseries} at flipped orders. Its collapsed limit has the four families $P^{3+2\sigma_1\nu_1+2\sigma_3\nu_3}$, with coefficients
\begin{equation}\label{eq:boxcorr}
\mathcal{C}^{\,\sigma_1\sigma_3}_N=\sum_{\epsilon_2,\epsilon_4=\pm}W_{\sigma_1\epsilon_2\sigma_3\epsilon_4}\;c_N(\sigma_1\nu_1,\epsilon_2\nu_2,\sigma_3\nu_3,\epsilon_4\nu_4)
\end{equation}
at every order. Summing the sixteen Schwinger--Keldysh assignments directly, with the branch pieces of the Feynman and cut kernels, gives the same coefficients to $10^{-15}$.

\paragraph{Helical lines.} The same steps go through for the helical vectors of Section~\ref{sec:chem}. For the pair-frequency signal a soft line keeps its same-branch pieces, products of $M_{-K,\sigma\nu}(2yt)$ at a vertex on the $+$ branch and $M_{K,\sigma\nu}(2yt)$ at a vertex on the $-$ branch, with the weights of Eq.~\eqref{eq:vecbeta} and the series of Eq.~\eqref{eq:en}, so the powers of $P$ advance in unit steps. At $\kappa\neq0$ the pieces with opposite branches at the two ends are not analytic in the line momentum either, and they give the further signals of Section~\ref{sec:chem.parity}. A hard line enters through its four Schwinger--Keldysh moments, and Eq.~\eqref{eq:Dfold} holds for each of them, with Eq.~\eqref{eq:helmoment} in the ordered ones and Gamma functions in the cut ones. For transverse lines the projectors $\Pi^h_{ij}(\hat{\mathbf{p}})=\frac12(\delta_{ij}-\hat p_i\hat p_j-\ii h\,\epsilon_{ijk}\hat p_k)$ add the trace $\mathrm{Tr}\prod_l\Pi^{h_l}(\hat{\mathbf{p}}_l)$ around the ring, and its monomials in $\bl$ shift the blocks of Eq.~\eqref{eq:C2J}. The collapsed limit of the helical box thus has a closed form at every order, for transverse and for longitudinal lines (Section~\ref{sec:chem.parity}). For one helical vector at $\mu=1$, with enhanced soft helicities $h_1=h_3=+1$ and summed over the transverse hard helicities, the coefficient of $P^{3+4\ii\mu}$ in the correlator grows relative to $\kappa=0$ by factors of $32$, $4.4\times10^3$, and $1.5\times10^7$ at $\kappa=0.5$, $1$, and $1.5$. The per-line amplitude $\sqrt{n_h(1+n_h)}$ of Eq.~\eqref{eq:vecweight}, squared, accounts for $24$, $1.1\times10^3$, and $3.0\times10^5$ of this, and the enhanced hard lines for the rest. The first correction grows as well. At $P=0.1|\mathbf{k}_1|$ it is $19\%$ of the leading term at $\kappa=0$ and $35\%$ at $\kappa=1$, because $e_1=-K/(1+2\nu)$ grows with $\kappa$.

\paragraph{Dirac lines.} The two ends of a Dirac line carry first indices that differ by zero or one (Section~\ref{sec:dirac.line}), so Eq.~\eqref{eq:helmoment} does not cover its ordered moments. They have closed forms of the same kind. At the folded point Kummer's transformation gives $e^{-t}M_{K,\nu'}(2t)=(2t)^{\frac12+\nu'}\,{}_1F_1\big(\tfrac12+\nu'+K;\,1+2\nu';\,-2t\big)$, a series in powers of the time alone. Integrating it term by term against the single moment of the other end, we find
\begin{multline}\label{eq:fermimoment}
\int_0^\infty\!\dd t_2\;t_2^{\,b}\,e^{-t_2}\,W_{K_1,\nu}(2t_2)\int_0^{t_2}\!\dd t_1\;t_1^{\,a}\,e^{-t_1}\,M_{K_2,\nu'}(2t_1)\\
=\frac{2^{\frac12+\nu'-P}\;\Gamma\big(P+\tfrac12+\nu\big)\,\Gamma\big(P+\tfrac12-\nu\big)}{\Gamma(P+1-K_1)\,\big(a+\tfrac32+\nu'\big)}\\
\times\,{}_4F_3\!\left[\begin{matrix}\tfrac12+\nu'+K_2,\ P+\tfrac12+\nu,\ P+\tfrac12-\nu,\ a+\tfrac32+\nu'\\[2pt] 1+2\nu',\ P+1-K_1,\ a+\tfrac52+\nu'\end{matrix}\,;-1\right],
\end{multline}
with $P=a+b+\frac52+\nu'$ and $\nu'=\pm\nu$, for any two first indices $K_1$ and $K_2$. The function is analytic at argument $-1$. Where its series does not converge there, we evaluate it by analytic continuation in the argument. The growing end of an ordered Schwinger--Keldysh function is a combination of $M_{-K,\pm\nu}$ with the branch coefficients of Appendix~\ref{app:helical}, so every ordered moment of the Dirac line is a sum of two such series. The cut moments are products of the single moments
\begin{equation*}
\int_0^\infty\!\dd t\;t^{\,x}\,e^{-t}\,W_{K,\nu}(2t)=\frac{\Gamma\big(x+\tfrac32+\nu\big)\,\Gamma\big(x+\tfrac32-\nu\big)}{2^{\,x+1}\,\Gamma(x+2-K)}\,.
\end{equation*}
Equation~\eqref{eq:fermimoment} agrees with a two-dimensional quadrature to $4\times10^{-22}$ at the indices and exponents of the fermion triangle. At $K_1=K_2$ its two orderings add up to Eq.~\eqref{eq:helmoment}, which we have checked to $2\times10^{-31}$ at Dirac and helical indices. With these moments the hard part of the fermion triangle (Section~\ref{sec:dirac.triangle}) is known in closed form. It agrees with a direct evaluation of the hard part on partially rotated contours to $2\times10^{-15}$ at $\kappa=5$ and $6\times10^{-16}$ at $\kappa=10$. Away from the folded point Eq.~\eqref{eq:Dfold} holds for each moment, as for the scalar and helical lines.

\paragraph{One soft leg.} When a single leg is soft instead, $k_3\to0$, the two lines at vertex 3 become soft together with its time, as for the triangle of Section~\ref{sec:triangle.shape}. The soft region is $\bl\simeq\mathbf{q}_2$ with $t_3\sim1/k_3$. There the soft lines contribute $K_\nu(yt_3)$ at vertex 3 and the series of $I_\nu$ at their hard ends, and the hard part is the two-line chain through vertex 1, with $y_1=x_2=|\mathbf{k}_2|$ and $y_4=x_4=|\mathbf{k}_4|$ (folded at both ends). For every monomial two applications of the first Barnes lemma again reduce the soft integral to Gamma functions, and at leading order
\begin{equation}\label{eq:boxsoftleg}
\psi_\Box\supset\frac{1}{8\pi^{3/2}\,4^{1+\nu_{23}}}\,\frac{\Gamma(\frac32+\nu_2)\,\Gamma(\frac32+\nu_3)\,\Gamma(r_3+\nu_{23})\,\Gamma(\frac{r_3-3-\nu_{23}}{2})\,\Gamma(\frac{r_3-2-\nu_{23}}{2})}{\Gamma(1+\nu_2)\,\Gamma(1+\nu_3)\,\Gamma(3+\nu_{23})\,\Gamma(r_3-1)}\;\mathcal{T}\,k_3^{\,3-r_3+\nu_{23}}\,,
\end{equation}
where $\nu_{23}=\nu_2+\nu_3$, and the chain is
\begin{equation}\label{eq:boxchain}
\mathcal{T}=\int\!\dd t_1\dd t_2\dd t_4\,t_1^{r_1-1}t_2^{\,r_2-1+\nu_2}t_4^{\,r_4-1+\nu_3}\,e^{-x_1t_1-x_2t_2-x_4t_4}\,\mathcal{G}_{\nu_1}(y_1;t_2,t_1)\,\mathcal{G}_{\nu_4}(y_4;t_1,t_4)\,.
\end{equation}
With the chain replaced by a single folded line this is the leading coefficient $c_0$ of the triangle~\cite{ref26}. The factor $1/\Gamma(r_3-1)$ removes the family for a conformally coupled soft leg, and it does so at every order, since the second lemma always leaves $1/\Gamma(r_3-1-n)$ with $n\ge0$. At $r_3=2$ only the $j=0$ terms of the analogue of Eq.~\eqref{eq:C2J} (with $\mathbf{k}_3$ in place of $\mathbf{P}$) survive, so in every monomial the soft integral replaces $\bl$ by a multiple of $\mathbf{k}_3$. The soft side therefore has a closed form at every order, while the hard side is the chain and its derivatives, a tree. We have checked the soft factor for general monomials against direct integration (at vertex powers where it converges) to $10^{-8}$. We have also checked Eq.~\eqref{eq:boxsoftleg} against a direct evaluation of the box at 22 points, for collinear momenta with $|\mathbf{k}_1|=0.9$, $|\mathbf{k}_2|=0.7$, and $0.002\le k_3\le0.06$ and the orders above. A free fit returns the leading coefficient to $10^{-6}$, and the first correction (from the first derivatives of $\mathcal{T}$ and the $\bl\cdot\hat{\mathbf{k}}_3$ term of the soft factor) to $3\times10^{-4}$.

\paragraph{The scalar triangle at general shape.} We obtain the coefficients $A_N$ of Eq.~\eqref{eq:scalarshape} from Eq.~\eqref{eq:pairshape} by three expansions, each of them finite at every order. We set $K=1$, so that $k_{1,2}=1\pm\epsilon$ with $\epsilon=\frac12R\delta$, and $y_\lambda^2=1+u$ with $u=2(1-2\lambda)\epsilon+\epsilon^2-\lambda(1-\lambda)R^2$. The two soft lines contribute their series in powers of $(\lambda Rt_1)^2$ and $((1-\lambda)Rt_2)^2$, with indices $i$ and $j$. The vertex energies are expanded around one, with $a$ and $b$ powers of the two times, and the energy of the hard line around one through Eq.~\eqref{eq:Dfold}. The integral over $\lambda$ is then a sum of Beta functions. With $\alpha=r_1-1+\nu_C+2i$ and $\beta=r_2-1+\nu_B+2j$ we find
\begin{multline}\label{eq:DNexplicit}
A_N(\delta)=\frac{\Gamma(2+\nu_{BC})\,\Gamma(-1-\nu_{BC})}{8\pi\,2^{\,\nu_{BC}}}\sum\frac{4^{-i-j}}{i!\,j!\,\Gamma(1+\nu_C+i)\,\Gamma(1+\nu_B+j)}\\
\times\frac{n!\;2^{n_1}(-1)^{n_3}}{n_1!\,n_2!\,n_3!}\Big(\frac{\delta}{2}\Big)^{s+n_1+2n_2}\;T_{s,n}\;\Lambda_{n_1n_3}\,,\qquad n=n_1+n_2+n_3\,,
\end{multline}
where the sum runs over the non-negative integers with $2i+2j+s+n_1+2n_2+2n_3=N$, and
\begin{align}
T_{s,n}&=\sum_{a+b=s}\frac{(-1)^a}{a!\,b!}\;\tau_n(\alpha+a,\beta+b)\,,\nonumber\\
\Lambda_{n_1n_3}&=\sum_{c=0}^{n_1}\binom{n_1}{c}(-1)^c\,B\big(\tfrac32+\nu_C+2i+n_3+c,\ \tfrac32+\nu_B+2j+n_3+n_1-c\big)\,.
\end{align}
Here $\tau_n(\alpha,\beta)$ is the coefficient of $u^n$ in Eq.~\eqref{eq:Dfold} at $x_0=x_1=1$ and $y=(1+u)^{1/2}$, a combination of the moments $\mathcal{H}_{\nu_A}(\alpha+p_0,\beta+p_1)$ with $p_0+p_1\le n$. All of these moments follow from the first one by Eq.~\eqref{eq:Hrec}.

%% file: appC.tex
\section{Details of the helical-vector loops}\label{app:helical}

In this appendix we collect the formulas behind Section~\ref{sec:chem}, in the order in which that section uses them.

\paragraph{Helical single exchange.} With the kernel of Eq.~\eqref{eq:helBarnes}, the exchange
\begin{equation*}
\psi_2^h=\int_0^\infty\dd t\,\dd t'\,t^{r_1-1}t'^{\,r_2-1}e^{-x_1t-x_2t'}\mathcal{K}_h(y;t,t')
\end{equation*}
equals, for $x_2-y>x_1+y$,
\begin{equation}\label{eq:helexch}
\begin{aligned}
(2y)^{r_1+r_2}\,\psi_2^h&=\Phi_h(u)\,T_h(v)+C_h(u,v)\,,\qquad u=\frac{2y}{x_1+y}\,,\quad v=\frac{2y}{x_2-y}\,,\\
\Phi_h(u)&=\frac{\Gamma(r_1+\nu)\,\Gamma(r_1-\nu)}{\Gamma(r_1+\frac12-K)}\;u^{\,r_1+\nu}\;{}_2F_1\bigl(a_+,\,r_1+\nu;\,r_1+\tfrac12-K;\,1-u\bigr)\,,\\
T_h(v)&=\frac{\Gamma(a_+)\,\Gamma(r_2+\nu)}{\Gamma(1+2\nu)}\;v^{\,r_2+\nu}\;{}_2F_1\bigl(1-a_-,\,r_2+\nu;\,1+2\nu;\,-v\bigr)\,,\\
C_h(u,v)&=-\frac{\Gamma(r_1+r_2)}{r_1^2-\nu^2}\,v^{\,r_1+r_2}\\
&\quad\times\sum_{j,k\ge0}\frac{(r_1+r_2)_{j+k}\,(r_1+\nu)_j\,(r_1-\nu)_j\,(\frac12+K+r_1+j)_k}{(1+r_1+\nu)_{j+k}\,(1+r_1-\nu)_{j+k}\;j!}\,\Bigl(-\frac{v}{u}\Bigr)^{j}(-v)^k\,.
\end{aligned}
\end{equation}
The double series $C_h$ is a Kamp\'e de F\'eriet function~\cite{ref82,ref85}. For $x_1-y>x_2+y$ the mirrored identity gives the same form with the vertices exchanged.

\paragraph{Branch coefficients and weights.} In Eq.~\eqref{eq:vecbranch},
\begin{equation}\label{eq:vecAB}
B_\pm=\frac{N\,\Gamma(\mp2\nu)}{\Gamma(\bar a_\pm)}\,,\qquad A_\pm=\frac{N\,e^{-\ii\pi(\frac12\pm\nu)}\,\Gamma(\mp2\nu)}{\Gamma(a_\mp)}\,,\qquad N=\frac{e^{\pi h\kappa/2}}{\sqrt{2y}}\,,\qquad \bar a_\pm=\tfrac12\mp\nu+K\,.
\end{equation}
The per-line weights of Eq.~\eqref{eq:vecpair} are $w_{+-,\sigma}=w_{-+,\sigma}=B_\sigma\overline{B_{-\sigma}}\equiv\beta_\sigma$ and $w_{\pm\pm,\sigma}=e^{\mp\ii\pi(\frac12+\sigma\nu)}\beta_\sigma$, with
\begin{equation}\label{eq:vecbeta}
2y\,\beta_+=\frac{e^{\pi h\kappa}\,\Gamma(-2\ii\mu)^2}{\Gamma\big(\frac12-\ii(\mu+h\kappa)\big)\,\Gamma\big(\frac12-\ii(\mu-h\kappa)\big)}\,.
\end{equation}

\paragraph{The longitudinal line.} The two field components $A_0$ and $A_L$ of the longitudinal line are both expressed through the neutral line $g(\tau)=W_{0,\nu}(2\ii q\tau)/\sqrt{2q}$, with modes $v_0=-\tau(q/m)g$ and $v_L=(\ii/m)(\tau\partial_\tau-1)g$, where $m^2=\mu^2+\frac14$. On the branch series these act diagonally,
\begin{equation}\label{eq:vecL}
E_L[n]=\pm\frac{\ii}{m}\big(\nu+n-\tfrac12\big)E_T[n]\,,\qquad E_0[n]=\pm\frac{\ii}{2m}\,E_T[n-1]\,,
\end{equation}
with $E_T[n]$ the coefficient of the corresponding transverse-type end. A transverse line couples only to the $A_L$ component of a longitudinal partner, with weight $\frac12(1-c^2)$. Two longitudinal lines couple through both components, with weights $\{1,-c,-c,c^2\}$ on the channels $(00;00)$, $(00;LL)$, $(LL;00)$, and $(LL;LL)$. Summed over the monomials of one weight, the leading loop coefficient is rational in $\nu$ times the two blocks $C_2^e=\hat C_2(1{+}2\nu,1{+}2\nu)$ and $C_2^o=\hat C_2(2{+}2\nu,2\nu)$,
\begin{gather}
c_{\rm pol}\Big[\tfrac{(1\mp c)^2}{4}\Big]=\frac{4\nu^2+4\nu+3}{8\nu^2}\,C_2^e\mp\frac{\nu+1}{2\nu+1}\,C_2^o\,,\nonumber\\
c_{\rm pol}\Big[\tfrac{1-c^2}{2}\Big]=-\frac{4\nu+3}{4\nu^2}\,C_2^e\,,\nonumber\\
c_{\rm pol}[c^2]=\frac{2\nu^2+4\nu+3}{2\nu^2}\,C_2^e\,,\qquad c_{\rm pol}[-c]=-\frac{2(\nu+1)}{2\nu+1}\,C_2^o\,.\label{eq:cpol}
\end{gather}
On the principal series the two terms nearly cancel for equal transverse helicities. The ratio of the same-helicity to the opposite-helicity coefficient tends to $1/(2\nu^2)$ only at large $\mu$ (it is $0.42$ at $\mu=1$).

\paragraph{The late-time series.} The branch series $M_{K,\nu}(z)=z^{\frac12+\nu}\sum_ne_nz^n$ has the coefficients
\begin{equation}\label{eq:en}
\begin{gathered}
e_n(K,\nu)=\frac{(-\frac12)^n}{n!}\;{}_2F_1\big(-n,\;\tfrac12+\nu-K;\;1+2\nu;\;2\big)\,,\\
(n+1)(n+1+2\nu)\,e_{n+1}=-K\,e_n+\tfrac14\,e_{n-1}\,.
\end{gathered}
\end{equation}
Inserting them into Eq.~\eqref{eq:vecpair} and doing each vertex integral and the loop integral monomial by monomial gives the pair piece of one channel,
\begin{equation}\label{eq:vecrung}
\begin{aligned}
F^{\sigma_1\sigma_2}={}&\frac{e^{\pi(h_1+h_2)\kappa}}{4\pi^2}\sum_{a,b}\sum_{n_l,m_l\ge0}E^{(a)}_{1,n_1}E^{(a)}_{2,n_2}E^{(b)}_{1,m_1}E^{(b)}_{2,m_2}\,\frac{\Gamma(c_1{+}n)\,\Gamma(c_2{+}m)}{x_1^{c_1+n}x_2^{c_2+m}}\\
&\times2^{2\sigma_1\nu+2\sigma_2\nu+P}\sum_{\rm mono}\mathrm{coef}\;k^{\,3+2(\sigma_1+\sigma_2)\nu+P+p+d}\,\hat C_2\big(p_1',p_2'\big)\,,
\end{aligned}
\end{equation}
with $p_l'=1+2\sigma_l\nu+P_l+d_l$, $c_v=r_v+(\sigma_1+\sigma_2)\nu$, $n=n_1+n_2$, $m=m_1+m_2$, $P_l=n_l+m_l$, $P=n+m$, and $d=d_1+d_2$. Here $E^{(a)}_{l,n}$ is the branch coefficient of the end of line $l$ that sector $(a,b)$ places at vertex 1, times $e_n(\mp K_l,\sigma_l\nu)$ and the end factors of Eq.~\eqref{eq:vecL}. Cauchy's estimate for $e_n$ and Stirling's formula for the block bound the general term by a power of $1+P$ times $(k/x_1)^n(k/x_2)^m$, so the series converges absolutely for $k<\min(x_1,x_2)$, with radius exactly one in each $k/x_v$.

\paragraph{The soft moment.} With the full Mellin transform of the Whittaker function~\cite{ref29},
\begin{multline}\label{eq:GK}
G_K(s)=\int_0^\infty z^{s-1}W_{K,\nu}(z)\,\dd z\\
=\frac{\Gamma(\frac12+\nu+s)\,\Gamma(\frac12-\nu+s)}{\Gamma(1-K+s)}\;{}_2F_1\Big(\tfrac12+\nu+s,\tfrac12-\nu+s;1-K+s;\tfrac12\Big),
\end{multline}
whose poles are precisely the two branch ladders, the squeezed coefficients of Eq.~\eqref{eq:vectemplate} are
\begin{equation}\label{eq:vecMn}
\mathcal{M}_n=2^{-8-2\nu}\,\Gamma(4+2\nu+n)\sum_{h_1h_2}e^{\pi(h_1+h_2)\kappa}\sum_{\rm ch}\sum_{a,b}\sum_{n_1+n_2=n}E^{(a)}_{1,n_1}E^{(a)}_{2,n_2}\,\mathcal{J}^{(b)}_{\rm ch}(n_1,n_2)\,,
\end{equation}
with the soft moment
\begin{equation}\label{eq:vecJ}
\begin{aligned}
\mathcal{J}^{(b)}_{\rm ch}(n_1,n_2)={}&\frac{1}{4\pi^2}\int\frac{\dd\sigma\,\dd\sigma'}{(2\pi\ii)^2}\,\hat G^{(b)}_1(\sigma)\,\hat G^{(b)}_2(\sigma')\,2^{2\nu-1-\sigma-\sigma'}\,\Gamma(1{-}\sigma{-}\sigma')\\
&\times\sum_{\rm mono}\mathrm{coef}\;\hat C_2\big(\nu{+}\tfrac12{-}\sigma{+}d_1{+}n_1,\ \nu{+}\tfrac12{-}\sigma'{+}d_2{+}n_2\big)+\mathcal{J}_{\rm UV}\,.
\end{aligned}
\end{equation}
Here $\hat G^{(b)}$ is $G_{\mp K}$ dressed by the end factors. The term $\mathcal{J}_{\rm UV}$ collects the residues of the ultraviolet poles of the block at $\sigma+\sigma'=2\nu+P$, $P\ge1$, which cross the soft contours when the hard residues are taken. It is about a fifth of the transverse soft moment at $\mu=1$ and $\kappa=\frac12$. Undoing the loop block turns Eq.~\eqref{eq:vecJ} into a radial integral in position space,
\begin{equation}\label{eq:vecJpos}
\mathcal{J}^{(b)}_{\rm ch}(n_1,n_2)=\sum_{\rm mono}\mathrm{coef}\;\frac{2^{-4-d-n}\,\Gamma(c)}{\pi^3}\int_0^\infty\frac{\dd\rho}{\rho}\;\Phi_1(\rho)\,\Phi_2(\rho)\;\frac{(1-2\ii\rho)^{-c}-(1+2\ii\rho)^{-c}}{2\ii}\,,
\end{equation}
with $c=-2-2\nu-d-n$, over legs given by Gauss functions,
\begin{equation}\label{eq:vecleg}
\begin{aligned}
\Phi_l(\rho)&=\frac{\mathcal{L}_l(-\ii\rho)-\mathcal{L}_l(\ii\rho)}{2\ii}\,,\\
\mathcal{L}_l(p)&=\frac{\Gamma(\frac12{+}\nu{+}a_l)\,\Gamma(\frac12{-}\nu{+}a_l)}{\Gamma(1{-}K_l'{+}a_l)}\;{}_2F_1\Big(a_l{-}\nu{+}\tfrac12,\ a_l{+}\nu{+}\tfrac12;\ a_l{-}K_l'{+}1;\ \tfrac12-p\Big),
\end{aligned}
\end{equation}
with $a_l=\nu+\frac32+d_l+n_l$ and $K_l'=\mp K_l$ for $b=\pm$. An $A_0$ end raises $a_l$ by one, and an $A_L$ end replaces $\mathcal{L}(a)$ by $-(a{+}1)\mathcal{L}(a)+p\,\mathcal{L}(a{+}1)$. Replacing $\mathcal{J}$ by its own leading residue $\sigma,\sigma'\to-\frac12-\nu$ gives the late-time coefficient
\begin{equation}\label{eq:vecMlate}
\mathcal{M}^{(h_1h_2)}_{A,N=0}=-\frac{2^{-8+2\nu}}{\pi^2}\,\sinh^2(\pi\mu)\,\Gamma(4+2\nu)\,\Gamma(2+2\nu)\,\big(2y_1\beta_+^{(1)}\big)\big(2y_2\beta_+^{(2)}\big)\,c_{\rm pol}^{(h_1h_2)}(\nu)\,,
\end{equation}
times the end factors of Eq.~\eqref{eq:vecL} for longitudinal lines.

\paragraph{The soft moment as a Kamp\'e de F\'eriet function.} The soft moments of Eq.~\eqref{eq:vecJpos} also have a form with a single Gauss function in each term. Each Whittaker function is an Euler integral over $p\in(1,\infty)$ with the weight $w(p)=(p-1)^{\gamma-1}(p+1)^{-2\nu-\gamma}$, where $\gamma=\frac12-\nu-K$. The time and loop integrals of a monomial are then derivatives of one kernel, which is a difference of two terms with the pole $1/(p_1^2-p_2^2)$. Integrating each term over one variable as a principal value leaves one integral. In the normalization of Eq.~\eqref{eq:Jpos}, for the monomial $(y_1y_2)^{\nu-3/2}$ and the time weight $t^{q-1}e^{-t}$ at the vertex, it is
\begin{equation}\label{eq:vecKdF}
\begin{aligned}
J&=\frac{2^{-q}\,\Gamma(q-1-2\nu)}{2\pi^2\,\Gamma(\gamma_1)\Gamma(\gamma_2)}\int_0^1\!\dd u\;\frac{u^{q+2\nu}}{2-u}\,\Big[(1-u)^{\gamma_2-1}\,\sigma_1(u)+(1-u)^{\gamma_1-1}\,\sigma_2(u)\Big]\,,\\
\sigma_l(u)&=-B(\gamma_l,2\nu)\,u^{-2\nu}\Big[{}_2F_1\big(1,1{-}\gamma_l{-}2\nu;1{-}2\nu;u\big)+{}_2F_1\big(1,\gamma_l;1{-}2\nu;u\big)\Big]\\
&\quad+\pi\cot(2\pi\nu)\,(1-u)^{\gamma_l-1}+\frac{\pi}{\sin2\pi\nu}\,(1-u)^{-2\nu-\gamma_l}\,,
\end{aligned}
\end{equation}
where $\gamma_l=\frac12-\nu-K_l$ and the powers of $1-u$ at $u=1$ are continued analytically. The integrand has the singular points $0$, $1$, and $\infty$ and one pole at $u=2$, which arises when the pole of the kernel is split into two simple poles. Expanding $1/(2-u)$ in powers of $u/2$ gives a series of ${}_3F_2$'s at unit argument whose terms fall as $2^{-n}$, which is a Kamp\'e de F\'eriet function~\cite{ref82,ref85} of arguments $(1,\frac12)$. Once each ${}_3F_2$ is transformed to the parametric excess $q+1+n+2\nu$, both sums converge absolutely at the physical exponents. We have written out this series for the monomial of Eq.~\eqref{eq:vecKdF}, where it agrees with the integral to $1.3\times10^{-28}$. Derivatives of the weights give the other monomials. An inverse square of a line momentum replaces the weight of that line by its primitive, an incomplete Beta function, which leads to series of the same kind with more parameters. For these monomials we have derived the form of the series, but we evaluate the integral directly. Equation~\eqref{eq:vecKdF} holds for any pair of first indices, and at $\kappa=\mu$ it requires no limit. In practice we split the integral at $u=\frac12$ and expand in $u$ and in $1-u$, where every series converges as $2^{-k}$. Single moments agree with Eq.~\eqref{eq:vecJpos} to $3\times10^{-26}$ or better at the benchmarks of Sections~\ref{sec:chem.bubble} and~\ref{sec:chem.hard}, for every kind of helicity pair and monomial. The coefficient $\mathcal{M}_A$ assembled from them agrees to $10^{-19}$ or better, which is the accuracy of the radial integral at our working precision. The series loses about four digits at $\mu=\kappa=1$ and about ten at $(\mu,\kappa)=(3,3.75)$, where $26$ digits give $\mathcal{M}_A$ to $10^{-15}$. It is about as fast as the radial integral, not faster.

\paragraph{The triangle at higher orders.} For the triangle of Section~\ref{sec:chem.hard} two things differ from the bubble. The first is the hard line away from the folded point. It carries $\mathbf{k}_1+\bl$, and its vertices have the energies $k_1$ and $k_2$. Equation~\eqref{eq:Dfold} holds for each of its four Schwinger--Keldysh moments, with $x_0=k_1$, $x_1=k_2$, and $y=|\mathbf{k}_1+\bl|$, so every order is a finite sum of the folded moments of Eq.~\eqref{eq:helmoment} at shifted exponents. The projectors of the hard line are expanded in the same way. After the average over the azimuth of $\bl$ around $\mathbf{k}_3$ the soft integrand is a sum of monomials in the two line energies, and each is one soft moment of the form of Eq.~\eqref{eq:vecJpos}. At $\mu=\kappa=1$ the orders $N\le6$ need $2500$ such moments. The expansion is natural in $k_3/k_1$ at fixed $\hat{\mathbf{k}}_1\cdot\hat{\mathbf{k}}_3$, where it converges only for $k_3<k_1$. With $k_3/k_1=R/(1+\frac12R\delta)$ and $\hat{\mathbf{k}}_1\cdot\hat{\mathbf{k}}_3=-(\delta+\frac12R)/(1+\frac12R\delta)$ it becomes Eq.~\eqref{eq:vecDseries}.

The second is a contact term. The time-ordered function of the temporal component of a Proca field is not the ordered product of its modes. The two differ by a term local in time,
\begin{equation}\label{eq:vecContact}
\langle T\,A_0(\tau,\mathbf{p})\,A_0(\tau',-\mathbf{p})\rangle'=\big[\text{ordered product of the modes}\big]+\frac{\ii\,\tau^2}{m^2}\,\delta(\tau-\tau')\,,
\end{equation}
which on the rotated contour is $+(t^2/m^2)\,\delta(t-t')$ for two vertices on the same branch. In the bubble no line joins two hard vertices, and in the triangle the temporal component of the hard line first contributes one order beyond the leading one. This term is thus absent from the leading coefficients of Sections~\ref{sec:chem.bubble} and~\ref{sec:chem.hard}, and it enters the triangle at order $R^2$. On $k_1=k_2$ at $\mu=\kappa=1$ it is $9\%$ of $\mathcal{M}^{\triangle}_2$ and $44\%$ of $\mathcal{M}^{\triangle}_4$. Its presence follows from the limit of a heavy hard line, in which every order must tend to twice the contracted bubble, and the direct integration of Section~\ref{sec:chem.hard} checks its expansion.

\paragraph{The collapsed trispectrum.} The hard pairs of Section~\ref{sec:chem.parity} are built as follows. A hard pair contributes the tensor
\begin{equation}\label{eq:Pitensor}
\Pi_{jk}(a,a')=\delta_{jk}\,Q_{\rm mono}+\big(\hat p_j\hat p_k-\tfrac13\delta_{jk}\big)Q_\Delta-\tfrac{\ii}{2}\,\epsilon_{jkl}\,\hat p_l\,Q_{\rm O}\,,
\end{equation}
where $\hat{\mathbf{p}}$ is $\hat{\mathbf{k}}_1$ or $\hat{\mathbf{k}}_3$ and the moments $Q$ (which depend on the branches $a,a'$ of the two cubic vertices) are the monopole, quadrupole, and parity-odd combinations of the three hard helicities. The diagram with a hard pair on both sides is a box, with coefficient
\begin{multline}\label{eq:Mbox}
\mathcal{M}_\square=\sum_{\lambda_1\lambda_2}\sum_{a_ib_i}g_{\lambda_1}(a_1,b_3)\,g_{\lambda_2}(a_2,b_4)\,\Pi^{(12)}_{jk}(a_1,a_2)\,\Pi^{(34)}_{mn}(b_3,b_4)\\
\times\int\!\frac{\dd^3\ell}{(2\pi)^3}\,(2y_1)^{2\nu}(2y_2)^{2\nu}\,P^{\lambda_1}_{jm}(\hat{\boldsymbol\ell})\,P^{\lambda_2}_{kn}(\widehat{\mathbf{s}-\boldsymbol\ell})\,,
\end{multline}
where $P^{\lambda}$ are the polarization projectors of the soft lines. The loop integral is a sum of the blocks of Eq.~\eqref{eq:C2} with polynomial numerators.

The transverse moments in $\Pi$ are Eq.~\eqref{eq:helmoment}. For the longitudinal line each end carries a time derivative, and moving it onto the vertex weight by parts reduces the ordered moments to the same integrals, together with one equal-time term fixed by the Wronskian $\mathcal{W}[M_{0,\ii\mu},W_{0,\ii\mu}]=-\Gamma(1+2\ii\mu)/\Gamma(\frac12+\ii\mu)$. The cut moments are Gamma functions. The quadrupole of the box is proportional to $\sin^2\alpha_1\sin^2\alpha_3$, and its part odd in $\beta$ vanishes at $\kappa=0$ to $10^{-16}$ numerically. For $\hat{\mathbf{k}}_1$ and $\hat{\mathbf{k}}_3$ perpendicular to $\mathbf{s}$, the ratio $|c_{-2}/c_{+2}|$ is $3$ at $\kappa=0.1$ and $160$ at $\kappa=0.5$ for $\mu=1$, and it saturates near $280$. At $\mu=2$ it reaches $3.5\times10^3$. In the same configuration the crossed box cancels the odd harmonics of the planar box and doubles the even ones. The parity-odd fraction is $0.7$ for $\kappa=2$ and $\mu=1$, and $2.0$ for $\kappa=3$ and $\mu=2$.

\paragraph{Further signals of the collapsed limit.} A line on opposite late-time branches at its two ends is analytic in its momentum for a scalar or a Proca field. For a helical vector at $\kappa\neq0$, however, the part of this line symmetric in the two times carries the weight $1+2n_h$, which differs between the two helicities. Its helicity-odd part then multiplies $\epsilon_{ijk}\hat\ell_k$, which is not analytic at $\bl=0$. One such line gives signals that scale as $s^{3\pm2\ii\mu}$, at half the pair frequency. Two such lines give terms $s^{3+N}$ without oscillation, with a factor $\ln s$ wherever $s^{3+N}$ multiplies a structure analytic in $\mathbf{s}$. None of them is small at $\kappa\simeq\mu$, and at $\mu=\kappa=1$ the non-oscillating monopole is five times that of the single-frequency signal. We have confirmed all three kinds by integrating the bubble between two quartic vertices directly, at $k_{12}=k_{34}$ and $0.01\le s/k_{12}\le0.35$. Fitted with an analytic background, the direct values return each kind with its predicted coefficient to $3\times10^{-3}$ or better, and without the single-frequency term the residual grows by a factor of $160$ or more.

\paragraph{The collapsed series.} The series of Section~\ref{sec:chem.parity} has a closed form at every order. We expand each soft line on its late-time branches at both ends with the series of Eq.~\eqref{eq:en}, and each hard pair around the doubly folded point with the scaling relation of Eq.~\eqref{eq:Dfold}. The temporal components of the lines enter at relative order $s/k$, and the contact term of the hard line at order $s^2/k^2$. Each order is again a finite sum of Gamma functions and ${}_3F_2$'s at unit argument. For equal external momenta, $|\mathbf{k}_i|=k$, the pair-frequency part of Eq.~\eqref{eq:triweights} is
\begin{equation}\label{eq:parityseries}
\lambda_1^2\,\frac{s^3}{k^{8}}\;2\Re\Big[r^{4\ii\mu}\sum_{N\ge0}r^N\sum_{n}c_n^{(N)}\,e^{\ii n\beta}\Big]\,,\qquad r=\frac{s}{k}\,,
\end{equation}
and Table~\ref{tab:parityseries} lists the first coefficients. Harmonics with $|n|=3$ first appear at $N=2$ (they cancel at equal momenta), and those with $|n|=4$ at $N=4$.

\begin{table}[t]
\centering
\begin{tabular}{crrr}
\toprule
 & $N=0$ & $N=1$ & $N=2$\\
\midrule
$10^{2}\,c^{(N)}_{-2}$ & $1.88-2.42\,\ii$ & $-10.55-4.10\,\ii$ & $0.69+18.10\,\ii$\\
$10^{2}\,c^{(N)}_{0}$ & $0.10-0.67\,\ii$ & $-1.54+0.52\,\ii$ & $1.66+1.21\,\ii$\\
$10^{2}\,c^{(N)}_{+2}$ & $0.011+0.008\,\ii$ & $-0.016-0.035\,\ii$ & $-0.002+0.055\,\ii$\\
\midrule
$10^{2}\,d^{(N)}_{-2}$ & $-2.35-2.63\,\ii$ & $-8.11+6.48\,\ii$ & $4.91+10.28\,\ii$\\
$10^{2}\,d^{(N)}_{0}$ & $7.49-4.36\,\ii$ & $-8.60-13.59\,\ii$ & $-8.86+8.87\,\ii$\\
$10^{2}\,d^{(N)}_{+2}$ & $0.14-0.14\,\ii$ & $-0.85+0.11\,\ii$ & $1.01+0.54\,\ii$\\
\bottomrule
\end{tabular}
\caption{Coefficients of the collapsed trispectrum at $\mu=\kappa=1$ and equal external momenta, with $\lambda_2^2/\lambda_1=-2m_A^2$. The $c_n^{(N)}$ are those of Eq.~\eqref{eq:parityseries}, and the $d_n^{(N)}$ are the corresponding coefficients of the single-frequency signals, which multiply $r^{2\ii\mu}$. The odd harmonics vanish in this configuration, and $|n|=4$ first appears at $N=4$.}
\label{tab:parityseries}
\end{table}

\begin{figure}[t]
\centering
\includegraphics{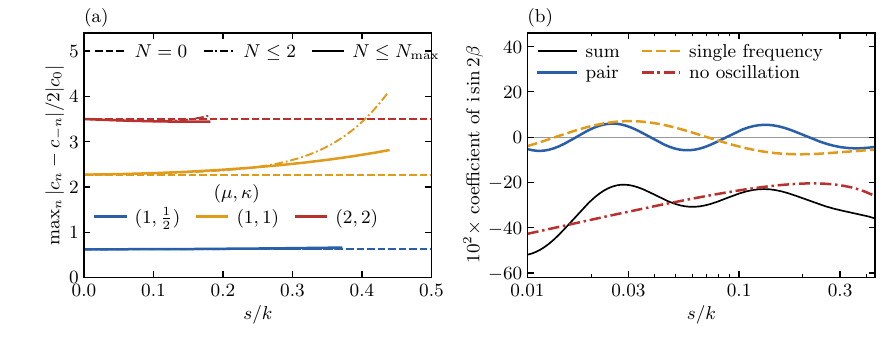}
\caption{The collapsed trispectrum of the helical loop beyond leading order, at equal external momenta. (a) The parity-odd part over the monopole of the pair signal against $r=s/k$, with the series through $N=0$ (dashed), $N\le2$ (dash-dotted), and $N\le N_{\max}$ (solid, with $N_{\max}=6$ at $\mu=\kappa=1$ and $4$ otherwise). Curves end where the last order exceeds three percent of the sum. (b) The coefficient of $\ii\sin2\beta$ in Eq.~\eqref{eq:triweights} at $\mu=\kappa=1$, in units of $\lambda_1^2s^3k^{-8}$, for the pair signal and the single-frequency signal (through $N=6$), and for the non-oscillating terms (through $N=4$, up to terms analytic in $\mathbf{s}$).}
\label{fig:parityseries}
\end{figure}

At $\mu=\kappa=1$ the moduli of $c_{-2}^{(N)}$ are $3.7$, $5.9$, $5.6$, $4.9$, $4.5$, and $3.4$ times that of the leading coefficient for $N=1$ to $6$, and at $\mu=\kappa=2$ they reach $26$ at $N=4$. For the monopole the leading term holds to $10\%$ up to $s=0.043\,k$, $0.105\,k$, and $0.028\,k$ at $(\mu,\kappa)=(1,1)$, $(1,0.5)$, and $(2,2)$. At $\mu=\kappa=1$ the modulus of the dominant harmonic stays within $10\%$ of its leading term up to $s=0.11\,k$. The series itself reaches much further. From $N=3$ on the successive ratios at $\mu=\kappa=1$ lie between $0.76$ and $0.94$, which is consistent with a radius of convergence near $s\simeq k$, where the hard line becomes soft. With the orders $N\le6$ the last one is below one percent of the sum for $s<0.36\,k$ there. With $N\le4$ the same holds for $s<0.28\,k$ at $(1,0.5)$ and for $s<0.14\,k$ at $(2,2)$. Over this range the parity-odd fraction is $2.27$ at leading order, $2.30$ at $r=0.1$, and $2.51$ at $r=0.3$ for $\mu=\kappa=1$ (Fig.~\ref{fig:parityseries}a). At $(1,0.5)$ it rises from $0.62$ to $0.65$ at $r=0.3$, and at $(2,2)$ it is $3.50$ at leading order and $3.45$ at $r=0.1$.

\paragraph{Checks of the collapsed series.} We have checked the series in five ways. First, the term $N=0$ reproduces Eq.~\eqref{eq:Mbox} to $2\times10^{-23}$. Second, exchanging the two legs of a side mixes the orders of the series, and the coefficients obey this exchange to $5\times10^{-24}$ through $N=6$ (and the exchange of the two sides to $10^{-24}$). Third, reversing $\kappa$ exchanges $c_n$ and $c_{-n}$ at every order. Fourth, when the hard line is made heavy, a hard pair tends to $-1/m_h^2$ times the quartic vertex at every order, with corrections that fall as $\mu_h^{-2}$ (for the box they are $1.0\%$ at $\mu_h=40$ and $4.0\%$ at $\mu_h=20$). Fifth, we have integrated the three kinds of diagram directly over the vertex times and the loop momentum, as in Fig.~\ref{fig:vecdirect}, and fitted an analytic background together with the series. For the bubble between two quartic vertices, at $k_{12}=k_{34}$ and $0.01\le s/k_{12}\le0.35$ (41 points, with the series through $N=6$), a free factor on the orders $N\ge1$ returns one to $2\times10^{-3}$ for every background and range we tried. For the diagram with one quartic vertex and one hard pair, at equal external momenta and $0.01\le s/k\le0.25$ (29 points, with the series through $N=4$), the normalization of the direct integral agrees with that of the series to $3\times10^{-5}$. The same factor then returns one to $2\times10^{-4}$ with a background of nine or more powers and logarithms of $s$, and to $2.3\times10^{-3}$ with eight. Without the orders $N\ge1$ the residual of the fits with nine or more terms grows by a factor of $10^{4}$ or more. With a background of even powers of $s$, the single-frequency and non-oscillating terms of this diagram are predictions as well, and their free factors return one to $10^{-4}$ and to $1.4\%$. For the box we integrate over the four vertex times and the loop momentum at equal external momenta and $0.01\le s/k\le0.35$ (32 points at each of two loop cuts, with the pair and single-frequency series through $N=6$). The normalization of the direct integral agrees with that of the series to $5\times10^{-6}$. With a background of the parity of each harmonic, a free factor on the orders $N\ge1$ returns one to $5\times10^{-5}$ for the monopole and to $3\times10^{-5}$ for the harmonic $n=-2$. Without these orders the residual grows by a factor of $10^{4}$ or more. The single-frequency and non-oscillating terms of the box are predictions as well, and their free factors return one to $5\times10^{-5}$ and to $10^{-4}$. The difference between the two loop cuts is a polynomial in $s$ of the parity of the harmonic to $10^{-12}$.

%% file: appDirac.tex
\section{Details of the Dirac loops}\label{app:diracline}

In this appendix we show how a Dirac line enters the representation of Section~\ref{sec:block}, and where its statistics sign comes from. After the spinor is rescaled by $a^{-3/2}$, the two chiral amplitudes of a Dirac fermion of mass $m$ obey a first-order system whose sums and differences are Bessel functions of the complex orders $\nu_\mp=\frac12\mp\ii\mu$, with $\mu=m$ (the $s=\frac12$ case of spin, with no offset)~\cite{ref20,AdsheadSfakianakis2015}. On the rotated contour an internal Dirac end thus contributes $t^2K_{\nu_\mp}(yt)$ in place of the bosonic $t^{3/2}K_{\ii\mu}(yt)$. Since the Barnes form of Eq.~\eqref{eq:GBarnes} holds for every order with $\Re\nu>-1$~\cite{ref26}, it applies without modification to each pair of components of the same order. The Dirac line then enters the representation through the scalar Gamma pair, with arguments shifted by one half,
\begin{equation}\label{eq:diracGamma}
\frac{\Gamma(-s_{v,l})}{\Gamma(\ii\mu+s_{v,l}+1)}\;\longrightarrow\;\frac{\Gamma(-s_{v,l})}{\Gamma\big(\frac32\mp\ii\mu+s_{v,l}\big)}\,,\qquad b_l=1+\nu_\mp+s_{v,l}+s_{v',l}\,.
\end{equation}
Consider a Yukawa vertex with one conformally coupled leg and one external spinor leg ($r_v=\frac32$). The two pieces then have $\rho\in\{-2-\ii\mu,\,-1+\ii\mu\}$, offset by an odd integer plus $2\ii\mu$ rather than by $2\ii\mu$ alone. This half-unit displacement is the entire source of the statistics sign. It flips the monodromy to $+e^{\mp2\pi\mu}$ and makes the cut weights real, $W=\pm2\sinh\pi\mu$. Correspondingly, the time-ordered kernel of the Dirac line has the Fermi--Dirac branch weights $(1-n_F,\,n_F)$, with $n_F=(e^{2\pi\mu}+1)^{-1}$, instead of the $(1+\bar n,\,-\bar n)$ of Eq.~\eqref{eq:GF}. These weights follow from the antisymmetric pairing conserved by the first-order system, together with the identity $I_\nu-(\ii/\pi)e^{\ii\pi\nu}K_\nu=(1-n_F)I_\nu+n_FI_{-\nu}$ at $\nu=\frac12+\ii\mu$.

\paragraph{The soft vertex in closed form.} For a Dirac loop the soft moments of Eq.~\eqref{eq:Jpos} have a closed form. With the Euler integral of each Whittaker function, the time and loop integrals of a monomial are elementary, and two integrals over the Euler parameters $z$ and $\xi$ remain. The first indices of the two soft lines are equal or opposite up to an integer ($K_1\mp K_2=\pm1$ for the triangle, and $K_1=\pm K_2$ for the bubble). The integrand then has only two branch points on each curve $z\xi=x$, and the integral along the curve is an incomplete Beta function. For equal first indices, the monomial $(y_1y_2)^{\nu-3/2}$, and the time weight $t^{-1}e^{-t}$ at the vertex, the moment is
\begin{equation}\label{eq:softclosed}
J=\frac{\Gamma(-1-2\nu)}{2\pi\,\Gamma(c)^2}\int\!\frac{\dd s}{2\pi\ii}\;\tan\frac{\pi s}{2}\;\Gamma(-s)\,\Gamma(2\nu+1+s)\,\frac{\Gamma(c-\frac12+\frac s2)}{\Gamma(c+\frac12+2\nu+\frac s2)}\,,\qquad c=\tfrac12-\nu-K\,,
\end{equation}
with $\nu=\ii\bar\mu$. The contour crosses the real axis between $-1$ and $0$, and it separates the poles of $\Gamma(-s)$ from those of $\Gamma(c-\frac12+\frac s2)$ and $\Gamma(2\nu+1+s)$. Every other monomial is a finite sum of Gamma functions, Gauss functions at argument $-1$, ${}_3F_2$ functions at unit argument, and Barnes integrals of this kind, each of which is a Meijer function $G^{3,4}_{4,4}$ at argument $-1$. As for the closed form of Ref.~\cite{Song2026}, these functions are evaluated on the contour, since their series do not converge at this argument. The moments agree with the radial integral of Eq.~\eqref{eq:Jpos} to $4\times10^{-29}$ or better for every monomial we tested, and the coefficient $C$ obtained from them agrees with Ref.~\cite{Song2026} to $3\times10^{-29}$ at $\kappa=1.3$ and to $9\times10^{-50}$ at $(\kappa,\mu)=(20,2)$. The same reduction applies to two transverse helical lines, to two longitudinal ones, and to a scalar line, for monomials without inverse squares of the line momenta. A transverse line paired with a longitudinal one, and the inverse squares, reduce to Kamp\'e de F\'eriet functions instead (Appendix~\ref{app:helical}).

%% file: appE.tex
\section{Coefficients of the fermion triangle}\label{app:fermicoef}

Table~\ref{tab:fermicoef} lists the coefficients of Eq.~\eqref{eq:fermishape} on $k_1=k_2$ for $\mu=1$, relative to $A_0=C$ (Table~\ref{tab:fermiC}). The coefficients of each power of $\delta$ (through $N=10$ at $\kappa=1.3$ and $5$, and through $N=8$ at $\kappa=10$) are listed in the code of Ref.~\cite{ref28}, which also computes them at other values of $\kappa$ and $\mu$.

\begin{table}[h]
\centering
\begin{tabular}{@{}cccc@{}}
\toprule
$N$ & $\kappa=1.3$ & $\kappa=5$ & $\kappa=10$\\
\midrule
$1$ & $1.02-2.17\,\ii$ & $0.500-5.11\,\ii$ & $0.659-9.99\,\ii$\\
$2$ & $-1.30-2.02\,\ii$ & $-13.0-1.53\,\ii$ & $-49.7-4.22\,\ii$\\
$3$ & $-0.600+0.0422\,\ii$ & $-1.35+21.5\,\ii$ & $-9.21+164\,\ii$\\
$4$ & $0.333-0.437\,\ii$ & $25.3-2.51\,\ii$ & $405-8.84\,\ii$\\
$5$ & $-0.164-0.286\,\ii$ & $-9.83-21.6\,\ii$ & $-120-787\,\ii$\\
$6$ & $0.0572+2.73\times10^{-3}\,\ii$ & $-12.2+15.8\,\ii$ & $-1.23\times10^{3}+430\,\ii$\\
$7$ & $0.0107-0.0856\,\ii$ & $15.7+1.76\,\ii$ & $1.00\times10^{3}+1.53\times10^{3}\,\ii$\\
$8$ & $-8.48\times10^{-3}+2.05\times10^{-3}\,\ii$ & $-5.52-9.83\,\ii$ & $1.45\times10^{3}-1.77\times10^{3}\,\ii$\\
\bottomrule
\end{tabular}
\caption{$A_N/A_0$ on $k_1=k_2$ ($\delta=0$) for $\mu=1$.}
\label{tab:fermicoef}
\end{table}